\documentclass[letterpaper,11pt]{article}
\pdfoutput=1

\usepackage{hyperref}

\usepackage{xspace}
\usepackage{xcolor}
\usepackage{slashed}
\usepackage{array,multirow}
\usepackage{graphicx}
\usepackage{graphics}
\usepackage{epsfig}
\usepackage{amsfonts}
\usepackage{amsmath}
\usepackage{amssymb}
\usepackage{dsfont}
\usepackage{color}
\usepackage{xcolor}
\usepackage{cite}
\usepackage{pdflscape}
\usepackage{pifont}
\usepackage{tikz} 
\usepackage{pgfplotstable}

\newcommand{\al}{\alpha}
\newcommand{\bt}{\beta}
\newcommand{\g}{\gamma}

\newcommand{\dt}{\delta}

\newcommand{\ka}{\kappa}

\newcommand{\vL}{\ensuremath{\mathcal{L}}}    
    \newcommand{\Dt}{\Delta}

\newcommand{\GA}{\Gamma}
\newcommand{\TG}{\tilde\Gamma}

\usepackage{bm}  %

\newcommand{\beq}{\begin{equation}}
\newcommand{\eeq}{\end{equation}}
\newcommand{\bea}{\begin{eqnarray}}
\newcommand{\eea}{\end{eqnarray}}
\newcommand{\ben}{\begin{eqnarray*}}
\newcommand{\een}{\end{eqnarray*}}

\newcommand{\boldsigma}{\mbox{\boldmath $\sigma$}}

\renewcommand{\vec}[1]{{\mathbf #1}} 

\newcommand{\bma}{\begin{pmatrix}}
\newcommand{\ema}{\end{pmatrix}}

\def\lixo#1{}

\def\slashchar#1{\setbox0=\hbox{$#1$}           
  \dimen0=\wd0                                    
  \setbox1=\hbox{/} \dimen1=\wd1                  
  \ifdim\dimen0>\dimen1                           
    \rlap{\hbox to \dimen0{\hfil/\hfil}}            
    #1                                             
  \else                                          
    \rlap{\hbox to \dimen1{\hfil$#1$\hfil}}        
    /                                           
 \fi}                                           %

\newcommand{\Or}{\mathcal O}

\newcommand{\vp}{\varphi}
\newcommand{\sq}{^{2}}

\newcommand\lsim{\lesssim}

\newcommand{\dslash}[1]{#1 \llap{/\kern-0.5pt}}
\newcommand{\Dslash}[1]{#1 \llap{/\kern+1.5pt}}
\newcommand{\DDslash}[1]{#1 \llap{/\kern+2.3pt}}
\newcommand{\dslashh}[1]{#1 \llap{/\kern+1pt}}

\newcommand{\nn}{\nonumber}

\newcommand{\NLDBD}{$0 \nu \beta \beta$}
\newcommand{\textoverline}[1]{$\overline{\mbox{#1}}$}

\definecolor{cadmiumgreen}{rgb}{0.0, 0.42, 0.24}
\definecolor{darkpastelgreen}{rgb}{0.01, 0.75, 0.24}
\definecolor{darkspringgreen}{rgb}{0.09, 0.45, 0.27}
\definecolor{forestgreen(web)}{rgb}{0.13, 0.55, 0.13}
\definecolor{forestgreen(traditional)}{rgb}{0.0, 0.27, 0.13}
\definecolor{cobalt}{rgb}{0.0, 0.28, 0.67}
\definecolor{darkblue}{rgb}{0.0, 0.0, 0.75}
\definecolor{darkred}{rgb}{0.55, 0.0, 0.0}
\definecolor{palatinatepurple}{rgb}{0.41, 0.16, 0.38}
\definecolor{burntorange}{rgb}{0.8, 0.33, 0.0}

\begin{document}

\begin{titlepage}

\begin{flushright}
LA-UR-18-24895\\
Nikhef 2018-023
\end{flushright}

\vspace{2.0cm}

\begin{center}
{\LARGE  \bf 
A neutrinoless double beta decay master formula\\ 
\vspace{3mm}
from effective field theory
\vspace{3mm}
}
\vspace{2cm}

{\large \bf  V. Cirigliano$^a$, W. Dekens$^{a,b}$, J. de Vries$^{c}$, \\ \vspace{3mm}M. L. Graesser$^a$, and E. Mereghetti$^a$ } 
\vspace{0.5cm}

\vspace{0.25cm}

{\large 
$^a$ 
{\it Theoretical Division, Los Alamos National Laboratory,
Los Alamos, NM 87545, USA}}

\vspace{0.25cm}
{\large 
$^b$ 
{\it 
New Mexico Consortium, Los Alamos Research Park, Los Alamos, NM 87544, USA
}}

\vspace{0.25cm}
{\large 
$^c$ 
{\it 
Nikhef, Theory Group, Science Park 105, 1098 XG, Amsterdam, The Netherlands
}}

\end{center}

\vspace{0.2cm}

\begin{abstract}
\vspace{0.1cm}

We  present a master formula describing the neutrinoless-double-beta decay 
(\NLDBD) 
rate  induced by 
lepton-number-violating  (LNV) operators up to dimension nine in the Standard Model Effective Field Theory. 
We provide an end-to-end framework connecting the possibly very high LNV scale to the nuclear scale, through a chain of effective field theories. Starting at the electroweak scale, we integrate out the heavy Standard Model degrees of freedom and we match to an $SU(3)_c\otimes U(1)_{\mathrm{em}}$ effective theory.
 After evolving the resulting effective Lagrangian to the QCD scale, we use chiral perturbation theory to derive the 
   lepton-number-violating chiral Lagrangian. 
 The chiral Lagrangian is used to derive the two-nucleon 
  \NLDBD\  
 transition operators 
 to leading order in the chiral power counting. Based on renormalization arguments we show that in various cases short-range 
 two-nucleon 
 operators need to be enhanced to leading order. 
 We show that all required nuclear matrix elements can be taken from existing calculations.  
 Our final result is a master formula  that describes 
the  \NLDBD\  rate  
in terms of phase-space factors, nuclear matrix elements,   hadronic low-energy constants, QCD evolution factors, and high-energy  LNV 
Wilson coefficients, 
including all the interference terms. 
 Our master formula can  be easily matched to any model where  
LNV 
originates at energy scales above the electroweak scale. 
 As an explicit example, we match our formula to the minimal left-right-symmetric model in which 
contributions  of operators of different dimension compete, 
and we discuss the resulting phenomenology.

\end{abstract}

\vfill
\end{titlepage}

\tableofcontents

\section{Introduction}
Neutrinoless double beta decay (\NLDBD) is the process where two neutrons inside an atomic nucleus are transmuted into two protons and two electrons without the emission of neutrinos. An observation of this process would indicate that lepton number (L) is not a good symmetry of nature and that the neutrino mass  has a Majorana component, implying that the mass eigenfields are self-conjugate. Current experimental limits are very stringent \cite{Gando:2012zm,Agostini:2013mzu,Albert:2014awa,Andringa:2015tza,KamLAND-Zen:2016pfg,Elliott:2016ble,Agostini:2017iyd,Aalseth:2017btx, Albert:2017owj,Alduino:2017ehq,Agostini:2018tnm, Azzolini:2018dyb}, e.g. $T^{0\nu}_{1/2}>1.07\times 10^{26}$yr for ${}^{126}$Xe
\cite{KamLAND-Zen:2016pfg},
with next-generation ton-scale experiments aiming for one to two orders of magnitude improvement.

The simplest
interpretation of $0 \nu \beta \beta$ experiments assumes that lepton-number violation (LNV) is due to the exchange of light Majorana neutrinos. 
However in various beyond-the-SM (BSM) scenarios other sources of LNV exist that can induce $0\nu\beta\beta$. For example, in left-right symmetric models, apart from the exchange of a light Majorana neutrino, there appear contributions  from the exchange of heavy neutrinos and charged scalars. 
While a single nonzero $0 \nu \beta \beta$ measurement can be attributed to any LNV interaction, 
in principle various LNV sources  can be disentangled by measurements of different isotopes, the angular or energy distributions of the outgoing electrons, or by correlating with collider observables, 
provided sufficient theoretical control can be achieved. 

In most BSM scenarios the LNV source responsible for $0\nu\beta\beta$ is induced at an energy scale $\Lambda$ well above the electroweak scale. This scale separation justifies an effective field theory (EFT) approach. Such an approach has the advantage that $0\nu\beta\beta$ and its correlation with collider observables can be described in a model-independent fashion. The Standard Model can be seen as the renormalizable part of an EFT that includes higher-dimensional operators which are suppressed by powers of the scale of BSM dynamics, $\Lambda$. Within this EFT, the $\Delta L=2$ operators have odd dimension \cite{Kobach:2016ami}. The first of $\Delta L=2$ term therefore appears at dimension five and provides a contribution to the neutrino Majorana mass \cite{Weinberg:1979sa}. In the standard type-I see-saw mechanism this dimension-five operator arises from integrating out heavy right-handed neutrinos at the typical GUT-scale $\Lambda \sim 10^{15}$ GeV. LNV operators with a dimension $> 5$ are then suppressed by powers of $v/\Lambda \simeq 10^{-13}$, where $v\simeq 246$ GeV is the Higgs vacuum expectation value, and can be safely neglected. In various models, however, the scale of  LNV new physics is much lower and the dimension-five operator  may be suppressed by loop factors and/or  small Yukawa couplings. For instance, in the above-mentioned left-right symmetric models the dimension-five operator  scales as $y^2 /\Lambda$, where $y$ is a Yukawa coupling  scaling as $y \sim m_e/v \sim 10^{-6}$. While dimension-seven and -nine LNV operators  are suppressed by additional powers of $\Lambda$, they can be suppressed by only one or even zero powers of $y$. As such, the dimension-seven and -nine operators can be competitive with the dimension-five operator, for $\Lambda$ in the $1-10$ TeV range.
Higher-dimensional LNV operators at the multi-TeV scale can be generated in radiative neutrino models \cite{Zee:1980ai, Zee:1985id, Babu:1988ki, Babu:1988ig, Babu:1988wk, Babu:2001ex} and have also been studied in Refs.\ \cite{Babu:2001ex,Prezeau:2003xn,deGouvea:2007qla,Lehman:2014jma,Graesser:2016bpz}.

In order  to describe $0\nu\beta\beta$ in a model-independent way, one should include all LNV operators up to dimension-nine in the SM-EFT\footnote{We are not aware of models where operators of dimension eleven or higher are competitive with lower-dimensional operators.}. In this work, we consider these operators and derive the form of the $SU(3)_c\times U(1)_{\rm em}$-invariant $\Delta L=2$ Lagrangian at a scale of a few GeV. We then construct the resulting chiral EFT Lagrangian that describes $\Delta L=2$ interactions between the low-energy degrees of freedom: pions, nucleons, electrons, and neutrinos. 
Armed with the chiral Lagrangian, we calculate the leading-order (LO) \NLDBD\  transition operator  or ``\NLDBD\ potential"  using a consistent power counting that was introduced in Refs.~\cite{Cirigliano:2017djv,Cirigliano:2017tvr,Cirigliano:2018hja}. The \NLDBD\  transition operator is the basis of many-body calculations of the $0\nu\beta\beta$ amplitude. We identify the nuclear matrix elements (NME) that need to be calculated in order to obtain the LO $0\nu\beta\beta$ decay rate for $0^+ \rightarrow 0^+$ transitions. Somewhat remarkably we find that the set of NMEs necessary to describe the LO $0\nu\beta\beta$ rate arising from LNV sources up to dimension nine is the same as the one required to describe the light
and heavy Majorana neutrino exchange contributions. As such, the required NMEs have already been studied extensively in the literature and we can compare results from different many-body methods.

Although the power counting employed here is consistent with (nonperturbative)
renormalization in the two-body sector, it has not been tested in the large nuclei
of experimental interest. To do so in a fully consistent manner, one would have to combine
the transition operators derived in this work with many-body wavefunctions obtained from
chiral potentials that are consistently renormalized. This has not been achieved for systems with more 
than a few nucleons. An intermediate approach is to use chiral wave functions from ab initio calculations \cite{Hagen:2013nca,Hagen:2013yba,Hagen:2016xjv} and the insertion of the neutrino transition operators derived in this work \cite{Pastore:2017ofx}. Still, it is not guaranteed that after renormalizing the strong interactions the 
neutrino transition matrix elements are correctly renormalized \cite{Cirigliano:2018hja}. Even this approach is limited to relatively light nuclei, but there is
a large ongoing effort to increase the reach to larger and denser systems which would allow for 
better tests of the chiral power counting. As the ab initio methods fall short for nuclei of experimental 
interests, we will use results obtained from several non-chiral many-body methods \cite{Hyvarinen:2015bda,Menendez:2017fdf,Barea:2015kwa,Barea} to estimate limits on the LNV operators.

\begin{figure}
\begin{center}
\includegraphics[trim= {2.8cm 1.8cm 3.6cm 1.2cm}, clip, width=\textwidth]{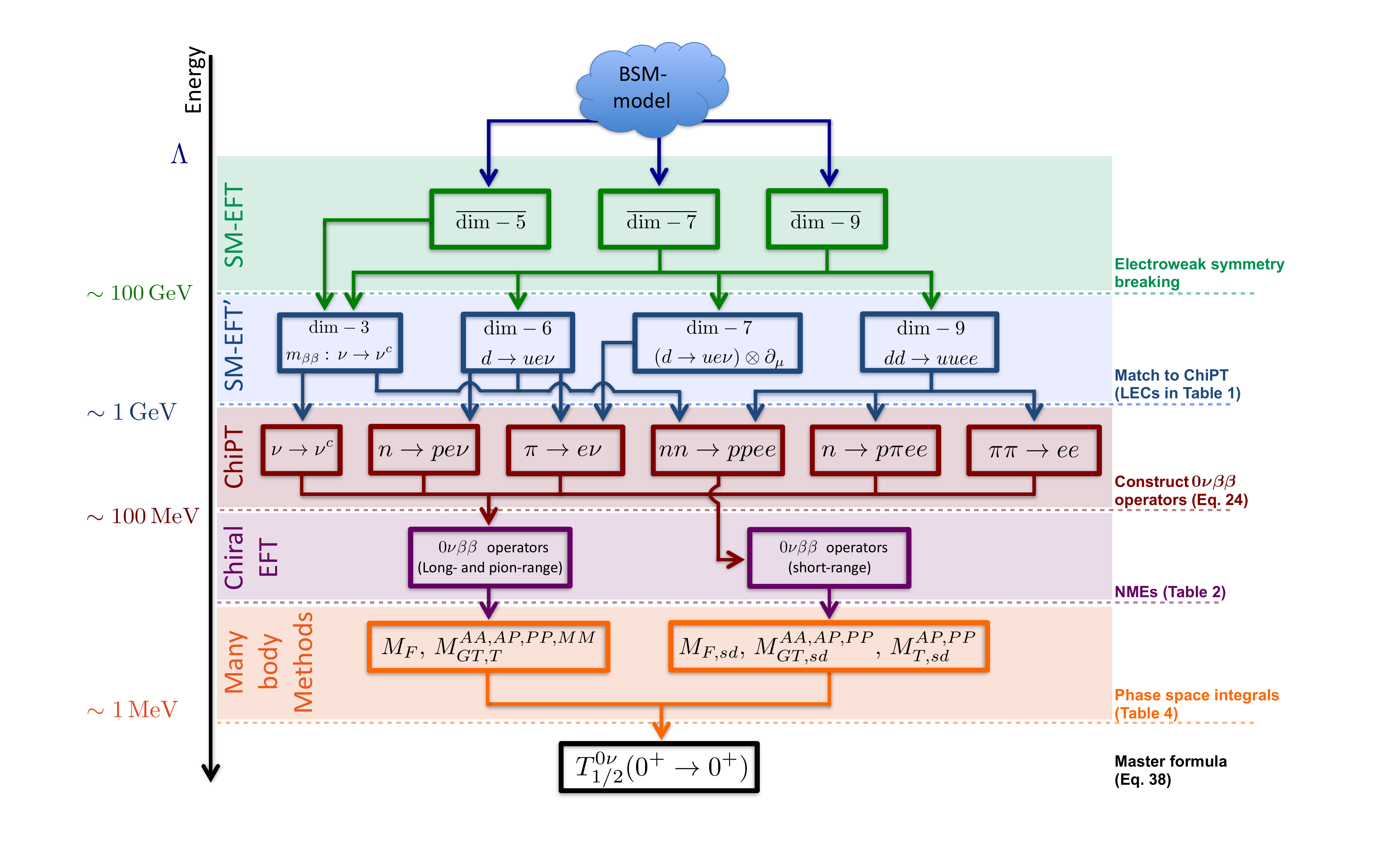}
\end{center}
\caption{A schematic overview of the effective field theory approach to evaluating 
 the $0\nu \beta \beta$-decay amplitude starting from high-scale  $\Delta L = 2$ dynamics. The different colors represent various effective field theories at different scales. See the main text for more details.   
 }
\label{landscape}
\end{figure}

Our main result is   a so-called ``Master formula"  for the \NLDBD\  decay rate which includes the contributions from all LNV sources up to dimension nine and takes into account all possible interference terms. The formula incorporates the matching and evolution of the effective LNV operators that appear at the electroweak  scale all the way to the final $0\nu\beta\beta$ decay rate by the application of several different EFTs, illustrated in Fig.~\ref{landscape}. This Master formula can be used to directly calculate the $0\nu\beta\beta$ decay rates for different isotopes in any given particle physics model after matching to the  SM-EFT  at the electroweak scale.  Our analysis also highlights the role played by uncertainties in hadronic and nuclear matrix elements, and we discuss how this is reflected on the limits we can set on BSM models. 

To illustrate the usefulness of the Master formula we study a specific high-energy model, the left-right symmetric model. We show that, after performing a matching calculation to the SM-EFT operators, the master formula can be directly used to set constraints on the model parameters. In particular, we investigate for what values of the right-handed scale the dimension-seven and -nine operators are competitive with respect to the dimension-five contributions.

In order to keep this paper somewhat compact, we heavily borrow from  
previous work that described in detail various technical aspects of the derivation \cite{Cirigliano:2017djv,Cirigliano:2017tvr,Cirigliano:2018hja}. In that work, we used the same SM-EFT and chiral EFT techniques but limited ourselves to LNV dimension-five and -seven operators. Here we extend the analysis to dimension-nine and derive the \NLDBD\  transition operators at LO in the chiral expansion. Nevertheless, the current paper is self-contained and all information to use the Master formula is given in the main text and accompanying appendices.

The organization of the paper is perhaps best described by referring to Fig.~\ref{landscape}. 
The prediction for the $0 \nu \beta \beta$ decay rate is  
obtained through  a sequence of effective field theories, in order to separate short-distance 
particle physics effects which can be treated perturbatively, from long-distance hadronic and nuclear physics that must be evaluated non-perturbatively. 
Here are the key elements:
\begin{itemize} 

\item   We assume  throughout that lepton number violation occurs at a scale $\Lambda \gg m_W$. This scale is shown in \textcolor{forestgreen(web)}{\textbf{green}} in Fig.~\ref{landscape}. At this scale there are a number of    $SU(3)_c\times  SU(2)_{L} \times U(1)_{Y}$-invariant   $\Delta L=2$  operators, beginning at dimension five (the usual Weinberg operator). In this paper we also consider  gauge-invariant  $\Delta L=2$ operators  in the SM-EFT with dimension seven and nine. 
We denote the  dimension of  SM-EFT operators  by \textoverline{dim-n}  with $n=5,7,9$.

\item At scales below $m_W$, after electroweak-symmetry breaking, we match to a new effective field theory in which $\Delta L=2$ manifests via operators of different dimension than the original \textoverline{dim-5}, \textoverline{-7}, \textoverline{-9} operators. This happens because once the Higgs field obtains its vacuum expectation value (vev) and the Higgs and $W$ boson are integrated out of the EFT, we generate at tree-level new dimension-3, -6 ,-7, and -9 operators. We evolve these operators to the GeV scale by considering the one-loop QCD renormalization of such operators, which captures the leading-order perturbative operator mixing and renormalization. This step is shown in {\color{darkblue} \textbf{blue}} in  Fig.~\ref{landscape}. 

\item The {\color{forestgreen(web)} \textbf{green}} and {\color{darkblue} \textbf{blue}} steps are described in Section \ref{sec:2}. The QCD renormalization group equations are given in Section \ref{sec:2} and solved in Appendix \ref{App:RG}.

\item The next step, shown in {\color{darkred} \textbf{red}}, occurs at the GeV scale, where the quark operators are matched to chiral EFT, describing the interactions of neutrons, protons, pions, and leptons. Here a number of effects occur: 

\begin{itemize} 
\item The dimension-3 Majorana neutrino mass generates the standard neutrino potential arising from $\nu \rightarrow \nu^c$ (antineutrino conversion into neutrino)
but also a contact operator $nn \rightarrow pp ee$. Both effects occur at leading order in the power counting~\cite{Cirigliano:2018hja}.    

\item Dimension-6 operators generate new interactions leading to $\Delta L=2$ beta decay, $n \rightarrow p e \nu$, and pion decay, $\pi \rightarrow e \nu$. When these are combined with the SM weak interactions, we find additional contributions to the  $nn \rightarrow pp ee$ transition operator. 
Renormalization also requires the introduction of  contact  $nn \rightarrow pp ee$ operators, discussed in Appendix \ref{app:neutrinopotentials567}. 

\item Dimension-7 operators can also generate $\Delta L=2$ pion decay ($\pi \rightarrow e \nu$)
 and consequently the $nn \rightarrow pp ee$ transition operator. 

\item Dimension-9 operators can generate both short-range contact operators, $nn \rightarrow pp ee$, as well as pion-range, $n \rightarrow p \pi ee$ and $\pi \pi \rightarrow ee$, interactions. When the pion-range interactions are combined with the strong interactions we generate additional contributions to the $nn \rightarrow pp ee$ transition operator. The three 
processes are shown in Figure~\ref{feyndiag}.

\end{itemize}

These effects are described in Section~\ref{ChiPT}. The hadronic input (low energy constants, or ``LECs'') required for our analysis is provided in Table \ref{Tab:LECs}. An important hadronic effect  is the non-perturbative renormalization of the Majorana neutrino mass \cite{Cirigliano:2018hja} and higher-dimension operators (see Figure \ref{AmpDim9}), due to the short-range nature of the nucleon-nucleon force. These effects 
are discussed in Appendix~\ref{NonWeinberg} and the expectations for the values of the new LECs are summarized in Table \ref{Tab:LECs}.

\item All these effects at the $\mathcal O(10-100 \hbox{ MeV})$ scale combine to form the $nn \rightarrow pp ee$ transition operator, to be used in  many-body 
nuclear calculations.    This step is shown in {\color{palatinatepurple} \textbf{purple}}   in  Fig.~\ref{landscape}. 

\item The next step,  shown in {\color{burntorange} \textbf{orange}} in  Fig.~\ref{landscape},  
is to evaluate the $0 \nu \beta \beta$ transition operators between the ground states of the relevant nuclei. 
The nuclear many-body matrix elements (see Table \ref{tab:comparison} and Appendix \ref{NME}), together with the phase space factors (see Table \ref{Tab:phasespace} and 
Appendix \ref{Phase}), and short-distance Wilson coefficients, combine to give the $0^+ \rightarrow 0^+$ decay rate. This final step is described in Section \ref{Master}.

\item Section~\ref{single} then applies our Master formula to obtain bounds on the {dim-3}, {-6},{-7}, and {-9} operators,   
assuming only a single operator dominates at a time. In Section~\ref{sec:LR} we consider an explicit example of a UV-complete model, namely the left-right symmetric model \cite{Pati:1974yy, Mohapatra:1974hk, Senjanovic:1975rk}, to illustrate the utility of our general formalism. In this model several operators of different dimension are generated. Further details on the left-right model (LRM) are provided in Appendix~\ref{App:LR}. 

\item 
In Section~\ref{LR-lightNuR} we 
discuss the case in which the right-handed neutrinos in the left-right symmetric model have masses close to the GeV scale. 
 Although the assumption $\Lambda\gg v$ no longer applies in this case, we show that the EFT framework can be straightforwardly  generalized to include LNV arising at intermediate scales.

\item We summarize our results and discuss future directions in Section~\ref{sec:conclude}. 

\end{itemize}

Finally, before diving into the details of our analysis, let us  comment on the relation of our work with prior literature.   First of all, the framework discussed here is similar in spirit to Refs.\ \cite{Prezeau:2003xn,Graesser:2016bpz}, whose results we generalize and extend here.
Secondly, a master formula for $0 \nu \beta \beta$ has previously appeared in the literature~\cite{Pas:2000vn,Pas:1999fc}, 
and it has been used in many subsequent studies. 
We note that we disagree with Refs.~\cite{Pas:2000vn,Pas:1999fc} in the treatment of the hadronization of quark operators. In particular, 
Ref.~\cite{Pas:2000vn} misses important hadronic physics, thereby underestimating the contributions of certain Wilson coefficients to the $0 \nu  \beta \beta$ {\em amplitude} by $\Or(16\pi\sq)$ compared to what we find here. One main difference is that Ref.\ \cite{Pas:2000vn} ignores the couplings to $\pi \pi$ -- which we now know with a fair amount of certainty \cite{Savage:1998yh,Nicholson:2016byl, Cirigliano:2017ymo,Nicholson:2018mwc} -- and to $\pi N$. The LECs of certain four-nucleon operators are also underestimated by $\Or(16\pi\sq)$, because non-perturbative renormalization is not considered. We further discuss these and other differences with Ref.\ \cite{Pas:2000vn} in Appendix \ref{comparison-to-lit}.

\section{Lepton number violation in the SM-EFT}\label{sec:2}
Lepton number is an accidental symmetry of the renormalizable part of the SM, which is violated by higher-dimensional operators. The $\Delta L=2$ operators relevant for $0\nu\beta\beta$ all have odd dimension \cite{Kobach:2016ami} and we focus on dimension-five, -seven, and -nine operators that, respectively, scale as $\Lambda^{-1}$, $\Lambda^{-3}$, and $\Lambda^{-5}$, where $\Lambda$ is the scale at which lepton number violation arises. At lower energies, after electroweak symmetry breaking (EWSB) and integrating out heavy SM fields (top, Higgs-, W-, and Z-bosons) the arising effective operators can have a different canonical dimension due to positive powers of  the Higgs vacuum expectation value, $v\simeq 246$ GeV (the SM-EFT approach assumes $\Lambda \gg v$). In particular, at energies around a few GeV the most important $\Delta L=2$ operators have canonical dimension three, six, seven, and nine. 
To avoid confusion, when discussing the original gauge-invariant SM-EFT $\Delta L=2$ operators, we denote their dimensions by \textoverline{dim-n}  with $n=5,7,9$. When discussing the operators after EWSB, which are only $SU(3)_c\times U(1)_{\rm em}$ invariant, we refer to them as dim-n operators (without the overline) where $n=3,6,7,9$. 

The structure of the gauge-invariant $\Delta L=2$ operators has been discussed in great detail in the literature \cite{Weinberg:1979sa,Babu:2001ex,Prezeau:2003xn,deGouvea:2007qla,Lehman:2014jma,Graesser:2016bpz}. The only \textoverline{dim-5} operator is the Weinberg operator \cite{Weinberg:1979sa} which, after EWSB, gives rise to the neutrino Majorana mass.  
The 12  operators that appear at \textoverline{dim-7} have been classified in Ref.~\cite{Lehman:2014jma} and were studied in the context of $0\nu\beta\beta$ in  detail in Ref.~\cite{Cirigliano:2017djv}. The complete set of \textoverline{dim-9} operators is currently unknown, but certain subclasses with particular field content have been constructed \cite{deGouvea:2007qla,Graesser:2016bpz}. For instance, Ref.~\cite{Graesser:2016bpz} identified the \textoverline{dim-9}  operators consisting of four quark fields and two electron fields, finding that only eleven such operators exist. However, additional classes of \textoverline{dim-9}  operators can be constructed that give rise to unsuppressed $0\nu\beta\beta$. Examples are operators involving two quark fields, two electron fields, and the combination of Higgs fields and derivatives $\tilde{\varphi}^{\dagger} D_\mu \varphi$. While these operators require the exchange of a $W$ boson to induce $0\nu\beta\beta$, the associated factor of $G_F$ is compensated by two powers of $v$ arising from the two $\varphi$ fields after EWSB. 

Here we do not list the gauge-invariant \textoverline{dim-n} operators but  refer to Refs.~\cite{Weinberg:1979sa,Babu:2001ex,Prezeau:2003xn,deGouvea:2007qla,Lehman:2014jma,Graesser:2016bpz,Cirigliano:2017djv} for more details. Instead we focus on the $\Delta L = 2$ Lagrangian after EWSB and integrating out the heavy SM degrees of freedom. At a scale slightly below the electroweak scale, the Lagrangian consists of $SU(3)_c \times  U(1)_{\rm em}$ operators of increasing dimension.
For applications to $0\nu\beta\beta$ it is convenient to organize the Lagrangian in operators that violate the number of charged leptons by zero, one, or two units
\begin{equation}\label{LagDeltaL2}
\mathcal L_{\Delta L = 2} = \mathcal L_{\Delta e = 0} +  \mathcal L_{\Delta e = 1} + \mathcal L_{\Delta e = 2}\,.  
\end{equation}
$\mathcal L_{\Delta e = 0}$ contains operators that violate lepton number in the neutrino sector, starting from the dim-3 Majorana mass of left-handed neutrinos
\begin{equation}\label{LagDeltaE0}
\mathcal L_{\Delta e = 0}   = - \frac{1}{2} (m_{\nu})_{ij} \, \nu^{T}_{L,\, i} C \nu_{L,\, j}  + \ldots
\end{equation}
The $SU(2)_L \times U(1)_Y$ invariance of the SM implies that the first contribution to $m_\nu$ arises from a \textoverline{dim-5} operator, such that $m_\nu \sim v^2/\Lambda$. 
The dots in Eq.\ \eqref{LagDeltaE0} denote operators of higher dimension, such as the dim-5 neutrino magnetic moment or dim-6 LNV neutral-current semileptonic operators. 
In order to induce $0\nu\beta\beta$, the two neutrinos in the operators in $\mathcal L_{\Delta e = 0}$ need to be converted into electrons via the SM weak interaction. 
The contributions to $0\nu\beta\beta$ from higher-dimensional operators in Eq.\ \eqref{LagDeltaE0} are thus  
suppressed 
at least 
by powers of
 $\Lambda_\chi^2 / v^2$
(if not  $m_\pi^2 / v^2$), 
 where $\Lambda_\chi \sim 1$ GeV is the 
chiral-symmetry-breaking scale~\cite{Manohar:1983md}, with respect to $m_\nu$.  
We therefore neglect these effects in this work.

A richer set of contributions
arises from $\mathcal L_{\Delta e = 1}$.  This Lagrangian contains  LNV operators with one charged lepton and one neutrino field.
In order to compensate the charge of the electron field, one needs at least an additional quark or lepton bilinear, making dim-6 the minimal dimension of these operators:
\begin{equation}\label{LagDeltaE1}
\mathcal L_{\Delta e = 1} = \mathcal L^{(6)}_{\Delta L = 2} + \mathcal L^{(7)}_{\Delta L = 2} + \ldots
\end{equation}
The operators most relevant to $0\nu\beta\beta$ are semileptonic four-fermion operators. At dim-6 we have  
\bea
\mathcal L^{(6)}_{\Delta L = 2}& =& \frac{2 G_F}{\sqrt{2}} \Bigg(
C^{(6)}_{\textrm{VL},ij} \,  \bar u_L \gamma^\mu d_L \, \bar e_{R,i} \,  \gamma_\mu \, C\bar \nu^T_{L,j} + 
C^{(6)}_{\textrm{VR},ij} \,  \bar u_R \gamma^\mu d_R \, \bar e_{R,i}\,  \gamma_\mu  \,C\bar\nu_{L,j}^T \label{lowenergy6}   \\
&  +& \!\!\!\!
C^{(6)}_{ \textrm{SR},ij} \,  \bar u_L  d_R \, \bar e_{L,i}\, C  \bar \nu^T_{L,j} + 
C^{(6)}_{ \textrm{SL},ij} \,  \bar u_R  d_L \, \bar e_{L,i} \, C  \bar\nu_{L,j}^T + 
C^{(6)}_{ \textrm{T},ij} \,  \bar u_L \sigma^{\mu\nu} d_R \, \bar e_{L,i}  \sigma_{\mu\nu}  \, C\bar\nu_{L,j}^T
\Bigg)  +{\rm h.c.}\nn
\eea
$\mathcal L^{(6)}_{\Delta L=2}$ contains all possible $\Delta L = 2$ dim-6 charged-current operators. At tree level, all operators in Eq.\ \eqref{lowenergy6} receive their first contributions from \textoverline{dim-7} operators \cite{Cirigliano:2017djv}, so that $C^{(6)}_i = \mathcal O(v^3/\Lambda^3)$. Beyond tree level, the operators in Eq.\ \eqref{lowenergy6} might also receive contributions of $\mathcal O(v/\Lambda)$ from the neutrino Majorana mass, but we neglect these loop corrections here.
Dim-7 operators in $\mathcal L_{\Delta e = 1}$ give rise to corrections that are suppressed by $\Lambda_\chi/v$ with respect to the dim-6 terms of  Eq.\ \eqref{lowenergy6}.
Here we consider only  the  subset of  $SU(3)_c\times U(1)_{\rm em}$ invariant dim-7 operators that receive tree-level matching coefficients at the EW scale from \textoverline{dim-7} operators \cite{Cirigliano:2017djv}
\bea
\mathcal L^{(7)}_{\Delta L = 2} &=& \frac{2 G_F}{\sqrt{2} v} \Bigg( 
C^{(7)}_{\textrm{VL},ij} \,  \bar u_L \gamma^\mu d_L \, \bar e_{L,i} \, C \,  i \overleftrightarrow{\partial}_\mu \bar \nu_{L,j}^T  +
C^{(7)}_{\textrm{VR},ij} \,  \bar u_R \gamma^\mu d_R \, \bar e_{L,i} \, C i \overleftrightarrow{\partial}_\mu \bar \nu^T_{L,j}  \Bigg)  +{\rm h.c.}\label{lowenergy7}
\eea
The coefficients of these operators  scale as $C^{(7)}_i = \mathcal O(v^3/\Lambda^3)$.  
Operators of higher dimension in Eq. \eqref{LagDeltaE1}, such as dim-8 dipole operators or dim-9 charged-current six-fermion operators, give rise to contributions that are more and more suppressed by powers of $\Lambda_
\chi/v$ and $v/\Lambda$.

The final class of operators are LNV operators with two electrons, which can directly contribute to $0\nu\beta\beta$ without additional SM weak interactions.
$U(1)_{\rm em}$ invariance forces these operators to be at least dim-9
\begin{equation}\label{LagDeltaE2}
\mathcal L_{\Delta e = 2}   =  \mathcal L^{(9)}_{\Delta L=2} + \ldots
\end{equation}
The set of $SU(3)_c\times U(1)_{\rm em}$ invariant four-quark two-lepton operators can be written as \cite{Graesser:2016bpz,Prezeau:2003xn}
\bea \label{eq:Lag}
\vL^{(9)}_{\Delta L =2} = \frac{1}{v^5}\sum_i\bigg[\left( C^{(9)}_{i\, \rm R}\, \bar e_R C \bar e^T_{R} + C^{(9)}_{i\, \rm L}\, \bar e_L C \bar e^T_{L} \right)  \, O_i +  C^{(9)}_i\bar e\g_\mu\g_5  C \bar e^T\, O_i^\mu\bigg],
\eea
where $O_i$ and  $O_i^\mu$ are four-quark operators that are Lorentz scalars and vectors, respectively. The scalar operators have been discussed in Refs.\ \cite{Graesser:2016bpz,Prezeau:2003xn} and can be written as
\bea\label{LagSca}
O_ 1  &=&  \bar{q}_L^\alpha  \gamma_\mu \tau^+ q_L^\alpha \ \bar{q}_L^\beta  \gamma^\mu \tau^+ q_L^\beta\,, \qquad O^\prime_ 1  =  \bar{q}_R^\alpha  \gamma_\mu \tau^+ q_R^\alpha \ \bar{q}_R^\beta  \gamma^\mu \tau^+ q_R^\beta      
\,\,,
\\
O_ 2  &=&  \bar{q}_R^\alpha  \tau^+ q_L^\alpha \  \bar{q}_R^\beta  \tau^+ q_L^\beta\,, \qquad \qquad     O^\prime_ 2  =  \bar{q}_L^\alpha  \tau^+ q_R^\alpha \  \bar{q}_L^\beta  \tau^+ q_R^\beta
\,\,,\\
O_ 3  &=&  \bar{q}_R^\alpha  \tau^+ q_L^\beta \  \bar{q}_R^\beta  \tau^+ q_L^\alpha\,, \qquad \qquad    O^\prime_ 3  =  \bar{q}_L^\alpha  \tau^+ q_R^\beta \  \bar{q}_L^\beta  \tau^+ q_R^\alpha
\,\,,\\
O_ 4  &=&  \bar{q}_L^\alpha  \gamma_\mu \tau^+ q_L^\alpha \  \bar{q}_R^\beta  \gamma^\mu \tau^+ q_R^\beta    
\,\,,\\
O_ 5  &=&  \bar{q}_L^\alpha  \gamma_\mu \tau^+ q_L^\beta \  \bar{q}_R^\beta  \gamma^\mu \tau^+ q_R^\alpha\,,
\eea
where $\tau^\pm = (\tau_1\pm i\tau_2)/2$ with $\tau_i$ the Pauli matrices and $\al$, $\bt$ are color indices. The $O_i'$ operators are related to the $O_i$ by parity. The vector operators take the form \cite{Graesser:2016bpz}
\bea \label{LagVec}
O_{6}^{\mu} &=& \left(\bar q_L \tau^+\g^\mu q_L\right)\left(\bar q_L \tau^+ q_R\right)\,\,,\qquad\qquad 
O_{6}^{\mu\, \prime} = \left(\bar q_R \tau^+\g^\mu q_R\right)\left(\bar q_R \tau^+ q_L\right)\,\,,
\nn\\
O_{7}^{\mu} &=& \left(\bar q_L t^a\tau^+\g^\mu q_L\right)\left(\bar q_L t^a\tau^+ q_R\right)\,\,,\qquad  \,\,
O_{7}^{\mu\, \prime} = \left(\bar q_R t^a\tau^+\g^\mu q_R\right)\left(\bar q_R t^a\tau^+ q_L\right)\,\,
,\nn\\
O_{8}^{\mu} &=& \left(\bar q_L \tau^+\g^\mu q_L\right)\left(\bar q_R \tau^+ q_L\right)\,\,,\qquad \qquad 
O_{8}^{\mu\, \prime} = \left(\bar q_R \tau^+\g^\mu q_R\right)\left(\bar q_L \tau^+ q_R\right)\,\
,\nn\\
O_{9}^{\mu } &=& \left(\bar q_L t^a\tau^+\g^\mu q_L\right)\left(\bar q_Rt^a\tau^+ q_L\right)\,\,,\qquad \,\,
O_{9}^{\mu \, \prime} = \left(\bar q_R t^a\tau^+\g^\mu q_R\right)\left(\bar q_Lt^a\tau^+ q_R\right)\,\,,
\eea
where the second column of operators is  related to the first column by a parity transformation. 

While  twenty-four dim-9 operators appear in Eq.~\eqref{eq:Lag}, not all of them are necessarily induced by the gauge-invariant \textoverline{dim-n} (for $n\leq 9$) operators. As discussed in  Ref.\ \cite{Cirigliano:2017djv} only three dim-9 operators, those with Wilson coefficients $C^{(9)}_{1\rm L}$, $C^{(9)}_{4\rm L}$ and $C^{(9)}_{5\rm L}$, receive tree-level matching from \textoverline{dim-7} operators. This implies that these coefficients can scale as $C^{(9)}_{\rm 1 L,\, 4L,\, 5L}= \mathcal O(v^3/\Lambda^3)$.

Ref.~\cite{Graesser:2016bpz} showed that the eleven dim-9 operators with Wilson coefficients $C^{(9)}_{\rm 2L,\, 3L,\, 4L,\, 5L}$, $C^{(9)\, \prime}_{\rm 2L,\, 3L}$,  $C^{(9) \prime}_{\rm 1R}$, 
and  $C^{(9)\, \prime}_{6,7,8,9}$  receive contributions from four-quark two-lepton \textoverline{dim-9} operators. As discussed above, by replacing  $\bar d_R \gamma^\mu u_R \rightarrow  \tilde{\varphi}^{\dagger} D_\mu \varphi$ in the \textoverline{dim-9} operators of Ref.~\cite{Graesser:2016bpz} we identify several additional dim-9 operators that can be induced by \textoverline{dim-9} operators, namely $C^{(9)}_{\rm 1 L}$, $C^{(9)}_{\rm 1R,\, 4R,\, 5R}\,$ and  $C^{(9)}_{6,7,8,9} $. To summarize, out of the twenty-four dim-9 operators in Eq.~\eqref{eq:Lag} we find that nineteen operators are actually induced by \textoverline{dim-n} (for $n\leq 9$) operators. 
$U(1)_Y$ invariance implies that there are no \textoverline{dim-9} operators that contribute to $C^{(9)\, \prime}_{\rm 1 L}$,  $C^{(9)}_{\rm 2R,\, 3R}$, and $C^{(9)\, \prime}_{\rm 2R,\, 3R}$.  These operators are then induced only at the \textoverline{dim-11} or higher level. 
Although this means these operators should be further suppressed, we include them for completeness, thereby keeping all operators in Eq.~\eqref{eq:Lag} in our analysis.  Finally, the effective $SU(3)_c\times U(1)_{\rm em}$ operators we consider  may not cover the contributions of all \textoverline{dim-9} terms as, unlike for \textoverline{dim-5} and \textoverline{dim-7} operators, a complete \textoverline{dim-9} basis is currently unavailable. Nevertheless, we expect the included $SU(3)_c\times U(1)_{\rm em}$ operators to capture the dominant $\Dt L=2$ effects in most, if not all, models of LNV.

The coefficients in Eqs.\ \eqref{lowenergy6}, \eqref{lowenergy7}, and \eqref{eq:Lag} need to be evolved from the matching scale $\mu \sim m_W$ to scales $\mu \sim 2$ GeV, where the matching to chiral perturbation theory and LQCD calculations is performed. The vector operators corresponding to  $C^{(6)}_{\textrm{ VL,\,VR}}$ and $C^{(7)}_{\textrm{VL,\,VR}}$ 
involve quark non-singlet axial and vector currents and therefore do not run in QCD.
The renormalization group equations (RGEs) of the scalar and tensor operators below $\mu=m_W$ are given by
\bea\label{RGE6}
\frac{d}{d\ln \mu} C^{(6)}_{\textrm{SL\,(SR)}} &=& -6 C_F\, \frac{\al_s}{4\pi}  C^{(6)}_{\, \textrm{SL\,(SR)}}\,,\quad
\frac{d}{d\ln \mu}  C^{(6)}_{\textrm T} = 2 C_F\, \frac{\al_s}{4\pi} C^{(6)}_{\textrm T}\,,
\eea
where $C_F=(N_c\sq-1)/(2N_c)$ and $N_c$ the number of colors.

The RGEs of the scalar dim-9 operators are given by
\bea\label{RGE9scalar}
\frac{d}{d\ln \mu} C^{(9)}_{1} &=& 6\left(1-\frac{1}{N_c}\right)\, \frac{\al_s}{4\pi}  C^{(9)}_{1}\,,\nn\\
\frac{d}{d\ln \mu}  \bma C^{(9)}_{2}\\C^{(9)}_{3}\ema  &=&  \frac{\al_s}{4\pi}\,\bma 8 + \frac{2}{N_c} - 6 N_c & -4 - \frac{8}{N_c} + 4 N_c \\4 - \frac{8}{N_c} &4 + \frac{2}{N_c} + 2 N_c \ema \bma C^{(9)}_{2}\\C^{(9)}_{3}\ema\, \,,\nn\\
\frac{d}{d\ln \mu}  \bma C^{(9)}_{4}\\C^{(9)}_{5}\ema  &=&  \frac{\al_s}{4\pi}\,\bma 6/N_c&0\\-6&-12 C_F\ema \bma C^{(9)}_{4}\\C^{(9)}_{5}\ema\,,
\eea
 in agreement with Refs.~\cite{Buras:2000if,Buras:2001ra}.
The RGEs do not depend on the lepton chirality, and we therefore omitted the subscripts $L$, $R$ in Eq.\ \eqref{RGE9scalar}.
The equations for the $C^{(9)\prime}_{1,2,3}$ coefficients are equivalent to those in Eq.~\eqref{RGE9scalar}, while the RGEs for the vector operators can be written as,
\bea\label{RGE9vector}
\frac{d}{d\ln \mu}  \bma C^{(9)}_{6}\\C^{(9)}_{7}\ema  &=&  \frac{\al_s}{4\pi}\,\bma - 2 C_F \frac{3 N_c -4}{N_c} &  2C_F\frac{(N_c+2)(N_c-1)}{N_c\sq} \\ 
4 \frac{N_c -2}{N_c}&  \frac{4 -  N_c + 2 N_c^2 + N_c^3}{N_c^2} \ema \bma C^{(9)}_{6}\\C^{(9)}_{7}\ema\, \,.
\eea
The evolution of $C^{(9)}_{8}$ and $C_{9}^{(9)}$, as well as that of the primed coefficients, are governed by RGEs of the same form. 
The RGEs for $C_{6,7,8,9}^{(9)}$ correct expressions that have previously appeared in the \NLDBD\ literature \cite{Gonzalez:2015ady}. 
The (numerical) solutions to the above RGEs are given in Appendix~\ref{App:RG} and a comparison to the literature is given in Appendix~\ref{comparison-to-lit}.

\section{The chiral Lagrangian}\label{ChiPT}
After obtaining the Lagrangian at the scale of a few GeV, the next step is to match the quark-level theory onto chiral perturbation theory ($\chi$PT) \cite{Gasser:1983yg,Bernard:1995dp}. $\chi$PT is the effective theory of QCD which describes interactions at low energy in terms of baryons, mesons, photons, and leptons. It relies on an expansion in $\epsilon_\chi\equiv p/\Lambda_\chi$ that is provided by the approximate chiral symmetry of QCD. Here 
$p\sim m_\pi$ is the typical momentum and $\Lambda_\chi\simeq 1$ GeV.
The interactions that are induced in the chiral Lagrangian can be derived by writing down all possible terms that have the same chiral symmetry properties under $SU(2)_L\times SU(2)_R$ as the quark-level operators. All such operators come with an unknown coefficient that parametrizes the non-perturbative nature of QCD.
In the mesonic and single nucleon sector of the theory,  these low-energy constants (LECs) can be estimated by naive dimensional analysis (NDA), 
and, as we will discuss, several of them can be extracted from  experimental data or existing lattice QCD calculations. 
The power counting of LNV nucleon-nucleon operators is complicated by the non-perturbative nature of the nuclear force, which causes NDA estimates of the coefficients of nucleon-nucleon operators to be unreliable \cite{Kaplan:1996xu,Nogga:2005hy, Valderrama:2014vra, Cirigliano:2018hja}. We will determine the scaling of nucleon-nucleon operators by requiring that the  LNV $nn\to pp ee$ scattering amplitude 
is properly renormalized \cite{Cirigliano:2018hja}.

The chiral Lagrangian induced by the neutrino Majorana mass and by the operators in Eqs.\ \eqref{lowenergy6} and \eqref{lowenergy7} has been discussed in Refs.\ \cite{Cirigliano:2017tvr,Cirigliano:2017djv,Cirigliano:2018hja}.
In Appendix \ref{app:neutrinopotentials567} we summarize the main results, and discuss several short-distance effects that were overlooked in Ref.\ \cite{Cirigliano:2017djv}.
We focus here on the dim-9 operators in Eq.\ \eqref{eq:Lag}, for which partial results are given in Refs.\ \cite{Prezeau:2003xn,Graesser:2016bpz}.
For the dim-9 operators under discussion three types of interactions will appear: purely pionic interactions, pion-nucleon couplings, and nucleon-nucleon ($N\!N$) terms. In what follows we construct the corresponding operators that give the LO contributions to the $nn\to pp ee$ amplitude.

\begin{figure}
\includegraphics[width=0.9\textwidth]{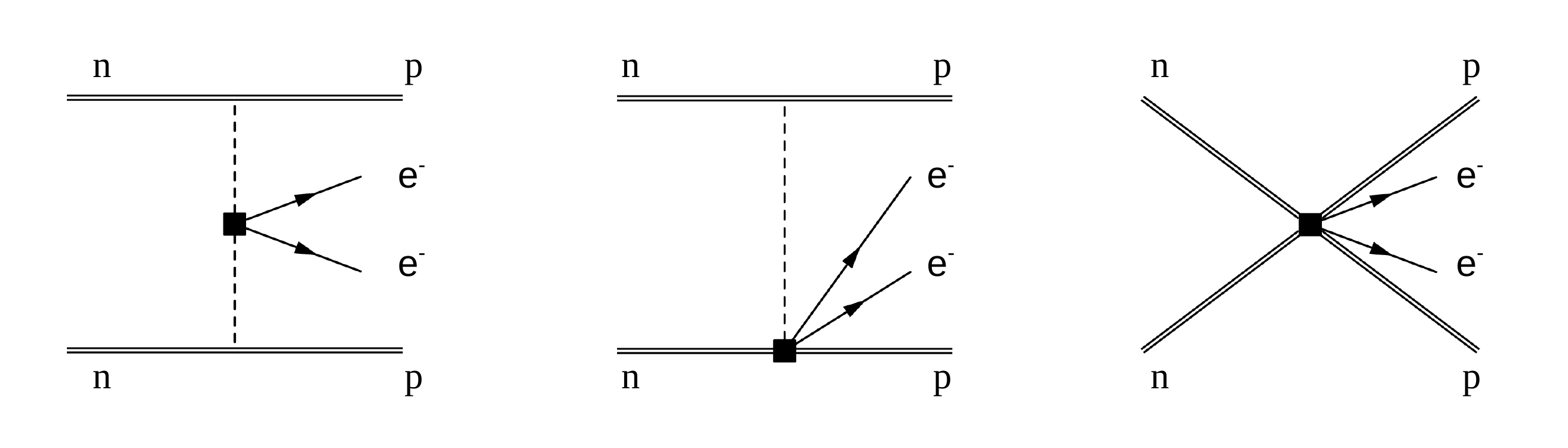}
\caption{The different contributions of dim-9 LNV operators to the \NLDBD\  potential, first discussed in   Refs.\ \cite{Pontecorvo:1968wp, Vergados:1981bm,Faessler:1996ph} \cite{Prezeau:2003xn}.
Double, dashed, and single lines denote, respectively, nucleon, pion, and lepton fields. The black square denotes $\Delta L=2$ $\pi\pi$, $\pi N$, and $N\!N$ operators, discussed in Sections
\ref{ChiPTscalar} and \ref{ChiPTvector}. The remaining vertices
are SM interactions between nucleons and pions.}\label{feyndiag}
\end{figure}

Unlike standard $\chi$PT applications, in the operators constructed below there appear explicit lepton fields. As such, operators can be constructed that depend on the electron mass and/or the momenta of the outgoing lepton fields. The typical Q-values, and the corresponding electron momenta, of experimentally measurable $0\nu\beta\beta$-transitions are small, $Q\sim\mathcal O(5\,\mathrm{MeV})$, compared to the typical momentum exchange between nucleons $q \sim k_F \sim m_\pi = \mathcal O(100\,\mathrm{MeV})$. In order to incorporate this new scale into the $\chi$PT power counting we assign $Q\sim m_e \sim m_\pi \epsilon_\chi^2$ \cite{Cirigliano:2017djv}.

\subsection{Scalar dim-9 operators}\label{ChiPTscalar}

The scalar operators $O_{1}$--$O_5$ generate the $\pi\pi ee $, $\pi N\!N e e$, and $N\!N\, N\!N\,ee$ LNV couplings shown in Fig.~\ref{feyndiag}.
The operators $O_{2,3,4,5}$ induce non-derivative pionic operators, while the first pionic operators induced by $O_{1}$ contain two derivatives and are therefore relatively suppressed.
The mesonic chiral Lagrangian\footnote{
The $\pi\pi$ couplings defined here are related to those of Refs.\ \cite{Cirigliano:2017ymo,Cirigliano:2017djv,Pastore:2017ofx} by
$g^{\pi\pi}_1 = g_{27 \times 1}$, $g^{\pi\pi}_2 = g_{6\times \bar 6}$, $g^{\pi\pi}_3 = g_{6\times \bar 6}^{\rm mix}$,
$g_4^{\pi\pi} = g_{8\times 8}$, $g_5^{\pi\pi} = g^{\rm mix}_{8\times 8}$, while for the  $\pi N$ and $NN$ couplings
we have $g^{\pi N}_1 = g^{\pi N}_{27 \times 1}$ and $g^{N N}_1 = g^{N N}_{27 \times 1}$.  The notation of Refs.\ \cite{Cirigliano:2017ymo,Cirigliano:2017djv,Pastore:2017ofx}
emphasizes the transformation properties under $SU(3)_L \times  SU(3)_R$.} for $O_{1,2,3,4,5}^{}$ is 
\bea\label{eq:dim9pipi}
\vL^{\rm scalar}_\pi &=&\frac{F_0^4}{4}\left[
\frac{5}{3}g^{\pi\pi}_{1} C_{\rm 1L}^{(9)} L_{21}^\mu L_{21\,\mu}
+\left(g^{\pi\pi}_{2}C_{\rm 2L}^{(9)} + g^{\pi\pi}_{3} C_{\rm 3L}^{(9)}\right) {\rm Tr}\left( U\tau^+U\tau^+\right)\right.\nn\\
&&\left.+
\left(g^{\pi\pi}_{4}C_{\rm 4L}^{(9)}+g_{5}^{\pi\pi}C_{\rm 5L }^{(9)}\right) {\rm Tr}\left( U\tau^+U^\dagger\tau^+\right)
\right]\frac{\bar e_L  C\bar e^T_L}{v^5}+(L\leftrightarrow R)\nn\\
&= &\frac{F_0^2}{2}\left[
\frac{5}{3}g^{\pi\pi}_{1} C^{(9)}_{\rm 1 L}  \,  \partial_\mu \pi^- \partial^\mu \pi^-   
+ \left( g^{\pi\pi}_{4} C^{(9)}_{\rm 4 L}  + g^{\pi\pi}_{5}  C^{(9)}_{\rm 5 L}-g^{\pi\pi}_{2} C^{(9)}_{\rm 2 L}  - g^{\pi\pi}_{3} C^{(9)}_{\rm 3 L}\right)  \pi^- \pi^- 
\right]\nn \\ & & 
\frac{\bar e_L  C\bar e^T_L}{v^5}+(L\leftrightarrow R)+\dots\,\,,\eea
where $U=u^2={\rm exp}\left(i \pi\cdot \tau/F_0\right)$ is the matrix of pseudo-Goldstone boson fields, $F_0$ is the pion decay constant in the chiral limit, and $L_{\mu} = iUD_\mu U^\dagger$. 
We use $F_\pi =92.2$ MeV for the physical pion decay constant.
By NDA the LECs  of the non-derivative pion operators are expected to be $g^{\pi\pi}_{2,3,4,5}  = \mathcal O(\Lambda^2_\chi)$,
while $g^{\pi\pi}_{1} = \mathcal O(1)$. These expectations are very well respected by the extractions of Ref.\ \cite{Nicholson:2016byl,Cirigliano:2017ymo,Nicholson:2018mwc} 
based on chiral symmetry and lattice QCD results. 
In  Table \ref{Tab:LECs} we give the value of the LECs at $\mu = 2$~GeV  in the $\overline{\rm MS}$ scheme, obtained in Ref.\ \cite{Nicholson:2018mwc}.
The physical amplitudes are scale and scheme independent provided one uses Wilson coefficients $C^{(9)}_i$   
evaluated at the same scale and in the same scheme as used for the $g_i^{\pi \pi}$.

The $\pi N$  terms are only  relevant for the $O_1$ operator and can be written as
\bea\label{eq:dim9PiN}
\vL_{\pi N}^{\rm scalar}& =& g_Ag^{\pi N}_{1}C_{\rm 1L}^{(9)}F_0^2\left[\,\bar N S^\mu u^\dagger\tau^+ u N\,{\rm Tr}\left(u_\mu u^\dagger \tau^+ u \right)\right]\frac{\bar e_L  C\bar e^T_L }{v^5}+(L\leftrightarrow R)\nn\\
&=&\sqrt{2}g_A g^{\pi N}_{1}C_{\rm 1L}^{(9)}F_0 \left[\bar p\, S\cdot (\partial \pi^-)n\right] \, \frac{\bar e_L  C\bar e^T_L}{v^5}+(L\leftrightarrow R)+\dots\,\,,
\eea
where $u_\mu=u^\dagger L_\mu u = i\left[u(\partial_\mu-ir_\mu)u^\dagger -u^\dagger(\partial_\mu-il_\mu)u \right]  $, $g_A \simeq 1.27$, $N = (p,\, n)^T$, and $S^\mu$ and $v^\mu$ are the nucleon spin and velocity. In the nucleon restframe  we have $S^\al = (0,\, \boldsigma/2)$ and $v^\mu = (1,\, \vec 0)$.
The LEC $g_1^{\pi N}$ is unknown, but expected to be $\mathcal O(1)$ by NDA.

In a power counting based on NDA, LNV four-nucleon interactions are relevant only for $O_1$,
 in which case they would compete with the $\pi \pi$ and $\pi N$ interactions $g^{\pi\pi}_{1}$ and $g^{\pi N}_{1}$.
However, the LNV potential induced by the non-derivative $\pi\pi$ operators in Eq.\ \eqref{eq:dim9pipi}
has the same short-distance behavior as the neutrino potential mediated by the neutrino Majorana mass,  $V(\vec q) \sim 1/\vec q^2$ at large $|\vec q|$.
Ref.\ \cite{Cirigliano:2018hja} showed that for these potentials the $nn \rightarrow p p e e$ scattering amplitude  has a logarithmic UV divergence, which can be absorbed by promoting the $N\!N$ operators 
stemming from  $O_{2,3,4,5}$ to leading order. The relevant  $N\!N$ interactions are
\bea\label{eq:dim9NN}
\vL_{NN}^{\rm scalar} &=& g_{1}^{NN} C_{\rm 1L}^{(9)}  \, (\bar N u^\dagger \tau^+ u N)(\bar N u^\dagger \tau^+ u N)\, \frac{\bar e_L  C\bar e^T_L}{v^5}\nn \\
& & +  \left(   g_{2}^{NN} C_{\rm 2L}^{(9)} +  g_{3}^{NN} C_{\rm 3L}^{(9)}  \right)\, (\bar N u^\dagger \tau^+ u^\dagger N)(\bar N u^\dagger \tau^+ u^\dagger N)\, \frac{\bar e_L  C\bar e^T_L}{v^5}  \nn \\
& & +  \left(  g_{4}^{NN} C_{\rm 4L}^{(9)}  +  g_{5}^{NN} C_{\rm 5L}^{(9)}  \right)\, (\bar N u^\dagger \tau^+ u N)(\bar N u \tau^+ u^\dagger N)\, \frac{\bar e_L  C\bar e^T_L}{v^5}
+(L\leftrightarrow R)
\nn\\
& =& \left(  g_{1}^{NN} C_{\rm 1L}^{(9)} +   g_{2}^{NN}  C_{\rm 2L}^{(9)} +  g_{3}^{NN} C_{\rm 3L}^{(9)}  +
 g_{4}^{NN} C_{\rm 4L}^{(9)}  +   g_{5}^{NN} C_{\rm 5L}^{(9)} \right) \left(\bar pn\right )\,\left(\bar pn\right )\, \frac{\bar e_L  C\bar e^T_L}{v^5}\nn \\ & & +(L\leftrightarrow R)+\dots\,.
\eea
The couplings $g_{i}^{NN}  = \mathcal O(1)$ in the Weinberg power counting, but need to be promoted to
$\mathcal O((4\pi)^2)$ in the case of $O_{2,3,4,5}$. The renormalization of the scattering amplitude does not require such enhancement for $g_{1}^{NN}$. 

The  $\pi\pi$, $\pi N$, and $N\!N$ Lagrangians for the $O_{1,2,3}'$ operators, which are related by parity to $O_{1,2,3}$, can  be obtained by replacing $C_{\rm 1L,\, 2L,\,3L}^{(9)}\to C_{\rm 1L,\,2L,\,3L}^{(9)\, \prime}$, $C_{\rm 1R,\, 2R,\,3R}^{(9)}\to C_{\rm 1R,\,2R,\,3R}^{(9)\, \prime}$, 
$S_\al \to  - (-1)^\al S_\al$, 
$u_\al \to  - (-1)^\al u_\al$,
$u\to u^\dagger$, and $U\to U^\dagger$ in Eqs.\ \eqref{eq:dim9pipi}-\eqref{eq:dim9NN}.~\footnote{We use the following standard notation:
$(-1)^\mu = 1$ for $\mu=0$ and 
$(-1)^\mu = - 1$ for $\mu=1,2,3$ \cite{Peskin:1995ev}.} 
This leads to $\pi\pi$, $\pi N$, and $NN$ Lagrangians of the same form (with $C_{1,2,3}^{(L,R)}\to C_{1,2,3}^{\prime\,(L,R)}$) after expanding the meson matrices $u$ (and $U$) to two, one, and zero pions, respectively. 

From Eq.~\eqref{eq:dim9NN} we see that all scalar operators in Eq.~\eqref{LagSca} induce a LNV four-nucleon operator that contributes to the  $nn \rightarrow pp e e$ amplitude at the same order as the pion-range contributions from the $\pi\pi ee$ operators. This happens either because the $\pi\pi ee$ interaction is suppressed by two powers in the chiral counting (as it is for $O_1$ and $O'_1$), or because of non-perturbative renormalization which enhances the four-nucleon operator to leading order ($O_{2,3,4,5}$ and $O'_{2,3}$). In other words, for all scalar operators the $\pi\pi ee$ and $NN$ interactions appear at the same order. For $O_1$ and $O'_1$ there appear additional contributions from the $\pi N$ interaction. More details on the renormalization of the $nn \rightarrow p p e e$ scattering amplitude and the non-perturbative RGE of $g_i^{NN}$ are given in Appendix~\ref{NonWeinberg}.

\subsection{Vector operators}\label{ChiPTvector}
The vector operators induce mesonic interactions involving a derivative on the pion fields, which, up to a total derivative, give rise to contributions proportional to $m_e$ \cite{Graesser:2016bpz,Prezeau:2003xn}. Instead, the contributions to the $nn\to pp ee$ amplitude from the $\pi N$ and $N\!N$ interactions are proportional to $|\vec q|\sim k_F$ and therefore larger than the contributions from the purely mesonic terms by one power of $1/\epsilon_\chi$.

The $\pi N$ Lagrangian induced by $O_{6,7,8,9}^{\mu}$ and $O_{6,7,8,9}^{\mu\, \prime}$ can be written as
\bea\label{eq:vectorPiN}
\vL^{\rm vector}_{\pi N}&=&g_A F_0\sq \bigg\{ 
{\rm Tr}\left(i UD^\al U^\dagger\tau^+\right)\left[ \left(g_6^{\pi N} C^{(9)}_{6} + g_{7}^{\pi N} C^{(9)}_{7} \right)\bar N u^\dagger \tau^+u^\dagger S_\al N \right. \nn \\ & &\left. 
+ \left(g_6^{\pi N} C^{(9)}_{8} + g_{7}^{\pi N} C^{(9)}_{9} \right) \bar N u\tau^+u S_\al N\right]
\nn\\
&&+
\bar N u^\dagger \tau^+u S_\al N\left[ \left( g_8^{\pi N} C^{(9)}_{6} +  g_{9}^{\pi N} C^{(9)}_{7} \right) {\rm Tr}\left(i D^\al U^\dagger \tau^+\right)
\right. \nn \\ & & \left. - \left( g_8^{\pi N} C^{(9)}_{8} +  g_{9}^{\pi N} C^{(9)}_{9} \right) {\rm Tr}\left(i D^\al U \tau^+\right)\right] 
\bigg\}v^\mu \frac{\bar e \g_\mu \g_5 C\bar e^T}{v^5}\,,
\eea
where $g_{ i}^{\pi N}$  are LECs of $\Or(1)$. The chiral representations of the operators $O_{6,7,8,9}^{\mu\, \prime}$ are related to the above Lagrangian by parity. The corresponding $\pi N$ terms can be obtained by replacing $L\leftrightarrow R$, $u\to u^\dagger$, $U\to U^\dagger$,  $D_\al \to  (-1)^\al  D_\al$,  and $S_\al\to - (-1)^\al S_\al$.
Combining all contributions and expanding up to one pion we have
\bea \label{eq:vectorPiN2}
\vL^{\rm vector}_{\pi N} = \sqrt{2}g_A F_0\,\bar p S\cdot (\partial \pi^-) n \bigg[ g_V^{\pi N}C^{(9)}_V+ \tilde{g}_{V}^{\pi N} \tilde{C}^{(9)}_V \bigg]v^\mu \frac{\bar e \g_\mu \g_5 C\bar e^T}{v^5} +\dots\,\,\,,
\eea
where $C_V^{(9)}\equiv C_{6}^{(9)}+C_{8}^{(9)}+C_{6}^{(9)\, \prime}+C_{8}^{(9)\, \prime}$,  
$\tilde{C}_V^{(9)} \equiv C_{7}^{(9)}+C_{9}^{(9)}+C_{7}^{(9)\, \prime}+C_{9}^{(9)\, \prime}$
and $g_{V}^{\pi N}=g_{6}^{\pi N}+g_{8}^{\pi N}$, $\tilde{g}_{V}^{\pi N}=g_{7}^{\pi N}+g_{9}^{\pi N}$.

For the $N\!N$ interactions induced by $O_{6,7,8,9}^{\mu}$ and $O_{6,7,8,9}^{\mu\, \prime}$ we obtain
\bea\label{eq:VectorNN}
\vL_{NN}^{\rm vector} &=&\bigg[
\left(g_6^{NN}C^{(9)}_{6}+g^{NN}_7 C^{(9)}_{7}\right)\bar Nu^\dagger \tau^+ u N\, \bar Nu^\dagger \tau^+ u^\dagger N
\nn\\
&&
+\left(g_6^{NN}C^{(9)}_{8}+g^{NN}_7 C_{9}^{(9)}\right)\bar Nu^\dagger \tau^+ u N\, \bar Nu\tau^+ u N\bigg] v^\mu \frac{\bar e \g_\mu \g_5 C\bar e^T}{v^5}\,\,,
\eea
where $g^{NN}_{6,7}$ are LECs of $\Or(1)$ within the Weinberg power counting.
The relevant Lagrangian for the  $O_{6,7,8,9}^{\mu\, \prime}$  operators is again related by parity, and can be obtained by replacing $L\leftrightarrow R$ and $u\to u^\dagger$ in Eq.\ \eqref{eq:VectorNN}. Expanding the $N\!N$ interactions  gives for terms without pions
\bea\label{vecNN}
\vL_{NN}^{\rm vector} &=&\bigg[g_6^{NN}C^{(9)}_{V}+g_7^{NN} \tilde C^{(9)}_{V}\bigg] (\bar p n)(\bar p n)\,v^\mu \frac{\bar e \g_\mu \g_5 C\bar e^T}{v^5}+\dots\,\,.
\eea

The non-perturbative renormalization of the scattering amplitude relates the $\pi N$ and $N\!N$ couplings through an RGE, see Appendix \ref{NonWeinberg}, but this does not change the power-counting expectations.

To summarize, the chiral Lagrangian from the dim-9 vector is rather simple and at LO consists of only the interactions in Eqs.~\eqref{eq:vectorPiN2} and \eqref{vecNN}. Unfortunately, very little is known about the corresponding LECs. 
\begin{table}
\center
\begin{tabular}{|c|cc||c|cc|}
\hline
 \multicolumn{3}{|c||}{ $n\rightarrow pe\nu$, $\pi \rightarrow e \nu$ } &  \multicolumn{3}{c|}{$\pi \pi \rightarrow e e$} \\
 \hline
 $g_A$ & $1.271\pm 0.002$ & \cite{Olive:2016xmw}      & $g^{\pi\pi}_{1}$   		& $  0.36 \pm 0.02 $             & \cite{Nicholson:2018mwc}  \\
 $g_S$ & $0.97\pm 0.13$ & \cite{Bhattacharya:2016zcn} & $g^{\pi\pi}_{2}$   		& $  2.0  \pm 0.2 $  \, GeV$^2$  & \cite{Nicholson:2018mwc}  \\
 $g_M$ &  $ 4.7$ &  \cite{Olive:2016xmw}              & $g^{\pi\pi}_{3}$ 	        & $ -(0.62 \pm 0.06)$  \, GeV$^2$  & \cite{Nicholson:2018mwc}  \\
 $g_T$ & $0.99\pm 0.06$ & \cite{Bhattacharya:2016zcn} & $g^{\pi\pi}_{4}$   		& $ -(1.9  \pm 0.2)$   \, GeV$^2$  & \cite{Nicholson:2018mwc}\\  
 $|g'_{T}|$ & $\mathcal O(1)$  	&		      & $g^{\pi\pi}_{5}$ 		& $ -(8.0  \pm 0.6)$   \, GeV$^2$  & \cite{Nicholson:2018mwc}  \\
 $B$      &    $2.7$~GeV       &     & $|g^{\pi\pi}_{ \rm T}|$     & $\mathcal O(1)$  & \\ 
 \hline
  \multicolumn{3}{|c||}{$n \rightarrow p\pi ee$} & \multicolumn{3}{c|}{$nn\rightarrow pp\, ee$}   
 \\
  \hline 
   $|g^{\pi N}_{1} |$       & $\mathcal{O}(1)$ &    & $|g^{N N}_1|$         & $\mathcal{O}(1)$ &  \\  
   $|g^{\pi N}_{6,7,8,9}|$ & $\mathcal{O}(1)$ &    & $|g_{6,7}^{N N}|$ & $\mathcal{O}(1)$ & \\
     $|g^{\pi N}_{\rm VL}|$ & $\mathcal{O}(1)$ &    &$ |g^{NN}_{\rm VL}|$ & $ \mathcal{O}(1) $ &     \\ 
           $|g^{\pi N}_{\rm T}|$ 			   & 		$ \Or(1)$     &    &
        $ |g_{\rm T}^{NN}|$ &$\Or(1)$&   
  \\ 
			   & 		      &    & $ |g^{NN}_{\nu}|$ & $ \mathcal{O}(1/F_\pi^2) $ &     \\ 
			   &&&$ |g_{VL,VR}^{E,m_e}|$ &$\Or(1)$&\\
			   			   &&&            $ |g^{NN}_{2,3,4,5}|$ & $ \mathcal{O}((4\pi)^2) $ &   
\\\hline
\end{tabular}
\caption{The low-energy constants relevant for the dim-3, dim-6, dim-7 and dim-9 operators. 
Whenever known,  we quote the values of the LECs at $\mu=2$ GeV in the $\overline{\rm MS}$ scheme.
The LECs $g_{1, \ldots, 5}^{\pi\pi}$ were first extracted in Refs.\ \cite{Savage:1998yh,Cirigliano:2017ymo} using 
$SU(3)$ relations between $\pi^+$-$\pi^-$, $K$-$\overline{K}$ and $K \rightarrow \pi \pi$  matrix elements. Ref.\ \cite{Cirigliano:2017ymo} found
$g^{\pi\pi}_{1} = 0.38 \pm 0.08$, $g^{\pi\pi}_{2} = 2.9 \pm 0.6$ GeV$^2$, $g^{\pi\pi}_{3} = -( 1.0 \pm 0.3)$ GeV$^2$, $g^{\pi\pi}_{4} = -(2.5 \pm 1.3)$ GeV$^2$, $g^{\pi\pi}_{5} = -(11 \pm 4)$ GeV$^2$,
in good agreement with the more precise results of Ref.\ \cite{Nicholson:2018mwc}.}  
\label{Tab:LECs}
\end{table}

\subsection{The $0\nu\beta\beta$ transition operator}
With the above chiral operators we can construct the LO two-nucleon operators  that induce $0\nu\beta\beta$, {to which we refer to as the ``\NLDBD\  transition operators" 
or ``\NLDBD\ potentials",  as commonly  done in the literature. 
 The calculation can be directly lifted from Ref.~\cite{Cirigliano:2017djv} as all chiral interactions constructed above also appear in the dim-7 chiral Lagragian but, of course, accompanied by different Wilson coefficients and LECs. We refer to Ref.~\cite{Cirigliano:2017djv} for details and give here the results. We define the $\Dt L=2$  transition operator or} potential as $V=-\mathcal A$ where $\mathcal A$ is the 
Born-level  amplitude of $nn\rightarrow pp\,ee$.

The LO $\Dt L=2$ \NLDBD\ potential from the scalar and vector dim-9 operators is
\bea\label{scalar}
V_{9}(\vec q^2) &=& -( \tau^{(1) +} \tau^{(2) + }) \, g_A\sq \,  \frac{4 G_F^2}{v} \,\bar u(k_1)P_R C\bar u^T(k_2) \nonumber \\
& \times &   \!\!\!\! 
\left[ -  
    \left(  \, \boldsigma^{(1)} \cdot \boldsigma^{(2)}-  \, S^{(12)}  \right) \left( \frac{C^{(9)}_{\pi \pi\,\rm L}}{6}   \frac{\vec q^2}{(\vec q^2 + m_\pi^2)^2}  -  \frac{ C^{(9)}_{\pi N\,\rm L}}{3} \frac{\vec q^2}{\vec q^2 + m_\pi^2} \right)
+ \frac{2}{g_A^2} C_{NN\, \rm L}^{(9)}  \right] \nn  \\  
& -& ( \tau^{(1) +} \tau^{(2) + }) \, g_A\sq \,  \frac{4 G_F^2}{v} \,\bar u(k_1)\gamma_0 \gamma_5 C\bar u^T(k_2) \nonumber \\
&\times &  \!\!\!\! 
\left[  \frac{g_V^{\pi N}C_V^{(9)}+ \tilde g_{V}^{\pi N} \tilde C_V^{(9)}}{3}
\left(  \, \boldsigma^{(1)} \cdot \boldsigma^{(2)}-  \, S^{(12)}  \right) \frac{\vec q^2}{\vec q^2  + m_\pi^2} + \frac{2}{g_A^2} \left(g_6^{NN}C^{(9)}_{V}+g_7^{NN} \tilde C^{(9)}_{V}\right)    \right] \nn \\ & & 
+ (L\leftrightarrow R)\, \,. 
\eea
Here the combinations $C^{(9)}_{\pi\pi,\, \pi N,\, NN}$ of scalar couplings are defined as 
\begin{eqnarray}
C^{(9)}_{\pi \pi\, \rm L} & = &  g^{\pi\pi}_{2} \left(C_{\rm 2L}^{(9)} + C_{\rm 2L}^{(9)\, \prime}\right) +    g^{\pi\pi}_{3}\left(C_{\rm 3L}^{(9)} + C_{\rm 3L}^{(9)\, \prime}\right) 
- g^{\pi\pi}_{4} C^{(9)}_{\rm 4L} -  g^{\pi\pi}_{5} C^{(9)}_{\rm 5L}   \nn \\ 
& & - \frac{5}{3} g^{\pi\pi}_{1} m_\pi^2  \left(C_{\rm 1L}^{(9)} + C_{\rm 1L}^{(9)\, \prime}\right)\,, \nn \\
C^{(9)}_{\pi N\, \rm L} &=& \left(g_{1}^{\pi N} - \frac{5}{6}g^{\pi\pi}_{1}  \right)  \left(C_{\rm 1L}^{(9)} + C_{\rm 1L}^{(9)\, \prime}\right)\,, \nn \\
C_{NN\, \rm L }^{(9)} &=&   g_{1}^{NN}\, \left(C_{\rm 1L}^{(9)} + C_{\rm 1L}^{(9)\, \prime}\right) +   g_{2}^{NN} \, \left(C_{\rm 2L}^{(9)} + C_{\rm 2L}^{(9)\, \prime}\right)  + g_{3}^{NN} \left(C_{\rm 3L}^{(9)} + C_{\rm 3L}^{(9)\prime}\right)   \nn \\
& & +   g_{4}^{NN} C_{\rm 4L}^{(9)} +  g_{5}^{NN} C_{\rm 5L}^{(9)}  \,,
\end{eqnarray}
and similarly for $C_{\{\pi\pi, \, \pi N,\, NN\}\,\rm R }^{(9)}$. 
In the above expressions we use $q^\mu = (q^0,\,\vec q) = (p-p')^\mu$, where $2p$ and $2p'$ are the relative momenta of the in- and outgoing nucleon pairs. We defined {$S^{(12)}= -(3 \vec  \boldsigma^{(1)}\cdot \hat{ \vec q} \,\vec  \boldsigma^{(2)}\cdot \hat{ \vec q}  - \vec  \boldsigma^{(1)}\cdot \vec  \boldsigma^{(2)})$} and $ \hat{ \vec q}  = \vec q/|\vec q|$. $k_{1,2}\sim Q$ are the outgoing electron momenta. 

The potential in Eq.~\eqref{scalar} can be expressed in terms of the Fermi (F), Gamow-Teller (GT), and Tensor (T) functions $h_F(\vec q\sq)$, $h^{AP,PP}_{GT}(\vec q\sq)$, and $h^{AP,PP}_{T}(\vec q\sq)$ defined in  Appendix~\ref{sec:app}, and, consequently, their matrix elements for experimentally interesting nuclei can be extracted from existing calculations \cite{Hyvarinen:2015bda,Menendez:2017fdf,Barea:2015kwa,Barea,Horoi:2017gmj}.

\section{$0\nu \beta \beta$ amplitudes and master formula for the decay rate}\label{Master}

\subsection{$0\nu \beta \beta$ amplitudes}
Now that we have derived the 
momentum-space  
\NLDBD\ potential, we can obtain an expression for the  inverse half-life for $0^+\rightarrow 0^+$ transitions. This requires one to insert the  two-nucleon potential between the initial- and final-state nuclear wave functions and sum over all nucleons. In the literature,  many-body matrix elements are usually calculated in coordinate space and we therefore define the 
amplitude
\bea \label{eq:FullAmp}
\mathcal A = \langle 0^+| \sum_{m,n} \int \frac{d^3\vec q}{(2\pi)^3} e^{i\vec q \cdot \vec r}V(\vec q\sq) |0^+\rangle\,\,,
\eea
where we sum over all nucleon pairs, $\vec q$ is the momentum transfer, and $\vec r = \vec r_n - \vec r_m$ is the distance between the $m^{\rm th}$ and $n^{\rm th}$ nucleon.  
The \NLDBD\ potential $V(\vec q\sq)$ is the sum of potentials induced by the low-energy $\Delta L=2$ dim-3, -6, -7, and -9 operators and we write
\begin{equation}
V(\vec q\sq) = V_{ 3}(\vec q\sq) + V_{ 6}(\vec q\sq) + V_{ 7}(\vec q\sq) + V_{9}(\vec q\sq)\,. 
\end{equation}
 $V_{3}(\vec q\sq)$, often called  the ``neutrino potential" in the literature, 
is the contribution from light Majorana-neutrino exchange. We give its expression in Eq. \eqref{nupot}. 
The potentials corresponding to higher-dimensional operators are 
$V_{6,7}(\vec q\sq)$, which were  derived in Ref.~\cite{Cirigliano:2017djv}
 and  are for convenience reproduced in Appendix \ref{app:neutrinopotentials567}, and $V_{ 9}(\vec q\sq)$,  given in Eq.~\eqref{scalar}.

It is useful to separate the total amplitude in Eq.~\eqref{eq:FullAmp} in various terms which differ by their leptonic structure. While $V_{ 9}(\vec q\sq)$ only consists of three different leptonic structures, two additional structures appear in $V_{3,6,7}(\vec q\sq)$ and we define
\bea\label{eq:TotAmp}
\mathcal A&=& \frac{g_A\sq G_F\sq m_e}{\pi R_A}\bigg[\mathcal{ A}_{\nu}\, \bar u (k_1)P_R C\bar u^T(k_2)+ \mathcal{ A}_{R}\, \bar u (k_1)P_L C\bar u^T(k_2)\\
&&+ \mathcal{ A}_{ E}\, \bar u (k_1)\g_0 C\bar u^T(k_2)\,\frac{E_1-E_2}{m_e}
+ \mathcal{ A}_{ m_e}\, \bar u (k_1) C\bar u^T(k_2)+ \mathcal{ A}_{ M} \,\bar u (k_1)\g_0 \g_5 C\bar u^T(k_2)\bigg]\,\,,\nn
\eea
where  $E_{1,2}$ ($k_{1,2}$) are the energies (momenta) of the electrons. Here we extracted an overall factor from the various sub-amplitudes $\mathcal{ A}_{i}$. In particular, a factor of $m_e/R_A$ is extracted,  where $m_e$ is the electron mass and $R_A =1.2\,A^{1/3}$ fm in terms of $A$, the  number of nucleons of the daughter nucleus. This normalization was chosen in order to align the definition of the various nuclear matrix elements with those appearing in the literature, but stress that in the final decay rate all the factors of $m_e/R_A$ will drop out.

The subamplitudes $\mathcal A_{i}$ depend on the Wilson coefficients of the $\Delta L=2$ operators,  
on hadronic matrix elements,  and nuclear matrix elements. The required LECs encoding hadronic matrix elements are listed in Table~\ref{Tab:LECs}. It turns out that all nuclear input that appears in Eq.\ \eqref{eq:TotAmp} can be expressed in terms of nine long-range NMEs ($M_F$, $M^{AA}_{GT}$, $M^{AP}_{GT}$, $M^{PP}_{GT}$, $M^{MM}_{GT}$, $M^{AA}_{T}$, $M^{AP}_{T}$, $M^{PP}_{T}$, $M^{MM}_{T}$) and six short-range matrix elements  ($M_{F,\, sd}$, $M^{AA}_{GT,\, sd}$, $M^{AP}_{GT,\, sd}$, $M^{PP}_{GT,\, sd}$, $M^{AP}_{T,\, sd}$, $M^{PP}_{T,\, sd}$). For the exact definitions we refer to Appendix~\ref{NME}. All NMEs, apart from one ($M_T^{AA}$), can be extracted from existing calculations of light- and heavy Majorana-neutrino exchange contributions. Furthermore, at LO in $\chi$PT the fifteen NMEs are related by five identities that can be used to further reduce the number of required many-body calculations or as a consistency check of the results \cite{Cirigliano:2017djv}. In Table~\ref{tab:comparison} we summarize several recent calculations of the NMEs, obtained by different groups applying different many-body methods. The NMEs often appear in certain linear combinations $\mathcal M_i$ that are defined below.

It is useful to further decompose the sub-amplitudes in terms of contributions from LNV operators of different dimension 
\begin{eqnarray}\label{MSM1}
\mathcal{ A}_{\nu} &=& \frac{m_{\beta \beta} }{m_e} \mathcal M^{(3)}_{\nu}  +  \frac{m_N}{m_e}  \mathcal M^{(6)}_{\nu} +    \frac{m^2_N}{m_e v} \mathcal M^{(9)}_{\nu}\,,
\nn \\
\mathcal A_R &=& \frac{m^2_N}{m_e v} \mathcal M^{(9)}_{R}\, , \nn\\
\mathcal A_{E} &=&  \mathcal M^{(6)}_{E,L} +  \, \mathcal M^{(6)}_{E,R} \, ,  \nn \\
\mathcal A_{m_e} &=& \mathcal M^{(6)}_{m_e,L} +  \,\mathcal M^{(6)}_{m_e,R} \,,\nn \\
\mathcal A_{ M} &=&  \frac{m_N}{m_e}    \, \mathcal M^{(6)}_{M} +  \frac{m_N^2}{m_e v} \mathcal M^{(9)}_{M}\,.
\end{eqnarray}

The subamplitude $\mathcal A_\nu$  multiplies the leptonic structure that arises from light Majorana-neutrino exchange,  from several long-range dim-6 and dim-7 contributions, and from short-range dim-9 contributions. We have therefore decomposed it in a component proportional to the electron-neutrino Majorana mass $m_{\beta\beta}$, and the additional terms  $\mathcal M^{(6)}_{\nu}$
and $\mathcal M^{(9)}_{\nu}$, generated, respectively, by dim-6 and -7, and by dim-9 LNV operators.
The short-distance component $\mathcal M^{(9)}_{\nu}$ arises from $V_9$ and always involves an additional power of $1/v$ with respect to the contribution from light Majorana-neutrino exchange. To compensate for this factor and for the absence of the neutrino mass, we have factored out two powers of $m_N$ in Eq.~\eqref{MSM1}.
In terms of the standard building blocks  defined in Appendix~\ref{NME}, 
the combination of NMEs $\mathcal M_i$ are defined as
\begin{eqnarray}\label{MSMstandard}
\mathcal M^{(3)}_{\nu}  & = & - V_{ud}^2 \left(  - \frac{1}{g_A^2} M_{F} +   \mathcal M_{GT}   +   \mathcal M_{T}  + 2 \frac{m_\pi^2 \, g_\nu^{NN}}{g_A^2} M_{F,\, sd}  \right), \\
 \mathcal M^{(6)}_{\nu} &=&   V_{ud}
\left( \frac{B}{m_N} (C^{(6)}_{\rm SL}  - C^{(6)}_{\rm SR} ) + \frac{m^2_\pi}{m_N v} \left( C^{(7)}_{\rm VL} - C^{(7)}_{\rm VR}\right) \right)  \mathcal M_{PS} + V_{ud} C^{(6)}_{\rm T}   \mathcal M_{T6}\,,\\
 \mathcal M^{(9)}_{\nu} &=& - \frac{1}{2 m_N^2} C_{\pi\pi \,\rm L}^{(9)} \left(\frac{1}{2} M^{AP}_{GT, sd} + M^{PP}_{GT, sd}
+ \frac{1}{2} M^{AP}_{T, sd} + M^{PP}_{T, sd}\right) \nn \\ & &  +   \frac{m_\pi^2}{2 m_N^2} C^{(9)}_{\pi N\, \rm L}   \left(M^{AP}_{GT, sd} + M^{AP}_{T, sd} \right)
- \frac{2}{g_A\sq} \frac{m_\pi^2}{m_N^2}C_{NN\, \rm L}^{(9)}\,  M_{F,\,sd}\,,
\end{eqnarray}
where $g_\nu^{NN} \sim O(F_\pi^{-2})$ is a new leading-order low-energy constant~\cite{Cirigliano:2018hja},
defined in Eq.\ \eqref{gnuNN}, and $B \equiv - \langle \bar{q} q \rangle/F_\pi^2  \simeq 2.7$~GeV at $\mu=2$~GeV in the $\overline{\rm MS}$ scheme. 
Only $C^{(9)}_{\rm 1L}$ and $C^{(9)}_{\rm 4L,\,5L}$ in $ \mathcal M^{(9)}_{\nu}$ receive matching contributions from  \textoverline{dim-7} operators \cite{Cirigliano:2017djv}, while the remaining terms are at least  \textoverline{dim-9} \cite{Graesser:2016bpz}. 
In the above expressions we have defined
 \begin{eqnarray}
 \mathcal M_{GT} &=&  M^{AA}_{GT} + M^{AP}_{GT}  +M^{PP}_{GT} + M^{MM}_{GT}\, ,\nn \\
 \mathcal M_{T} &=&  M^{AP}_{T} + M^{PP}_{T}+ M^{MM}_{T}\, ,\nn \\
   \mathcal M_{PS} &=& \frac{1}{2} M^{AP}_{GT} + M^{PP}_{GT}  + \frac{1}{2} M^{AP}_{T} + M^{PP}_{T}\, ,\nn \\
 \mathcal M_{T6} &=& 2 \frac{g^\prime_T  - g^{NN}_{\rm T}}{g_A^2} \frac{m^2_\pi}{m_N^2} M_{F,\, sd}  - \frac{8 g_T}{ g_M} \left( M_{GT}^{MM} + M_T^{MM}\right) 
+ g^{\pi N}_{\rm T} \frac{m_\pi^2}{4m_N^2}    \left(M^{AP}_{GT, sd} + M^{AP}_{T, sd} \right) \nn \\
& &+  g_{ \rm T}^{\pi\pi}\frac{m_\pi^2}{4 m_N^2} \left(  M^{PP}_{GT,\, sd}  + M^{PP}_{T,\, sd} \right)\,,
 \end{eqnarray}
in terms of  matrix elements defined in Appendix~\ref{NME}. 
$g_{\rm T}^{\pi\pi}$, $g_{\rm T}^{\pi N}$ and $g_{\rm T}^{NN}$ are the LECs of  $\pi \pi  ee $, $\pi NN\, ee$ and $NN\, NN\, ee$ short-range operators induced by $C^{(6)}_{\rm T}$,  
defined in Appendix \ref{app:neutrinopotentials567}.  

The subamplitude $\mathcal A_R$ only receives contributions from the dim-9 scalar operators involving right-handed electrons. It is only induced by \textoverline{dim-9} operators and is proportional to 
\bea
  \mathcal M^{(9)}_{R} =  \mathcal M^{(9)}_{\nu}\big |_{L\rightarrow R}\,.
\eea

The subamplitudes $\mathcal A_E$ and $\mathcal A_{m_e}$ are not affected by dim-9 operators and their expressions are therefore the same as in Ref.~\cite{Cirigliano:2017djv}, apart from additional short-range contributions that were not included there. They depend on the NME combinations 
 \begin{eqnarray}
\mathcal M^{(6)}_{E,L} &=&- \frac{V_{ud} C^{(6)}_{\rm VL}}{3}  \left(  \frac{g^2_V}{g_A^2} M_F  + \frac{1}{3} \left( 2 M_{GT}^{AA} + M_T^{AA} \right) +  \frac{6 g_{VL}^{E}}{g_A^2} M_{F,sd} \right) \,,  \\
\mathcal M^{(6)}_{E,R}& =&- \frac{V_{ud} C^{(6)}_{\rm VR}}{3}\left(  \frac{g^2_V}{g_A^2} M_F  - \frac{1}{3} \left( 2 M_{GT}^{AA} + M_T^{AA} \right) +  \frac{6 g_{VL}^{E}}{g_A^2} M_{F,sd} \right) \,, \nn \\
 \mathcal M^{(6)}_{m_e,L} &=&  \frac{V_{ud} C^{(6)}_{\rm VL}}{6}\left(   \frac{g^2_V}{g_A^2} M_F  - \frac{1}{3} \left( M_{GT}^{AA} - 4 M^{AA}_{T}\right) - 3 \left( M^{AP}_{GT} + M^{PP}_{GT} + M^{AP}_{T} + M^{PP}_{T}\right) 
\right. \nn \\ & & \left. - \frac{12g_{VL}^{m_e} }{ g_A^2}  M_{F, sd} 
 \right) \, , \nn \\
 \mathcal M^{(6)}_{m_e,R} &=&  \frac{V_{ud} C^{(6)}_{\rm VR}}{6}\left(   \frac{g^2_V}{g_A^2} M_F  + \frac{1}{3} \left( M_{GT}^{AA} - 4 M^{AA}_{T}\right) +  3 \left( M^{AP}_{GT} + M^{PP}_{GT} + M^{AP}_{T} + M^{PP}_{T}\right)
 \right. \nn \\ & & \left. - \frac{12g_{VR}^{m_e} }{ g_A^2}  M_{F, sd} 
 \right) \, ,\nn
\end{eqnarray}
where $g_{VL,VR}^{E,m_e}$ are defined in Appendix \ref{app:neutrinopotentials567}.

The final subamplitude, $\mathcal A_M$, receives a contribution from the dim-6 operator $C^{(6)}_{\rm VL}$ and from the dim-9 vector operators in Eq.\ \eqref{eq:Lag} which are, respectively, proportional to
\begin{eqnarray}\label{eq:vectorNMEs}
\mathcal M^{(6)}_M  &=&  V_{ud} C^{(6)}_{\rm VL } \Bigg[ 2\frac{g_A}{g_M}  \left(M_{GT}^{MM} + M_{T}^{MM} \right)
\nn\\
&&+ \frac{m^2_\pi}{m_N^2} \left( - \frac{2}{g_A^2}g^{NN}_{\rm VL}  M_{F,\, sd} 
+ \frac{1}{2} g^{\pi N}_{\rm VL} ( M^{AP}_{GT,\, sd} + M^{AP}_{T,\, sd})\right)
\Bigg]\,,\nn \\ 
\mathcal M^{(9)}_{M} & = & \frac{m_\pi^2}{m_N^2}  \bigg[  - \frac{2}{g_A^2} \left(g_6^{NN}C^{(9)}_{V}+ g_7^{NN} \tilde C^{(9)}_{V}  \right)  \, M_{F,\, sd} 
\nonumber \\
&&+ \frac{1}{2} \left(g_V^{\pi N}C_V^{(9)}+\tilde g_{V}^{\pi N} \tilde C_V^{(9)} \right)
\left( M^{AP}_{GT, \, sd}  + M^{AP}_{T, \, sd} \right) \bigg] \,.
\end{eqnarray}
The $\pi NN ee$ and $NN\, NN\, ee$ couplings  $g_{\rm VL}^{\pi N}$ and $g_{\rm VL}^{NN}$ are defined in Appendix \ref{app:neutrinopotentials567}.

The contributions from the various LNV operators to the subamplitudes $\mathcal A_i$ can be organized according to their scaling in the $\chi$PT expansion parameter $\epsilon_\chi = m_\pi/\Lambda_\chi$, where $\Lambda_\chi \sim 4 \pi F_\pi$.
The scaling is summarized in Table \ref{TabPC}. The Table indicates that low-energy dim-7 and -9 contributions are suppressed by at least one power of $\Lambda_\chi/ v$ compared to the dim-6 contributions. Because the dim-9 operators $C^{(9)(\prime)}_{2,3,4,5}$ can induce the pionic operators in Eq.~\eqref{eq:dim9pipi}, their contribution is enhanced by two powers of $\epsilon_\chi$ with respect to the dim-7 operators and the dim-9 scalar operators $C^{(9)(\prime)}_{1}$ and all vector operators $C_V^{(9)}$.

In Table \ref{tab:comparison} we report recent evaluations of the NMEs for experimentally interesting $0\nu\beta\beta$ emitters, 
obtained with three different many-body methods: the quasi particle random phase approximation (QRPA) \cite{Hyvarinen:2015bda}, the shell model  \cite{Menendez:2017fdf}, and the interacting boson model (IBM) \cite{Barea:2015kwa,Barea}.  
Refs.\  \cite{Hyvarinen:2015bda,Menendez:2017fdf,Barea:2015kwa,Barea}
were chosen as representative of each method because 
they organized their results in terms of the nine long-range 
($M_F$, \ldots, $M^{MM}_{T}$) and six short-range  ($M_{F,\, sd}$, \ldots, $M^{PP}_{T,\, sd}$) NMEs, while
computing the NMEs for light- and heavy-Majorana neutrino exchange. These 15 NMEs are sufficient to constrain all effective operators up to dim-9.
In the chiral EFT  power counting all NMEs are expected to be $\mathcal O(1)$, with the exception of $M^{MM}_{GT}$ and $M^{MM}_T$ that are formally suppressed by $\epsilon_\chi^2$.
From Table \ref{tab:comparison} we see that these expectations are well respected by the Fermi and Gamow-Teller NMEs. $M^{MM}_{GT}$ is larger than expected, which can be understood by taking into account that
the $\epsilon_\chi^2$ suppression is partially compensated by the large isovector nucleon magnetic moment, $g_M \simeq 4.7$. 
The tensor matrix elements are usually smaller because the tensor operator $S^{(12)}$ vanishes between $nn$ pairs in the $^1S_0$ channel, which is the dominant two-nucleon component  
\cite{Simkovic:2007vu}. Part of this suppression might be an artifact of the applied many-body methods, as Variational Monte Carlo calculations in lighter nuclei, such as $^{12}$Be and $^{12}$C,
show the ratio $M^{AP}_{T}/M^{AP}_{GT}$ to be roughly $25\%$ \cite{Pastore:2017ofx}.

\begin{table}
\center
$\renewcommand{\arraystretch}{1.5}
\begin{array}{l||rrr|rr|rr |rr}
 \text{NMEs} & \multicolumn{3}{c|}{\text{}^{76} \text{Ge}} & \multicolumn{2}{c|}{\text{}^{82} \text{Se}} & \multicolumn{2}{c|}{ \text{}^{130} \text{Te}} & \multicolumn{2}{c}{  \text{}^{136} \text{Xe}}  \\
& \text{\cite{Hyvarinen:2015bda}} &   \text{\cite{Menendez:2017fdf}}  
& \text{\cite{Barea:2015kwa,Barea}} & \text{\cite{Hyvarinen:2015bda}} &    \text{\cite{Menendez:2017fdf}} & \text{\cite{Hyvarinen:2015bda}} &   \text{\cite{Menendez:2017fdf}} & \text{\cite{Hyvarinen:2015bda}} &   \text{\cite{Menendez:2017fdf}} \\
 \hline
 M_F 			   & $-$1.74    &  $-$0.59 	& $-$0.68 	& $-$1.29 	&  $-$0.55	& $-$1.52    	&   $-$0.67	& $-$0.89  	&   $-$0.54 \\
 M_{GT}^{AA} 		   & 5.48       &  3.15	    	& 5.06 		& 3.87 		&  2.97		& 4.28    	&   2.97	& 3.16   	&   2.45 \\
 M_{GT}^{AP} 		   & $-$2.02    & $-$0.94	& $-$0.92 	& $-$1.46    	& $-$0.89   	& $-$1.74 	&  $-$0.97	& $-$1.19   	&  $-$0.79  \\
 M_{GT}^{PP} 		   & 0.66  	&  0.30		& 0.24 		& 0.48          &  0.28 	& 0.59   	&   0.31   	& 0.39   	&   0.25\\
 M_{GT}^{MM} 		   & 0.51 	& 0.22		& 0.17 		& 0.37       	& 0.20 		& 0.45 		&  0.23 	& 0.31 		&  0.19 \\
 M_T^{AA} 	   	   &  -    	& - 		& - 		&  -         	& -		&  -     	&  -      	&  -     	&   -	\\
 M_T^{AP} 	   	   & $-$0.35 	& $-$0.01	& $-$0.31 	& $-$0.27    	& $-$0.01 	& $-$0.50	&     0.01	& $-$0.28 	&     0.01	\\
 M_T^{PP} 	 	   & 0.10     	&    0.00	& 0.09  	& 0.08       	& 0.00 		& 0.16 		&  $-$0.01 	& 0.09 		&  $-$0.01 	 \\
 M_T^{MM}           	   & $-$0.04	&   0.00	& $-$0.04	& $-$0.03    	&   0.00 	& $-$0.06 	&    0.00	& $-$0.03 	&    0.00  \\\hline
 M_{F,\, sd}  	   	   & $-$3.46 	& $-$1.46	& $-$1.1  	& $-$2.53   	& $-$1.37 	& $-$2.97 	&  $-$1.61 	& $-$1.53   	&  $-$1.28	\\
 M^{AA}_{GT,\, sd}  	   &    11.1 	& 4.87		& 3.62		& 7.98    	& 4.54 		& 	10.1 	&  5.31 	&    5.71    	&  4.25  \\
M^{AP}_{GT,\, sd}	   & $-$5.35 	& $-$2.26 	& $-$1.37 	& $-$3.82    	& $-$2.09 	& $-$4.94 	&  $-$2.51 	& $-$2.80  	&  $-$1.99  \\
M^{PP}_{GT,\, sd}	   & 1.99 	& 0.82		& 0.42		& 1.42      	& 0.77 		& 1.86 		&  0.92 	& 1.06  	&   0.74\\
M^{AP}_{T,\, sd}	   & $-$0.85 	&  $-$0.05	&  $-$0.97	& $-$0.65    	&  $-$0.05	& $-$1.50 	&   0.07	& $-$0.92  	&   0.05		\\  
M^{PP}_{T,\, sd} 	   & 0.32 	&  0.02		&  0.38		& 0.24       	&  0.02		& 0.58 		&   $-$0.02	& 0.36  	&   $-$0.02 \\
\end{array}$
\caption{
Comparison of NMEs computed in the quasi particle random phase approximation  \cite{Hyvarinen:2015bda}, shell model  \cite{Menendez:2017fdf}, and interacting boson model  \cite{Barea:2015kwa,Barea}
for several nuclei  of experimental interest.  
All NMEs are expected to be $\mathcal O(1)$ in the chiral EFT power counting, except for $M^{MM}_{GT}$ and $M^{MM}_T$ which are formally suppressed by $\epsilon_\chi^2$. This suppression is partially compensated by large isovector nucleon magnetic moment $g_M \simeq 4.7$. The power counting expectations agree fairly well with the actual numerical values for the Fermi (F) and Gamow-Teller (GT) NMEs, while the tensor (T) NMEs are usually smaller.
The NMEs are defined in Eq.\ \eqref{MEdef} of Appendix \ref{NME}.}
\label{tab:comparison}
\end{table}

\begin{table}
\center
\footnotesize
\begin{tabular}{||c|ccccccccccc||}
 \hline \hline 
			    & $d$=3  &  $C^{(6)}_{\rm SL,\, SR}$       & $C^{(6)}_{\rm T}$ 		  &  $C^{(6)}_{\rm VL}$  & $C^{(6)}_{\rm VR}$  & $C^{(7)}_{\rm VL, \, VR}$  & $C^{(9)\,(\prime)}_{\rm 1R}$ & $C^{(9)(\prime)}_{\rm 1L}$ & $C^{(9)(\prime)}_{\rm 2R-5R}$ & $C^{(9)(\prime)}_{\rm 2L-5L}$ & $C^{(9)}_{\mathrm{vector}} $ \\
\hline
$m_e \mathcal A_{\nu}$   & $m_{\beta\beta}$ & $\Lambda_{\chi}$  & $\Lambda_{\chi} \epsilon^2_\chi$ &  $-$  &  $-$ & $ \frac{\Lambda_\chi^2}{v} \, \epsilon_\chi^2$  & $-$ & $\frac{\Lambda_\chi^2}{v} \, \epsilon_\chi^2$ & $ -$ & $\frac{\Lambda_\chi^2}{v} $ & $-$\\
$m_e \mathcal A_{R}$   & $-$ & $-$  & $-$ &  $-$  &  $-$ &   $-$  & $ \frac{\Lambda_\chi^2}{v} \, \epsilon_\chi^2$ & $-$ & $ \frac{\Lambda_\chi^2}{v}$ & $-$ & $-$ \\
$m_e \mathcal A_{M}$        & $-$ 	       & $-$ 		   & $-$ &  $\Lambda_{\chi} \epsilon^2_\chi$ & $-$ & $-$ & $-$ & $-$  & $-$ & $-$ &  $ \frac{\Lambda_\chi^2}{v} \epsilon_\chi^2$   \\
$m_e \mathcal A_{E}$        & $-$ 	       & $-$  		   & $-$ &  $\Lambda_{\chi} \epsilon^3_\chi$ & $\Lambda_{\chi} \epsilon^3_\chi$ & $-$ & $-$ & $-$  &$-$ & $-$  & $-$  \\
$m_e \mathcal A_{me}$       & $-$ 	       & $-$  		   & $-$ &  $\Lambda_{\chi} \epsilon^3_\chi$ & $\Lambda_{\chi} \epsilon^3_\chi$ & $-$ & $-$ & $-$  & $-$ & $-$  & $-$ \\
\hline
\end{tabular}
\caption{
Power-counting estimates  of the contribution of low-energy dim-3, -6, -7, and -9 operators to the amplitudes in Eq.~\eqref{eq:TotAmp}, 
in terms of  $m_{\beta \beta}$,  the Higgs vev  $v$,  and  $\epsilon_\chi \equiv m_\pi/\Lambda_\chi$, where $\Lambda_\chi \sim m_N \sim 1$ GeV. 
We take the electron mass and energies  to scale as $E_1 \sim E_2 \sim m_e \sim \Lambda_\chi \, \epsilon_\chi^3$. 
This  Table assumes the NMEs to follow the chiral EFT power counting. 
$C_{\mathrm{vector}}^{(9)}$ indicates any of the vector operators in Eq.~\eqref{eq:Lag}.
Finally, note that  to estimate the overall scaling of the amplitudes one needs to take into account that, 
up to  insertions of  dimensionless couplings, 
the Wilson coefficients  scale as follows:  
$m_{\beta \beta} = \mathcal O(v^2/\Lambda)$,      $C^{(6,7)}_i = \mathcal O(v^3/\Lambda^3)$,  
 $C^{(9)}_{\rm 1 L,\, 4L,\, 5L}= \mathcal O(v^3/\Lambda^3)$ or $\mathcal O(v^5/\Lambda^5)$ (depending on the underlying model), and 
$C^{(9)}_i = \mathcal O(v^5/\Lambda^5)$ for the remaining dim-9 operators. }\label{TabPC}
\end{table}

\subsection{Master formula for the $0\nu\beta\beta$ decay rate}
Using the amplitude in Eq.\ \eqref{eq:FullAmp}, the expression for the inverse half-life becomes \cite{Doi:1985dx,Bilenky:2014uka},
\bea \label{eq:InvHalfLife}
\left(T^{0\nu}_{1/2}\right)^{-1} = \frac{1}{8 \ln 2}\frac{1}{(2\pi)^5 }\int \frac{d^3k_1}{2E_1}\frac{d^3k_2}{2E_2} |\mathcal A |\sq F(Z,E_1)F(Z,E_2)\dt(E_1+E_2+E_f-M_i)\,\,.
\eea
Here $M_i$ is the mass of the decaying nucleus, while $E_{1,2}$  and $E_f$ are 
 the energies of the electrons and final daughter nucleus  in the rest frame of the decaying nucleus.
The functions $F(Z,E_i)$ are defined in Appendix \ref{Phase} and take into account the fact that the emitted electrons feel the Coulomb potential of the daughter nucleus and are therefore not plane waves. 

Using the decomposition of the amplitude in Eq.\ \eqref{eq:TotAmp} to separate the different leptonic structures, we obtain the final expression
\bea\label{eq:T1/2}
\left(T^{0\nu}_{1/2}\right)^{-1} &=& g_A^4 \bigg\{ G_{01} \, \left( |\mathcal A_{\nu}|\sq + |\mathcal A_{R}|\sq \right)
- 2 (G_{01} - G_{04}) \textrm{Re} \mathcal A_{\nu}^* \mathcal A_{R} 
+ 4G_{02} \,|\mathcal A_{E}|\sq \nn \\ & & + 2 G_{04} \left[|\mathcal A_{m_e}|\sq+{\rm Re} \left(\mathcal A_{m_e}^* (\mathcal A_{\nu} + \mathcal A_{R})\right)\right]
-2 G_{03}\,{\rm Re}\left[ (\mathcal A_{\nu} + \mathcal A_{R} )\mathcal A_{E}^*+2\mathcal A_{m_e} \mathcal A_{E}^*\right]
\nn\\
&&+ G_{09}\, |\mathcal A_{M}|\sq + G_{06}\, {\rm Re}\left[ (\mathcal A_{\nu} - \mathcal A_{R} )\mathcal A_{M}^*\right] \bigg\}\,. 
\eea
This `Master-formula'  describes the $0\nu\beta\beta$ decay rate up to \textoverline{dim-9} operators in the SM-EFT. 
It includes all contributions from the low-energy $\Delta L=2$ operators in Eq.~\eqref{LagDeltaL2} and takes into account all interference terms. It should provide a useful tool to constrain any model of high-scale LNV,  
using the most up-to-date hadronic and nuclear input.  
 A differential version of Eq.~\eqref{eq:T1/2} is given in Appendix \ref{Phase}. 
The various components in Eq.~\eqref{eq:T1/2}  can be obtained as follows: 
\begin{itemize}
\item  $G_{0i}$ are phase space factors defined in Appendix~\ref{Phase} and  their numerical values are given in Table~\ref{Tab:phasespace}. 

\item The five  sub-amplitudes ${\mathcal A}_\alpha$  ($\alpha \in \{ \nu,R,E,m_e,M \}$) corresponding to different leptonic bilinears are decomposed 
in Eq.~\eqref{MSM1}  in terms of contributions from LNV operators of different dimension,  generically denoted as ${\mathcal M}_{\alpha}^{(d)}$  with $d=3,6,9$. 

\item  Expressions for  ${\mathcal M}_{\alpha}^{(d)}$  can be found in Eqs.~\eqref{MSMstandard}-\eqref{eq:vectorNMEs}.  
Each  ${\mathcal M}_{\alpha}^{(d)}$ is given by a linear combination of terms that are products of: 
(i) short-distance Wilson coefficients,  which depend on the underlying LNV model;
(ii)  hadronic LECs,   whose current knowledge is summarized in Tab.~\ref{Tab:LECs};
(iii)  nuclear matrix elements  defined in Appendix~\ref{NME},    whose numerical values from different many-body methods  can be found in Tab.~\ref{tab:comparison}.

\item 
Several hadronic LECs are at the moment unknown. An assessment  of the ensuing theoretical uncertainty can
be obtained by varying the LECs in a range around the values of Table \ref{Tab:LECs}.  We stress that for all the operators in Eqs.\ \eqref{lowenergy6}, \eqref{lowenergy7}, and \eqref{eq:Lag}, with the exception of  $O^{\mu (\prime)}_{6,\ldots,9}$, the long-range component of the amplitude is reliably known, providing a solid estimate of the order of magnitude of the contribution of each operator.
The unknown LECs should affect such estimates by $\mathcal O(1)$ factors, but should not change the order of magnitude.

\end{itemize}

On a more technical note, it should be stressed that the decay-rate formula is expressed in terms of the Wilson coefficients at a low-energy scale $\mu \simeq 2$ GeV. In order to match the formula to specific BSM theories, some additional steps are required. At the high-energy scale $\Lambda$ where any beyond-the-SM fields are integrated out, we need to perform a matching calculation to gauge-invariant \textoverline{dim-5}, \textoverline{dim-7}, and \textoverline{dim-9} operators. The resulting operators need to be evolved down to the electroweak scale where they are matched to the operators in Eq.~\eqref{LagDeltaL2}. This procedure was completed  in Ref.~\cite{Cirigliano:2017djv} for the \textoverline{dim-7} operators, and below we study it for a particular BSM model where also relevant \textoverline{dim-9} operators are induced.   Finally, the low-energy EFT operators are evolved to the low-energy scale using the RGEs in Eqs.~\eqref{RGE6}-\eqref{RGE9vector}. The numerical factors of the last step are given in Appendix~\ref{App:RG}.
All the steps leading from  a generic LNV Lagrangian at scale $\Lambda$  to Eq.~\eqref{eq:T1/2} are illustrated in Fig.~\ref{landscape}.

\begin{table}
\center
\begin{tabular}{|c|cccc|}
\hline
\hline
\cite{Horoi:2017gmj}	    & $^{76}$Ge & $^{82}$Se & $^{130}$Te & $^{136}$Xe \\ 

\hline
$G_{01}$    & 0.22 & 1. & 1.4 & 1.5 \\
$G_{02}$    & 0.35 & 3.2 & 3.2 & 3.2 \\
$G_{03}$    & 0.12 & 0.65 & 0.85 & 0.86 \\
$G_{04}$    & 0.19 & 0.86 & 1.1 & 1.2 \\
$G_{06}$    & 0.33 & 1.1 & 1.7 & 1.8 \\
$G_{09}$    & 0.48 & 2. & 2.8 & 2.8 \\\hline
\hline
$Q/{\rm MeV} $ \cite{Stoica:2013lka} & 2.04& 3.0&2.5 & 2.5 \\
\hline\hline
\end{tabular}
\caption{Phase space factors in units of $10^{-14}$ yr$^{-1}$ taken from Ref.~\cite{Horoi:2017gmj}. More details are given in Appendix \ref{Phase}. The last row shows the $Q$ values for the different isotopes, where $Q = M_i - M_f -2m_e$.}
\label{Tab:phasespace}
\end{table}

\section{Single-coupling constraints }\label{single}
We now investigate the constraints from $0\nu\beta\beta$ limits on the low-energy $\Delta L=2$ operators in Eq.~\eqref{LagDeltaL2}.  
In particular, we apply the experimental limits \cite{Agostini:2018tnm,KamLAND-Zen:2016pfg,Alduino:2017ehq} (all at $90\%$ c.l.)
\begin{equation}
T^{0\nu}_{1/2}({}^{76}\mathrm{Ge}) > 8\cdot10^{25}\,\mathrm{yr}\,, \qquad T^{0\nu}_{1/2}({}^{130}\mathrm{Te}) > 1.5\cdot10^{25}\,\mathrm{yr}\,,\qquad T^{0\nu}_{1/2}({}^{136}\mathrm{Xe}) > 1.1\cdot10^{26}\,\mathrm{yr}\, .
\end{equation}
For operators of dimension six and higher, we interpret the limit as a lower bound on the scale of new $\Delta L=2$ physics, $\Lambda$. The operators in the left column of Table~\ref{tab:limits}  can be induced by \textoverline{dim-7} operators and we assume $C^{(d)}_i(\mu) =( v/\Lambda^{(d)}_i)^3$. The operators in the right column can only be induced by \textoverline{dim-9} operators and here we assume $C^{(d)}_i(\mu) = (v/\Lambda^{(d)}_i)^5$. As such, the probed scale in  the left column is typically significantly higher, $\mathcal O(100\,\mathrm{TeV})$,  than in the right column,  $\mathcal O(5\,\mathrm{TeV})$. However, in cases where the \textoverline{dim-7} operators predominantly induce dim-7 operators, the additional suppression of $\epsilon_\chi^2 \Lambda_\chi/v$ in Table~\ref{TabPC} makes the probed scale much lower and  comparable  to that in the right column.

We give the bounds in two cases. The top panel of Table \ref{tab:limits} shows  the limits obtained assuming that only one operator is active at the scale $\mu = 2$ GeV.
To highlight the impact of the QCD evolution, in the lower panel of Table \ref{tab:limits} we show the limits in the assumption that the operators are turned on at $\mu = m_W = 80.4$ GeV.
We can see that the QCD running gives $\mathcal O(1)$ corrections to the bounds. We thus find that the RGEs have a far milder effect than the $\Or(10^3)$ effects that were found for some operators in Ref.\ \cite{Gonzalez:2015ady}. The origin of this discrepancy is discussed in more detail in Appendix \ref{comparison-to-lit}.

In order to set these limits we had to make several assumptions. 
Firstly, we used the NMEs from  Ref.~\cite{Menendez:2017fdf}. Results from other groups and many-body methods
roughly differ by factor of 2 to 3, depending on the NME under consideration. In particular, for the light Majorana-neutrino exchange the relevant NME is $\mathcal M_\nu^{(3)}$ which differs by roughly a factor 2 between Refs.~\cite{Hyvarinen:2015bda,Menendez:2017fdf,Barea:2015kwa,Horoi:2017gmj} and this impacts the limit on $m_{\beta\beta}$ by the same amount. For the operators that scale as $ v^3/\Lambda^3$ or $v^5/\Lambda^5$ the NME uncertainties give an uncertainty on $\Lambda$ of roughly a factor $3^{1/3}\simeq 1.5$ and $3^{1/5}\simeq1.25$, respectively. 

The remaining uncertainty arises from the size of the LECs, in particular those associated with the $\Delta L=2$ pion-nucleon and nucleon-nucleon couplings. For the light Majorana-neutrino exchange in Weinberg's counting there appear no LO nucleon-nucleon LECs and there would only be a small uncertainty from higher-order chiral corrections. However, as demonstrated in 
Ref.~\cite{Cirigliano:2018hja}, renormalization requires that a  $\Delta L=2$ nucleon-nucleon operator is promoted to LO. Currently the contribution from this term has not been incorporated consistently in calculations for the heavy nuclei under considerations but estimates for light nuclei show that the nucleon-nucleon terms can alter the total amplitude by $\mathcal O(1)$ corrections \cite{Cirigliano:2018hja}, and the associated uncertainty is as large as the NME uncertainty.

 Similar uncertainties affect the limits on the \textoverline{dim-7} and \textoverline{dim-9} operators. For the limits in Table~\ref{tab:limits} we assumed $g^{NN}_{1} = g^{NN}_{6} = g^{NN}_{7} =  1$ and $g^{\pi N}_{1} = g^{\pi N}_V = \tilde g^{\pi N}_{V} =  g_{\rm T}'=1$. We furthermore assumed that the pion-exchange contributions saturate the amplitude for the operators which induce the pionic $\Delta L=2$ operators in Eq.~\eqref{eq:dim9pipi}. That is, we assumed Weinberg's counting and neglected the nucleon-nucleon contributions, $g_\nu^{NN}=g^{NN}_{2,3,4,5}   = g_{VL,VR}^{E,m_e}=0$, even though these terms are enhanced to LO by renormalization arguments. In addition, for the short-distance operators induced by the dimension-six operators, we assumed $g_{\rm T}^{\pi\pi, \pi N,NN}=g_{\rm VL}^{\pi N,NN}=0$.
Our limits are therefore affected by $\mathcal O(1)$ uncertainties and should be revisited once more is known about the LECs.

\begin{table}
\center
$
\renewcommand{\arraystretch}{1.5}\footnotesize
\begin{array}{|c||ccc||c||ccc|}
\hline
& \text{}^{76} \text{Ge} & \text{}^{130} \text{Te} & \text{}^{136} \text{Xe} & &  \text{}^{76} \text{Ge} & \text{}^{130} \text{Te} & \text{}^{136} \text{Xe}  \\\hline
m_{\beta\beta}(\mathrm{eV}) & 0.25  & 0.23 & 0.1 & & & & \\
 \hline  \hline
 \multicolumn{8}{|c|} { \mu = 2\, \text{GeV}} \\
 \hline
  \Lambda_{\text{SL}}^{(6)}  (\mathrm{TeV})  & 210  & 220 & 290 &  \Lambda_{\rm 1L}^{(9)(\prime)}    (\mathrm{TeV})   	    & 2.4  & 2.5 & 3.0\\
\Lambda_{\text{SR}}^{(6)} & 210  & 220 	& 290 	&  
\Lambda_{\rm 2L,\, 2R}^{(9)(\prime)}	    & 4.3& 4.4& 5.2 \\
\Lambda_{\text T}^{(6)}   & 200 & 210& 270 	&  
\Lambda_{\rm 3L, 3R}^{(9)\,(\prime)}	    & 3.4  & 3.5& 4.1\\
\Lambda_{\text{VL}}^{(6)} & 150  & 150 	& 200 	&  
\Lambda_{\rm 4L, 4R}^{(9)} 	    	    & 4.3  & 4.4& 5.1\\
\Lambda_{\text{VR}}^{(6)} & 28   & 30 	& 39    &  
\Lambda_{\rm 5L, 5R}^{(9)}	    		    & 5.7  & 5.9& 6.8 \\
\Lambda_{\text{VL}}^{(7)} & 6.5    & 7 		& 8.9   &  
\Lambda^{(9)}_{6,7,8,9}	    	    & 2.5 & 2.5& 3.0   \\
\Lambda_{\text{VR}}^{(7)} & 6.5    & 7 		& 8.9	& & & & \\
\Lambda_{\rm 1L}^{(9)}           & 11   	& 12 	& 16 & & & & \\
\Lambda_{\rm 4L}^{(9)}           & 29	& 30& 38& & & & \\
\Lambda_{\rm 5L}^{(9)}           & 47 	& 49	& 62  & & & & \\
\hline \hline
 \multicolumn{8}{|c|} { \mu = 80.4\, \text{GeV}} \\
 \hline
\Lambda_{\text{SL}}^{(6)} (\mathrm{TeV}) & 240  & 250 & 340     & 
 \Lambda_{\rm 1L}^{(9)(\prime)}   (\mathrm{TeV})   	  	    & 2.3  & 2.4 & 2.9\\
\Lambda_{\text{SR}}^{(6)} 	   & 240 	& 250 	& 340 	&  
\Lambda_{\rm 2L,\, 2R}^{(9)(\prime)}	    & 4.8 & 4.9& 5.7 \\
\Lambda_{\text T}^{(6)}   	   & 190  	& 200	& 260 	&  
\Lambda_{\rm 3L, 3R}^{(9)\,(\prime)}	    & 3.7& 3.8& 4.4 \\
\Lambda_{\text{VL}}^{(6)} 	   & 150  	& 150 	& 200 	& 
 \Lambda_{\rm 4L, 4R}^{(9)} 	    	    & 5.3  & 5.4 & 6.3 \\
\Lambda_{\text{VR}}^{(6)} 	   & 28   	& 30 	& 39    & 
 \Lambda_{\rm 5L, 5R}^{(9)}	    		    & 6.7  & 6.9 & 8.0 \\
\Lambda_{\text{VL}}^{(7)} 	   & 6.5    	& 7 	& 8.9   &  
\Lambda^{(9)}_{6,8}	  &2.6 & 2.7& 3.1   \\
\Lambda_{\text{VR}}^{(7)} 	   & 6.5    	& 7 	& 8.9	& 
\Lambda^{(9)}_{7,9}	   &2.2 & 2.3& 2.7\\
\Lambda_{\rm 1L}^{(9)}             & 10   	& 11 	& 15 &
  & & & \\
\Lambda_{\rm 4L}^{(9)}             & 41   	& 42& 54  & & & & \\
\Lambda_{\rm 5L}^{(9)}             & 61   	& 64& 82  & & & & \\
\hline
\end{array}$
\caption{
The Table shows the upper limits on $|m_{\bt\bt}|$ 
and lower limits on the scales, $\Lambda^{(d)}_i$, related to the dim-6, dim-7, and dim-9 operators from the GERDA \cite{Agostini:2018tnm}, CUORE \cite{Alduino:2017ehq}, and KamLAND-Zen \cite{KamLAND-Zen:2016pfg} experiments.
In the left column, we assume $C^{(d)}_i(\mu) = ( v/\Lambda^{(d)}_i)^3$ for the dim-6, dim-7, and dim-9 operators that receive contributions from gauge-invariant \textoverline{dim-7}  operators at the electroweak scale. For the remaining dim-9 operators we assume $C^{(9)}_i(\mu) = (v/\Lambda^{(9)}_i)^5$ in the right column.  We use the nuclear matrix elements of Ref.\ \cite{Menendez:2017fdf} and the values of the LECs described in the text.  
To illustrate the effect of the QCD running, we show the limits in two cases. In the upper panel of the Table we assume that the operators are turned on one at a time at the scale $\mu= 2$ GeV,
while in the lower panel we take $\mu = m_W$.
}\label{tab:limits}
\end{table}

\section{An explicit example: the minimal left-right symmetric model}\label{sec:LR}

The formula for the total decay rate in Eq.~\eqref{eq:T1/2} is given in terms of Wilson coefficients of effective operators. This formula can be matched to any model of heavy BSM physics that contributes to $0\nu\beta\beta$. In what follows we demonstrate this by considering an explicit BSM model: the minimal left-right symmetric model (mLRSM) \cite{Pati:1974yy, Mohapatra:1974hk, Senjanovic:1975rk}. The mLRSM has been studied in the context of $0\nu\beta\beta$ in great detail \cite{Prezeau:2003xn,Bambhaniya:2015ipg,Dev:2014xea,Tello:2010am,Nemevsek:2011aa,Barry:2013xxa}.  
This scenario is interesting because 
it provides an elegant explanation of P- or C-violation through spontaneous symmetry breaking and 
it allows for the generation of neutrino masses at a relatively low scale,  within reach of the LHC or possible future colliders. Furthermore, within the model there appear \textoverline{dim-5}, \textoverline{dim-7}, and \textoverline{dim-9} contributions to $0\nu\beta\beta$. As such, the mLRSM is particularly well suited to demonstrate that Eq.~\eqref{eq:T1/2} is able to capture all of these effects. 
We start by giving a brief overview of the model, and refer to Refs.~\cite{Pati:1974yy, Mohapatra:1974hk,Senjanovic:1975rk,Senjanovic:1978ev, Mohapatra:1986uf} and Appendix~\ref{App:LR} for a more detailed discussion. We stress that we do not aim to perform a full study of the model. Our main goal here is to illustrate   the EFT framework and in particular the use of  Eq.~\eqref{eq:T1/2}.

The model is based on the gauge group  $SU(3)_c\times SU(2)_L \times SU(2)_R \times U(1)_{B-L}$ and we adopt the version of the model in which charge conjugation is conserved at high energies~\footnote{This assumption gives rise to a relatively simple expression for the Dirac mass matrix, see Eq.\ \eqref{Eq:MD}, and somewhat simplifies our numerical analysis.  In principle, one could perform a similar analysis for the P-symmetric case, however, in what follows we will only briefly mention the differences with the C-symmetric case.}, see e.g.\ Ref.\ \cite{Maiezza:2010ic}. 
The fermions are assigned to representations of the above gauge group as follows
\bea 
Q_L &=&\bma u_L\\d_L\ema \in (3,2,1,1/3)\,, \qquad Q_R = \bma u_R\\d_R\ema\in (3,1,2,1/3)\,,\nn\\
L_L &=&\bma \nu_L\\l_L\ema \in (1,2,1,-1)\,, \qquad L_R = \bma \nu_R\\l_R\ema\in (1,1,2,-1)\,.
\eea
The introduction of  right-handed neutrinos, $\nu_R$, is required by the symmetries of the model.

In addition to the fermions, the mLRSM involves several scalar fields, namely a bidoublet transforming as $\phi\in (1, 2,2^*,0)$, as well as two triplet fields, $\Delta_{L,R}$ assigned to $(1,3,1,2)$ and $(1, 1,3,2)$, respectively.
These fields can be written as 
\bea \phi = \bma \phi_1^0 & \phi_2^+\\ \phi_1^- & \phi_2^0 \ema ,\qquad
\Delta_{L,R} = \bma \delta^+_{L,R}/\sqrt{2} & \delta^{++}_{L,R} \\ \delta^0_{L,R} & -\delta^+_{L,R}/\sqrt{2} \ema ,
\label{scalars}\eea
and transform as $\phi\to U_L\phi U_R^\dagger$, $\Dt_{L,R}\to U_{L,R}\Dt_{L,R} U_{L,R}^\dagger$ under $SU(2)_{L,R}$ transformations. The neutral components of these fields obtain the following vacuum expectation values (vevs)
\bea\label{vevs}
\langle \phi^0_1 \rangle = \kappa/\sqrt{2}\,,\qquad\langle \phi^0_2\rangle = \kappa' e^{i\al}/\sqrt{2}\,,\qquad 
\langle \delta_{L}^0 \rangle = v_{L}e^{i\theta_L}/\sqrt{2}\,,\qquad \langle \delta_{R}^0 \rangle = v_R/\sqrt{2}\,.
\eea 
In the first step of symmetry breaking the vev of the right-handed triplet, $v_R$, breaks the charge-conjugation symmetry or parity and $SU(2)_L \times SU(2)_R \times U(1)_{B-L}$ down to $SU(2)_L \times U(1)_{Y}$. This vev  defines the high scale of the model 
and gives the main contribution to the masses of the
right-handed gauge bosons, the right-handed neutrinos, and the heavy Higgs fields. The vevs of the bidoublet, $\ka$ and $\ka'e^{i\al}$,  break $SU(2)_L \times U(1)_{Y}$ to $U(1)_{\text{em}}$, and are of the order of the electroweak scale. $v_L$ contributes to the masses of the light neutrinos and to the $\rho$ parameter, and is therefore required to be much smaller than the other vevs. 

Apart from the kinetic terms  and the scalar potential, the Lagrangian contains interactions between fermions and scalars that give rise to the $\Dt L=2$ operators of interest 
\bea \label{eq:LRLagrangian}
\vL &=& 
-\bar Q_L\big( \GA \phi + \TG \tilde \phi \big)Q_R - \bar L_L\big( \GA_l \phi + \TG_l \tilde \phi \big)L_R- \left[
\overline{L_L^c}  \,  i\tau_2\Dt_L M_L L_L+ (L\rightarrow R)\right]+\text{h.c.}\,,\eea
 in terms of $\tilde \phi=\tau_2 \phi^* \tau_2$
 (which also transforms as $\tilde \phi \to U_L \tilde \phi U_R^\dagger$), two quark Yukawa matrices, $\Gamma$ and $\tilde \Gamma$, two lepton Yukawa matrices, $\Gamma_l$ and $\tilde \Gamma_l$, and the symmetric $3\times 3$ matrices   $M_{L}$ and $M_{R}$. Charge-conjugation (parity) invariance implies that the latter two matrices are related by   $M_L=M_R^\dagger$ ($M_L= M_R)$, while the Yukawa matrices are forced to be symmetric (hermitian). After the scalar fields obtain their vevs, the Yukawa terms induce the Dirac masses for the quarks and leptons, while the $M_{L,R}$ terms induce neutrino Majorana masses and $\Dt L=2$ interactions.  
In the case of C-invariant Yukawa couplings,  there is a relation between the left- and right-handed CKM matrices, $V_L$ and $V_R$, namely, $V_L =  K_u V_R^* K_d$, where $K_{u,d}$ are diagonal matrices of phases. Instead, if one assumes parity to be preserved at high energies, $V_R$ can be expressed in terms of quark masses, $V_L$,  $\ka'/\ka$, and $ \sin\al$ \cite{Senjanovic:2015yea}.

\subsection{Matching to SM-EFT operators}
We integrate out the heavy fields (with masses $\sim v_R$)  at a scale just below the one at which
$\Dt_R$ obtains a vev, so  that $SU(2)_R$ is broken while $SU(2)_L$ is intact. The heavy fields that contribute to $\Dt L=2$ interactions are the right-handed neutrinos  and gauge fields, $\nu_R$ and  $W_R$, as well as the heavy scalar fields, $\vp_H$, $\dt_R^{++}$, and $\Dt_L$~\footnote{Here $\vp_H$ refers to the combination of $SU(2)_L$ doublets in $\phi$ that obtains a mass of $\Or(v_R)$, while we identify the remaining doublet with the SM Higgs doublet, denoted by $\vp$, see Appendix \ref{App:LR} for details. Note that the  scalar fields  $\dt_R^+$ and $\dt_R^0$ do not mediate $\Dt L=2$ effects since they are absorbed in the now massive $SU(2)_R$ gauge bosons.}. 
At the scale $\mu\simeq m_{W_R}$, the mLRSM then gives matching contributions to the following gauge-invariant effective Lagrangian
\bea\label{eq:LREFT}
\mathcal L &=&  \epsilon_{kl} \epsilon_{mn}\left(L_k^T  \mathcal C^{(5)}C L_m \right)\varphi_l \varphi_n \nn \\
&& +   \epsilon_{ij}(L_i^T C \gamma_\mu e)\varphi_j\left[ \mathcal C^{(7)}_{Leu\bar d  \varphi}\,\bar d_R \gamma^\mu u_R+ \mathcal  C^{(7)}_{L\varphi De}\epsilon_{mn} \varphi_m (D^\mu \varphi)_n\right] \\
&& + \bar e_R C \bar e_R^T\,\left[\mathcal C_{eeud}^{(9)} \bar u_R \gamma_\mu d_R\,  \bar u_R \gamma^\mu d_R + \mathcal  C_{ee\varphi u d}^{(9)} \bar u_R \gamma_\mu d_R 
( (i D_\mu \varphi)^\dagger  \tilde \varphi)
+ \mathcal C_{ee\varphi D}^{(9)}( (i D_\mu \varphi)^\dagger  \tilde \varphi)^2\right],\nn
\eea
where the first, second, and third lines correspond to $\Delta L=2$ \textoverline{dim-5}, \textoverline{dim-7}, and \textoverline{dim-9} operators, respectively. We followed Ref.~\cite{Cirigliano:2017djv} for the definition of the \textoverline{dim-7} operators. 

The matching condition for the  \textoverline{dim-5} operator is
\bea\label{eq:LRdim5}
\mathcal C^{(5)}& =&\left({\frac{1}{2}} M_D^T M_{\nu_R}^{-1}M_D-\frac{\sqrt{2}v_Le^{i\theta_L}}{v\sq}M_L\right)\,,
\eea
where  $M_{\nu_R} = \sqrt{2}v_R M_R^\dagger$ is the mass matrix for the right-handed neutrinos and $M_D =\frac{\Gamma^\dagger_l+\xi e^{i\al}\tilde \Gamma^\dagger_l}{\sqrt{1+\xi\sq}}$ is the Dirac Yukawa matrix, with $\xi=\ka'/\ka$. Here the first term arises from integrating out the heavy right-handed neutrinos, corresponding to the usual type-I see-saw mechanism. The second term arises from a type-II see-saw mechanism and is induced by integrating out the $\Dt_L$ fields. This contribution is proportional to $v_R \bt_i/m_{\Dt_L}\sq$, where $\bt_i$ are the parameters in the scalar potential that couple $\Dt_L$ to the SM Higgs doublet. By use of the minimum equations  this can be written in terms of the vev $v_L$, as was done in Eq.\ \eqref{eq:LRdim5}, see Appendix \ref{App:LR} for details.

The \textoverline{dim-7} operators are induced by combining the right-handed neutrino Majorana mass term with the right-handed charged-current interaction. The $W_R$ boson can either couple to right-handed quarks or to $(\tilde \vp^\dagger D_\mu \vp)$, which leads to mixing with $W_L$ proportional to $\xi$ after EWSB. In total, we have
\bea\label{dim7match}
\mathcal C^{(7)}_{Leu\bar d \vp}  ={\frac{1}{v_R^2}} \left(V_R^{ud}\right)^* \left( M_D^TM_{\nu_R}^{-1} \right)_{ee} ,\qquad \mathcal C^{(7)}_{L\vp De}  =\frac{2i\xi e^{i\al}}{1+\xi^2}\frac{\mathcal C^{(7)}_{Leu\bar d \vp}}{\left(V_R^{ud}\right)^*}\,.
\eea

The \textoverline{dim-9} operators result from diagrams in which either $\nu_R$ or $\dt^{++}_R$ couples to two right-handed electrons and two $W_R$ bosons. Again, the $W_R$ bosons can couple to right-handed quarks, or mix with the $W_L$ boson and we obtain
\bea\label{dim9match}
\mathcal C^{(9)}_{eeud} &=&-\frac{1}{2 v_R^4} \left(V_{R}^{ud}\right)^2\left[\left(M_{\nu_R}^\dagger\right)^{-1}+\frac{2}{m_{\Delta_R}\sq} M_{\nu_R}\right]_{ee},\nn\\
\mathcal C^{(9)}_{ee\varphi u d}&=&-4\frac{\xi e^{-i\al}}{1+\xi\sq}\frac{\mathcal C^{(9)}_{eeud}}{V_{R}^{ud}}\,,\qquad 
\mathcal C_{ee \varphi D}^{(9)}=4\frac{\xi\sq e^{-2i\al}}{(1+\xi\sq)\sq}\frac{\mathcal C^{(9)}_{eeud}}{\left(V_{R}^{ud}\right)\sq}\,.
\eea
The above matching conditions hold at the scale $m_{W_R}$, so that the QCD running between the right-handed and the electroweak scale has to be taken into account before integrating out the heavy SM fields. Of the induced operators in Eq.\ \eqref{eq:LREFT}, only  $ \mathcal C_{eeud}^{(9)}$ is affected by QCD evolution. The other operators do not involve colored particles or consist of currents such that their QCD anomalous dimensions vanish. 
$\mathcal C_{eeud}^{(9)}$ follows the same RGEs as $C_1$ in Eq.\ \eqref{RGE9scalar}, which leads to,
\bea
\mathcal C_{eeud}^{(9)}(m_W) =\left( \frac{\al_s(m_{t})}{\al_s(m_{W})}\right)^{6/23}\left( \frac{\al_s(m_{W_R})}{\al_s(m_{t})}\right)^{2/7}\mathcal C_{eeud}^{(9)}(m_{W_R})\,, 
\eea
This gives $\mathcal C_{eeud}^{(9)}(m_W) =0.88\,\mathcal C_{eeud}^{(9)}(m_{W_R})$, for the value $m_{W_R}=4.5$ TeV used below. 

The  \textoverline{dim-7} and  \textoverline{dim-9} operators are, respectively, suppressed by $(v/v_R)^2$ and $(v/v_R)^4$ compared to the  \textoverline{dim-5} operator. However,  the small masses of the SM neutrinos imply that the Dirac neutrino couplings, $M_D$, should be small $\sim m_e/v\simeq 10^{-6}$ for a right-handed scale in the $1-10$ TeV range. This implies that the \textoverline{dim-7} and  \textoverline{dim-9} operators actually scale as $M_D^{-1}(v/v_R)^2$ and $M_D^{-2}(v/v_R)^4$, with respect to the 
 \textoverline{dim-5} operator. The smallness of $M_D$ can therefore compensate for the powers of $(v/v_R)^2$. These estimates suggest that the \textoverline{dim-7} and  \textoverline{dim-9} cannot be neglected for right-handed scales within reach of collider experiments.
 
 \subsection{Matching to $SU(3)_c \times U(1)_{\rm em}$-invariant operators and the $0\nu\beta\beta$ decay rate}
After integrating out the heavy SM fields at the electroweak scale the $SU(2)_L$-invariant operators induce the following effective neutrino mass
\bea \label{eq:LRmatchmw5}
m_{\bt\bt}& =&-v^2 \left(\mathcal  C^{(5)}\right)_{ee}.
\eea
At dim-6 we have the following contributions \cite{Cirigliano:2017djv},
\bea\label{eq:LRmatchmw7}
C_{\rm VL}^{(6)}=-iV_L^{ud }\frac{v^3}{\sqrt{2}} \left(\mathcal C^{(7)}_{L\vp De}\right)^* ,\qquad C_{\rm VR}^{(6)}=\frac{v^3}{\sqrt{2}} \left(\mathcal C^{(7)}_{Leu\bar d \vp}\right)^*  .
\eea
Finally, at dim-9 we obtain
\bea
\label{eq:LRMatchmw}
C_{\rm 1R}^{(9)\,\prime}(m_W) &=& v^5\,\mathcal C_{eeud}^{(9)}(m_W),
\nn \\
 C_{\rm 4R}^{(9)}(m_W) &=& -v^5\,V_L^{ud}\mathcal C_{ee\varphi u d}^{(9)}(m_W),
 \nn \\
C_{\rm 1R}^{(9)}(m_W) &=&  v^5\,\left(V_L^{ud}\right)\sq C_{ee\varphi D}^{(9)}(m_W).
\eea
These operators still have to be evolved to $\mu\simeq 2$ GeV. The running of the dim-9 operators is discussed in Section\ \ref{sec:2} and induces $C_{\rm 5R}^{(9)}$ in addition to the couplings in Eq.\ \eqref{eq:LRMatchmw}, see Appendix \ref{App:RG} for explicit solutions to the RGEs. Instead, $m_{\bt\bt}$, $C_{\rm VL}^{(6)}$, and $C_{\rm VR}^{(6)}$ do not evolve under QCD. 

\subsection{Matching in case of light right-handed neutrinos}\label{LR-lightNuR}
So far we have been assuming that the right-handed neutrinos have $\Or(v_R)$ masses  and can be integrated out simultaneously with the right-handed $W$ boson. In principle, it is possible for the right-handed neutrinos to have masses well below the right-handed scale if the $M_R$ couplings in Eq.\ \eqref{eq:LRLagrangian} are small. In this case the right-handed neutrinos need to be integrated out separately from the BSM particles with $\Or(v_R)$ masses. Scenarios with  $m_{W_R}>m_{\nu_R}>\Lambda_\chi$ can straightforwardly be included in our framework by slightly modifying the matching discussed above.

At the scale $\mu=m_{W_R}$ we now match to an EFT in which we have integrated out the $W_R$, $\vp_H$, $\dt_R^{++}$ and $\Dt_L$ fields, but not $\nu_R$. The resulting EFT contains apart from the SM-EFT Lagrangian, additional gauge-invariant operators that involve the right-handed neutrinos which are now relevant low-energy degrees of freedom.
This induces several additional operators with respect to those in the SM-EFT. Those relevant to our analysis are \cite{Cirigliano:2012ab}
\bea\label{eq:EFTnuR}
\vL \supset \bar e_R \gamma_\mu \nu_R \left[C_R^{(6)}\, \bar  u_R\gamma^\mu d_R + C_L^{(6)}\, i(D^\mu \vp)^\dagger \tilde \vp\right]+{\rm h.c.}\,.
\eea
The matching at $\mu=m_{W_R}$ is then given by
\bea\label{eq:LRmatchNuR}
\mathcal C^{(5)}& =&-\frac{\sqrt{2}v_Le^{i\theta_L}}{v\sq}M_L\,,\qquad
\mathcal C^{(7)}_{Leu\bar d \vp}  =0,\qquad \mathcal C^{(7)}_{L\vp De} = 0\,,\nn\\
C_{R}^{(6)}&=&-\frac{1}{v_R\sq}V_R^{ud},\qquad C_{L}^{(6)} = 2\frac{\xi e^{-i\al}}{1+\xi\sq} \frac{C_R^{(6)}}{V_R^{ud}}\,,\nn\\
\mathcal C^{(9)}_{eeud} &=&-\frac{1}{v_R^4} \left(V_{R}^{ud}\right)^2\frac{1}{m_{\Delta_R}\sq} \left(M_{\nu_R}\right)_{ee},\nn\\
\mathcal C^{(9)}_{ee\varphi u d}&=&-4\frac{\xi e^{-i\al}}{1+\xi\sq}\frac{\mathcal C^{(9)}_{eeud}}{V_{R}^{ud}}\,,\qquad 
\mathcal C_{ee \varphi D}^{(9)}=4\frac{\xi\sq e^{-2i\al}}{(1+\xi\sq)\sq}\frac{\mathcal C^{(9)}_{eeud}}{\left(V_{R}^{ud}\right)\sq}\,.
\eea
Apart from the contributions to $C_{L,R}^{(6)}$, this is equivalent to the matching in Eqs.\ \eqref{eq:LRdim5}, \eqref{dim7match}, and \eqref{dim9match}, without the contributions induced by the right-handed neutrinos. Note that the  $C_{L,R}^{(6)}$ coefficients do not evolve at one loop in QCD.

There are now two cases we can consider, namely, $m_W<m_{\nu_R}<m_{W_R}$ and  $\Lambda_\chi<m_{\nu_R}<m_{W}$. Starting with the former case, one can use the RGEs in Eq.\ \eqref{RGE9scalar} to evolve the $\mathcal C^{(9)}_{eeud}$ coefficient from $m_{W_R}$ to $m_{\nu_R}$, at which point we move from a theory involving right-handed neutrinos to the SM-EFT. The corresponding matching equations are given by
\bea
\mathcal C^{(5)}(m_{\nu_R}^-)& =&
\mathcal C^{(5)}(m_{\nu_R}^+)+{\frac{1}{2}} M_D^T M_{\nu_R}^{-1}M_D\,,\nn\\
\mathcal C^{(7)}_{Leu\bar d \vp} & =&-\left(M_D^T M_{\nu_R}^{-1}\right)_{ee}\left(C_R^{(6)}\right)^*,\qquad \mathcal C^{(7)}_{L\vp De} = \frac{2i\xi e^{i\al}}{1+\xi\sq}\frac{\mathcal C_{Leu\bar d\vp}^{(7)}}{\left(V_R^{ud}\right)^*}\,,\nn\\
\mathcal C^{(9)}_{eeud}(m_{\nu_R}^-) &=&
\mathcal C^{(9)}_{eeud}(m_{\nu_R}^+)-\frac{1}{2}\left(C_R^{(6)}\right)\sq \left(M_{\nu_R}^\dagger\right)^{-1}_{ee}
,\nn\\
\mathcal C^{(9)}_{ee\varphi u d}(m_{\nu_R}^-)&=&\mathcal C^{(9)}_{ee\varphi u d}(m_{\nu_R}^+)+C_L^{(6)}C_R^{(6)} \left(M_{\nu_R}^\dagger\right)^{-1}_{ee}
\,,\nn\\
\mathcal C_{ee \varphi D}^{(9)}(m_{\nu_R}^-)&=&\mathcal C_{ee \varphi D}^{(9)}(m_{\nu_R}^+)-\frac{1}{2}\left(C_L^{(6)}\right)\sq \left(M_{\nu_R}^\dagger\right)^{-1}_{ee}\,,
\eea
where $m_{\nu_R}^+$ and $m_{\nu_R}^-$ indicate scales just above and below the $m_{\nu_R}$ threshold.
Below this threshold one can employ the RGEs in Eq.\ \eqref{RGE9scalar} to evolve the couplings to the electroweak scale, where the matching conditions in Eqs.\ \eqref{eq:LRmatchmw5}, \eqref{eq:LRmatchmw7}, and \eqref{eq:LRMatchmw} still apply, and use the RGEs discussed in Section\ \ref{sec:2} to evolve the dim-9 operators to $\Lambda_\chi$. 

In the case that $m_{\nu_R}$ lies below the electroweak scale the matching is again slightly different. One now evolves the couplings in Eq.\ \eqref{eq:LRmatchNuR} directly to the electroweak scale. Here, the same matching conditions as in Eqs.\ \eqref{eq:LRmatchmw5}, \eqref{eq:LRmatchmw7}, and \eqref{eq:LRMatchmw} apply. One can then  use the RGEs discussed in Section\ \ref{sec:2} to evolve the dim-9 operators to $m_{\nu_R}$, where the right-handed neutrinos are finally integrated out. At this scale we obtain the following matching conditions
\bea
m_{\bt\bt}(m_{\nu_R}^-) &=& m_{\bt\bt}(m_{\nu_R}^+) - {\frac{v\sq}{2}} \left(M_D^T M_{\nu_R}^{-1}M_D\right)_{ee}\,,\nn\\
C_{\rm VL}^{(6)}&=&V_L^{ud }\frac{v^3}{\sqrt{2}} \left(M_D^T M_{\nu_R}^{-1}\right)_{ee}^*  C_L^{(6)},\qquad C_{\rm VR}^{(6)}=-\frac{v^3}{\sqrt{2}} \left[\left(M_D^T M_{\nu_R}^{-1}\right)_{ee}C_R^{(6)}\right]^*  \,,\nn\\
C_{\rm 1R}^{(9)\,\prime}(m_{\nu_R}^-) &=&C_{\rm 1R}^{(9)\,\prime}(m_{\nu_R}^+) - \frac{v^5}{2}\left(C_R^{(6)}\right)\sq \left(M_{\nu_R}^\dagger\right)^{-1}_{ee}\,,
\nn \\
 C_{\rm 4R}^{(9)}(m_{\nu_R}^-) &=&  C_{\rm 4R}^{(9)}(m_{\nu_R}^+) -v^5\,V_L^{ud}C_L^{(6)}C_R^{(6)} \left(M_{\nu_R}^\dagger\right)^{-1}_{ee},
\qquad  C_{\rm 5R}^{(9)}(m_{\nu_R}^-) =  C_{\rm 5R}^{(9)}(m_{\nu_R}^+)\,, \nn \\
C_{\rm 1R}^{(9)}(m_{\nu_R}^-) &=& C_{\rm 1R}^{(9)}(m_{\nu_R}^+)-\frac{v^5}{2}\left(V_L^{ud}C_L^{(6)}\right)\sq \left(M_{\nu_R}^\dagger\right)^{-1}_{ee}.
\eea
Although the matching conditions are somewhat different for the three cases, $m_{\nu_R}\sim m_{W_R}$, $m_{\nu_R}< m_{W_R}$, and $m_{\nu_R}< m_{W}$, in practice these differences only affect the running factors associated with the contributions from the right-handed neutrinos. Since the RGEs lead to factors of $\Or(1)$, the numerical impact of the differences in the matching are also  $\Or(1)$.

The above matching shows that it is in principle straightforward to include new light degrees of freedom, such as right-handed neutrinos, in the EFT framework. In case of the right-handed neutrinos in the left-right model (LRM) this  requires the inclusion of the two new operators in Eq.\ \eqref{eq:EFTnuR}, involving $\nu_R$, with respect to those that appear in the SM-EFT. Compared to the case of heavy right-handed neutrinos only several RG factors need to be changed. It should be stressed that the above matching does not allow one to describe even lighter fields, below $\Lambda_\chi$. In this case one would have to keep the new light fields as degrees of freedom in the chiral EFT. Although the construction of the chiral Lagrangian will be analogous to the case discussed here, we leave the $m_{\nu_R}<\Lambda_\chi$ scenario to future work and only discuss the $m_{\nu_R}>\Lambda_\chi$ possibility in what follows.

\subsection{Discussion}

Irrespective of the mass of the right-handed neutrinos, after accounting for the QCD evolution and matching, we can apply the formula for the inverse half-life: Eq.~\eqref{eq:T1/2}.
Looking at the amplitudes discussed in Section\ \ref{Master}, we infer that the   \textoverline{dim-5} term induces the standard light-neutrino exchange contribution which is proportional to $\mathcal A_\nu$. The   \textoverline{dim-7} terms give rise to $\mathcal A_E$ and $\mathcal A_{m_e}$ (via $C_{\rm VL}^{(6)}$ and $C_{\rm VR}^{(6)}$) and to $\mathcal A_M$ (via $C_{\rm VL}^{(6)}$ only). Finally,  the  \textoverline{dim-9} operators induce $\mathcal A_R$. 

The hierarchy between the \textoverline{dim-5}, \textoverline{dim-7}, and \textoverline{dim-9} contributions can be estimated 
by  noting  that  $C^{(7)}/C^{(5)}\sim (v/v_R)^2 M_D^{-1}$ and $C^{(9)}/C^{(5)}\sim (v/v_R)^4 M_D^{-2}$ 
and  using Table~\ref{TabPC}.   
It is instructive to go through this analysis at various levels of increasing complexity:

\begin{itemize}

\item  A first naive estimate can be obtained by neglecting any chiral suppression (i.e.\ taking the largest possible contributions of the \textoverline{dim-5,-7,-9} couplings in Table~\ref{TabPC}). This leads to a value of $(v/v_R)^2(\Lambda_\chi/v)(1/M_D)$ for the ratio of \textoverline{dim-7} to \textoverline{dim-5} contributions to $0\nu\beta\beta$ decay, and $[(v/v_R)^2(\Lambda_\chi/v)(1/M_D)]^2$ for the ratio of \textoverline{dim-9} to \textoverline{dim-5} contributions. These ratios are both $\mathcal O(1)$ for reasonable lepton Yukawa values $M_D \sim m_e/v$ and right-handed scales $v_R \sim 10$ TeV. This would imply that for such relatively low right-handed scales, the \textoverline{dim-5}, \textoverline{dim-7}, and \textoverline{dim-9} contributions are all of the same order.

\item  
However, the contributions of the induced \textoverline{dim-7} and \textoverline{dim-9} operators are chirally suppressed by 
up to  three and two powers of $\epsilon_\chi$, respectively, see Table \ref{TabPC}. 
As a result, the  \textoverline{dim-9} contributions  only compete with the \textoverline{dim-5} terms for relatively small values of $m_{\bt\bt}$, and they generally dominate the  \textoverline{dim-7} terms for $v_R$ in the $1-10$ TeV range. 

\item 
The above  counting holds in the limit of no mixing between left- and right-handed $W$-bosons, 
$\xi = 0$, and assumes $m_{\nu_R}\sim m_{W_R}$. In this case, the \textoverline{dim-7} and \textoverline{dim-9} operators only give rise to interactions between right-handed  fields at low-energies, $C_{\rm VR}^{(6)}$ and
$C_{\rm 1R}^{(9)\,\prime}$,
 whose contributions are suppressed by chiral symmetry,  see Table \ref{TabPC}.  On the other hand, the terms that involve left-handed fields, $C_{\rm VL}^{(6)}$ and  $C^{(9)}_{\rm 4R}$, which depend linearly on $\xi$ are suppressed by fewer powers of $\epsilon_\chi$ and  can  therefore be important even for relatively small values of $\xi$.

\item 
The suppression of the  \textoverline{dim-7} contributions compared to the  \textoverline{dim-9} terms can in principle be avoided by taking $M_D$ larger than $\sim m_e/v$. However, for large $M_D$ the Type-I seesaw term in Eq.\ \eqref{eq:LRdim5} gives a large contribution to the neutrino masses. This implies that a cancellation, and a certain measure of fine tuning, between the Type-I and Type-II seesaw mechanisms is needed  to reproduce the light neutrino masses $m_\nu\sim 0.1$ eV. 

\item
In the case of light $m_{\nu_R}$, the contributions to the \textoverline{dim-5}, -\textoverline{7}, and -\textoverline{9} operators induced by the exchange of right-handed neutrinos become enhanced by a factor of $m_{W_R}/m_{\nu_R}$. 
For the \textoverline{dim-5}  (\textoverline{dim-7}) operators this enhancement is (in part) mitigated by the fact that the \textoverline{dim-5} terms have to reproduce the usual expression for $m_{\bt\bt}$, which has an upper bound of $m_{\bt\bt}<0.1$ eV for reasonable values of $m_\nu^{\rm lightest}<0.1$ eV. As a result, the enhancement of the type-I seesaw contribution has to be compensated; either by a smaller Dirac mass matrix or by a cancellation with the type-II seesaw term. In the case of the C-symmetric LRM it is the Dirac mass matrix that compensates, see Eq.\ \eqref{Eq:MD},  which now roughly scales as $M_D\sim \sqrt{m_{\nu_R}}$. This leads to \textoverline{dim-5} terms that reproduce the usual expression for $m_{\bt\bt}$, while the \textoverline{dim-7} terms scale as $\mathcal C^{(7)}\sim M_D M_{\nu_R}\sim m_{\nu_R}^{-1/2}$ and therefore are enhanced by a factor of $\sqrt{m_{W_R}/m_{\nu_R}}$ compared to the heavy right-handed neutrino case.
In contrast, the elements of the $M_R$ matrix become smaller since they scale like  $m_{\nu_R}/m_{W_R}$. The same holds for $M_L$ in the case of a C (or P) symmetry. These couplings appear in the $\dt^{++}_R$ contributions to the dim-9 operators, see Eq.\ \eqref{dim9match}, as well as in the type-II seesaw contributions to the dim-5 operator in Eq.\ \eqref{eq:LRdim5}. These two types of contributions are therefore suppressed by a factor of $m_{\nu_R}/m_{W_R}$, compared to the heavy right-handed neutrino scenario. And to reiterate, \textoverline{dim-9} operators 
are enhanced by a factor of $m_{W_R}/m_{\nu_R}$.

\end{itemize}

To see whether the above expectations hold up and to assess the implications  of the TeV-scale mLRSM on $0\nu\bt\bt$, 
we discuss the inverse half-life for representative regions of parameter space in the next subsection.

\subsection{Phenomenology}
In order to make the number of free parameters manageable we will assume a certain flavor structure for $M_{\nu_R}$. In particular, we assume that the mixing matrix of the right-handed neutrinos is equal to that of the left-handed neutrinos. We diagonalize the  left- and right-handed neutrino mass matrices as follows
\bea
M_\nu &=& -v^2 C^{(5)} = U^\dagger_{\rm PMNS} m_\nu U^*_{\rm PMNS}\,,\qquad m_\nu=  {\rm diag}\,(m_{\nu_{1}},\,m_{\nu_{2}},\,m_{\nu_{3}})\,,\nn\\
M_{\nu_R} &=& U^\dagger m_{\nu_R} U^*\,,\qquad\qquad  \qquad   \qquad   m_{\nu_R} = {\rm diag}\,(m_{\nu_{R_1}},\, m_{\nu_{R_2}},\, m_{\nu_{R_3}})\,.
\eea
The assumption $U=U_{\rm PMNS}$ combined with the  charge-conjugation symmetry, allows us to express the Dirac mass matrix as \cite{Nemevsek:2012iq}
\bea
\frac{v}{\sqrt{2}}  M_D = U^\dagger_{\rm PMNS} \, m_{\nu_R} \, \sqrt{\frac{v_L e^{i\theta_L}}{v_R}-m_{\nu_R}^{-1}m_\nu^{}} \  U_{\rm PMNS}^*\,.\label{Eq:MD}
\eea
The inverse half-life can then be expressed in terms of the mixing angles and masses of the light neutrinos and several model parameters: $v_R$, $v_L e^{i\theta_L}$, $\xi e^{i\al}$, $m_{\Dt_R}$, and $m_{\nu_R}$.

\begin{figure}[t]
\begin{center}
\includegraphics[width=0.49\textwidth]{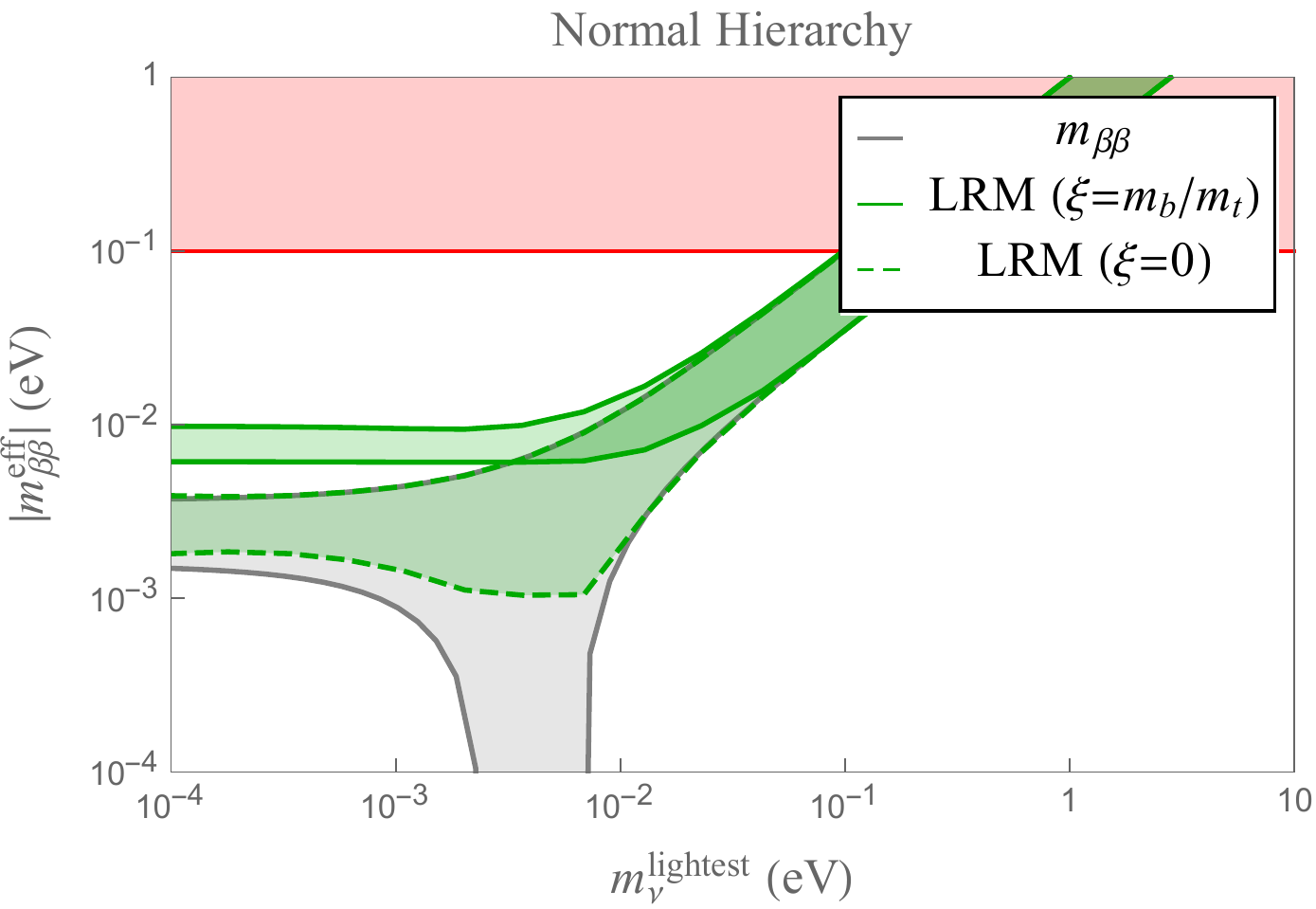}
\includegraphics[width=0.49\textwidth]{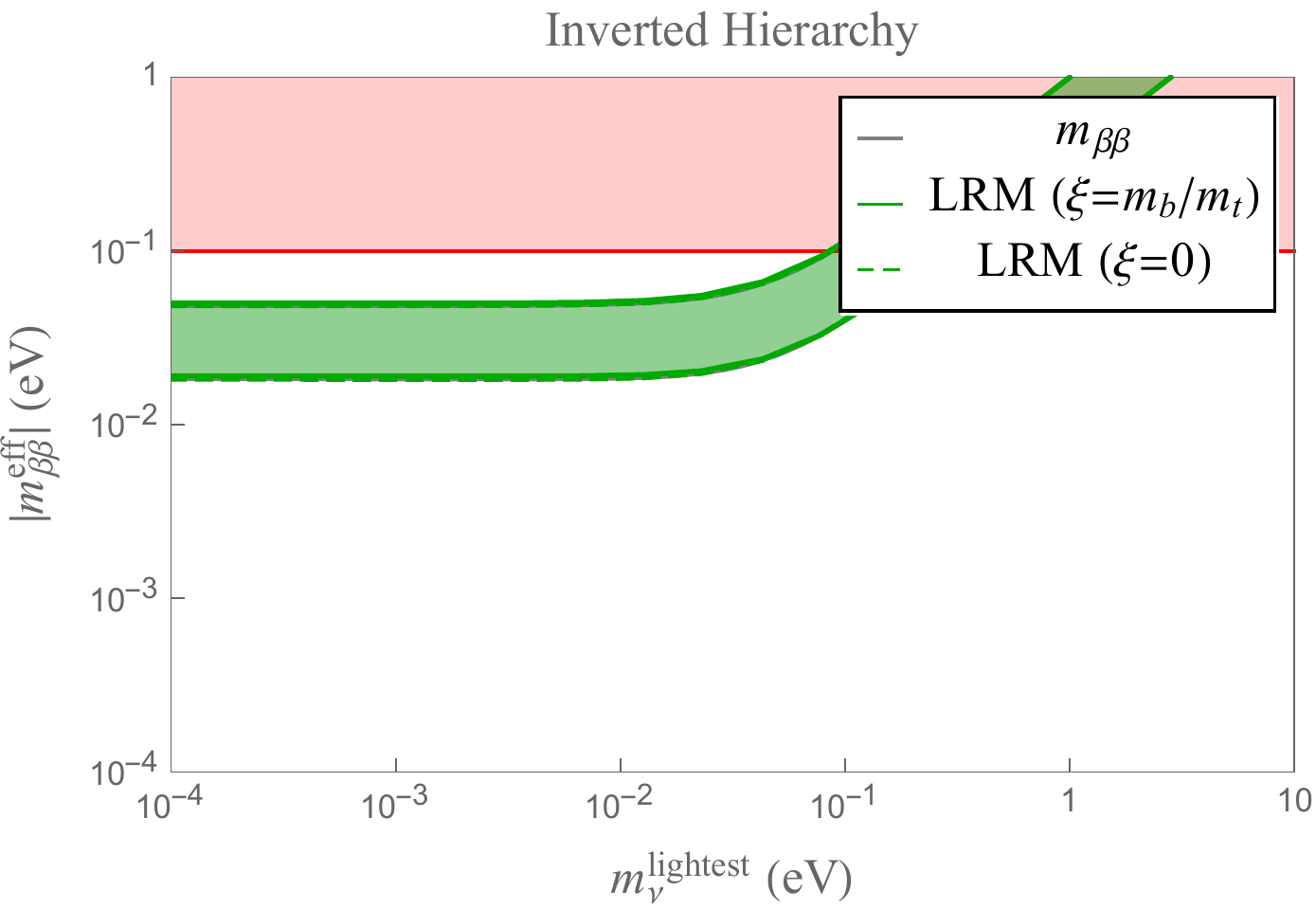}
\includegraphics[width=0.49\textwidth]{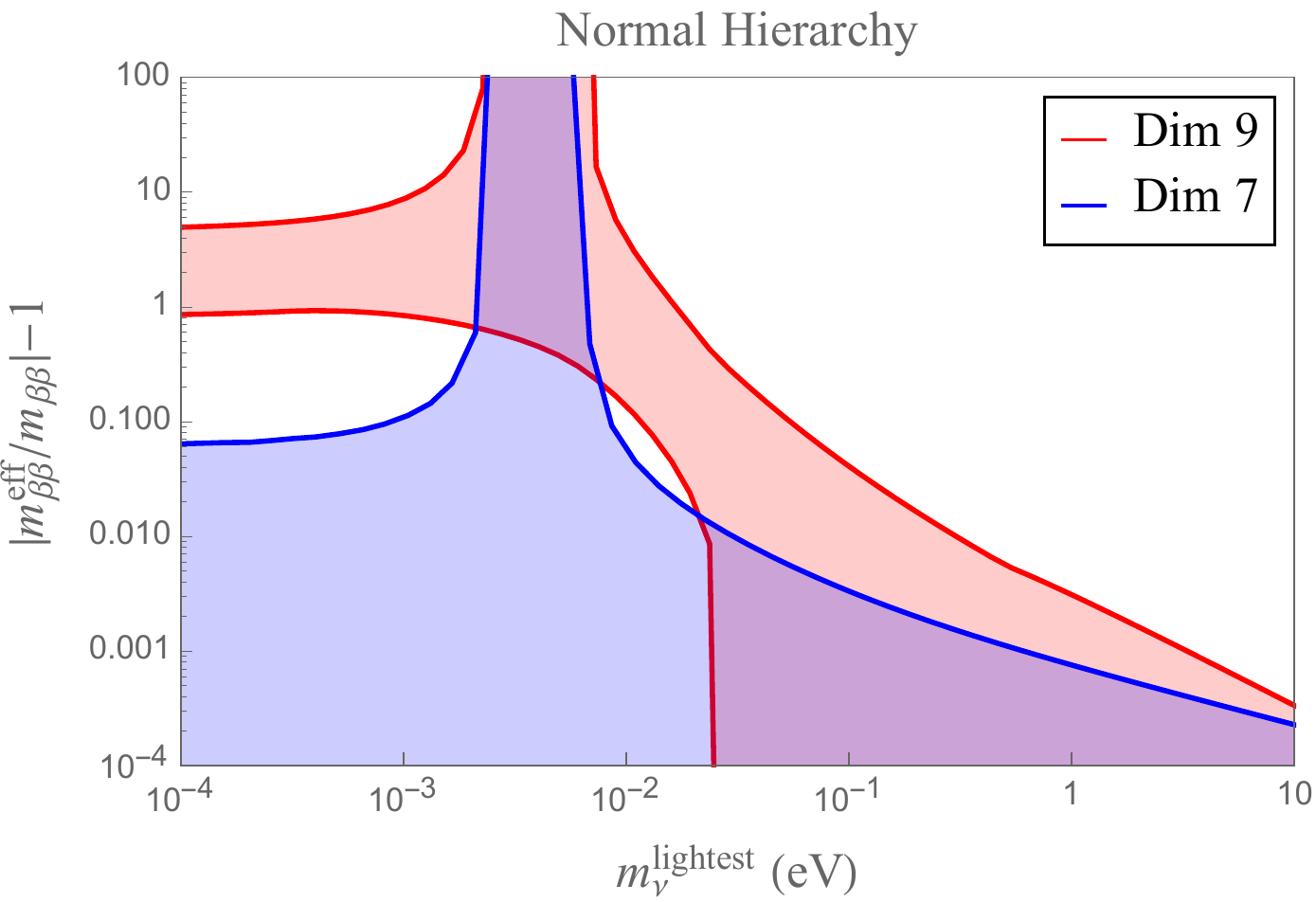}
\includegraphics[width=0.49\textwidth]{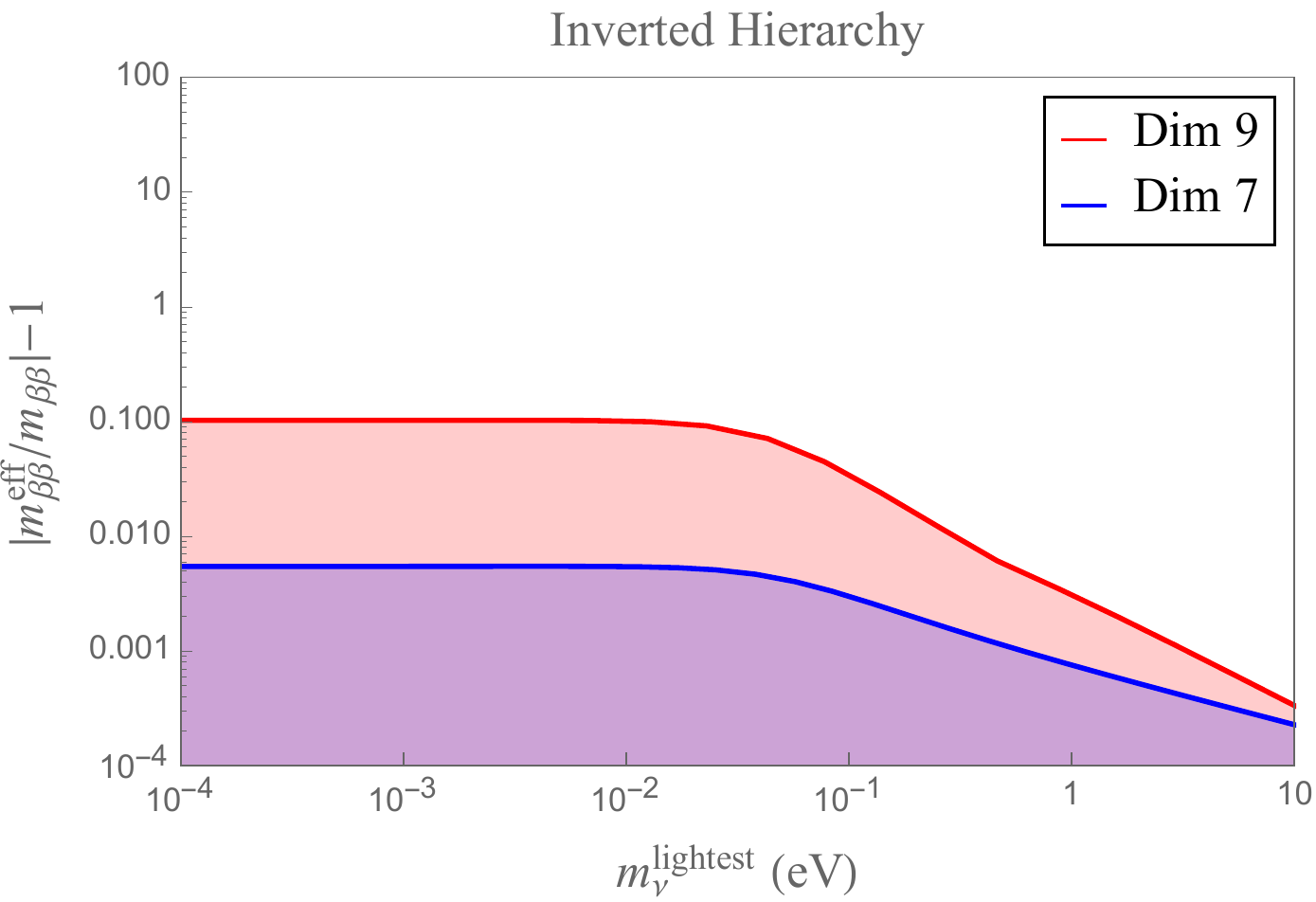}
\vspace{-0.75cm}
\end{center}
\caption{
The upper-left and -right panels show $m_{\bt\bt}^{\rm eff}$ as a function of the lightest neutrino mass for the normal and inverted hierarchy, respectively. The gray bands depict the contribution from $m_{\bt\bt}$ alone, while the (dashed) green bands show the full mLRSM contribution for $\xi=m_b/m_t$ ($\xi=0$) with the choices of model parameters in Eq.\ \eqref{eq:LRvalues}. The current experimental limit on $|m_{\bt\bt}|$ from $^{136}$Xe, derived using the NMEs of Ref.\ \cite{Menendez:2017fdf} and $g_\nu^{NN}=0$, is depicted by the red line.
The lower-left and -right panels show the combination $|m_{\bt\bt}^{\rm eff}/m_{\bt\bt}|-1$, for the normal and inverted hierarchy, respectively. Here we take $\xi=m_b/m_t$ and show the results when turning on only the \textoverline{dim-5}  and \textoverline{dim-7}  (\textoverline{dim-5}  and \textoverline{dim-9}) terms in $m_{\bt\bt}^{\rm eff}$ in blue (red).
}\label{LRplot1}
\end{figure}

To show the impact of the higher-dimension operators on $0\nu\beta\beta$ within the mLRSM, we plot the effective parameter
\bea
m_{\bt\bt}^{\rm eff } = \frac{m_e}{g_A\sq V_{ud}\sq  \mathcal{M}^{(3)}_\nu G_{01}^{1/2}}\left(T_{1/2}^{0\nu}\right)^{-1/2}\,, 
\eea
 as a function of the lightest neutrino mass in Fig.\ \ref{LRplot1}. Here we picked the following values for the model parameters
\bea\label{eq:LRvalues}
v_L&=&0.1 \, {\rm eV}\,, \quad v_R=10\, {\rm TeV}\,,\quad V_R^{ud}=V_L^{ud}\,,\quad m_{\Dt_R} = 4\, {\rm TeV}\,,\nn\\
\quad m_{\nu_{R_1}}&=&10\,{\rm TeV}\,,\quad m_{\nu_{R_2}}=12\,{\rm TeV}\,,\quad m_{\nu_{R_3}}=13\,{\rm TeV}\,.
\eea
These values correspond to $m_{W_R}\simeq 4.5$ TeV, see Appendix \ref{App:LR}, and are consistent with direct collider searches for $W_R$ bosons \cite{CMS:2017xrr,Aaboud:2017yvp}, heavy Majorana neutrinos
\cite{Aad:2015xaa,Sirunyan:2018pom}, and doubly-charged scalars \cite{CMS:2017pet,Aaboud:2017qph}. In addition, 
we take the central values for the mixing angles in the PMNS matrix \cite{Olive:2016xmw}, and marginalize over the Majorana phases as well as $\theta_L$ and $\al$. Finally, to show the impact of the parameter $\xi$, we show results with two values, namely, $\xi=0$ and $\xi=m_b/m_t$. The latter value is inspired by LR models with a P symmetry which requires $\xi \sin\al \lesssim m_b/m_t$, in order to reproduce the hierarchy between the top and bottom masses \cite{Maiezza:2010ic}.

The upper panels in Fig.\ \ref{LRplot1} show that the \textoverline{dim-7} and \textoverline{dim-9} contributions are subdominant for the inverted hierarchy, while they can have a significant impact in the normal hierarchy. By comparing the solid ($\xi=m_b/m_t$) and dashed lines ($\xi=0$), it is confirmed that the contributions proportional to $\xi$ can be dominant even for modest values of $\xi$ due to the chiral suppression of the terms independent of $\xi$.

The lower panels of Fig.\ \ref{LRplot1} show the value of $r\equiv|m_{\bt\bt}^{\rm eff}/m_{\bt\bt}|-1$. The blue (red) area corresponds to $r$ if only  \textoverline{dim-5} and \textoverline{dim-7} (\textoverline{dim-9}) contributions to $m_{\bt\bt}^{\rm eff}$ are considered. The bottom-left panel illustrates that, for the chosen values of the model parameters, the  \textoverline{dim-9} contributions dominate over the \textoverline{dim-5} and \textoverline{dim-7} terms for light neutrino masses in the normal hierarchy. This confirms the power-counting expectations from the previous section. In the inverted hierarchy the \textoverline{dim-9} contributions are also larger than the \textoverline{dim-7} contributions, but both play a marginal role.

\begin{figure}[t]
\begin{center}
\includegraphics[width=0.49\textwidth]{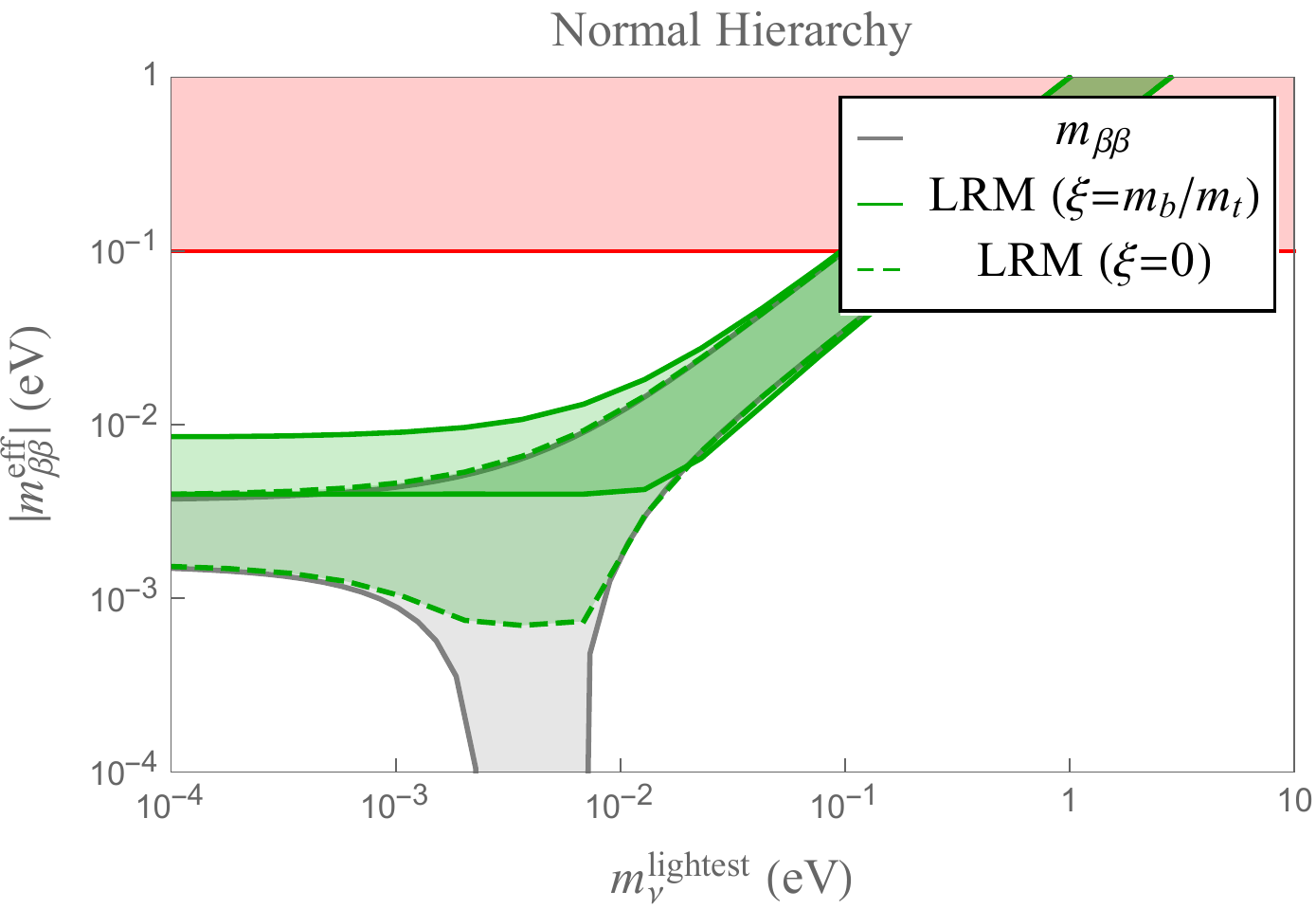}
\includegraphics[width=0.49\textwidth]{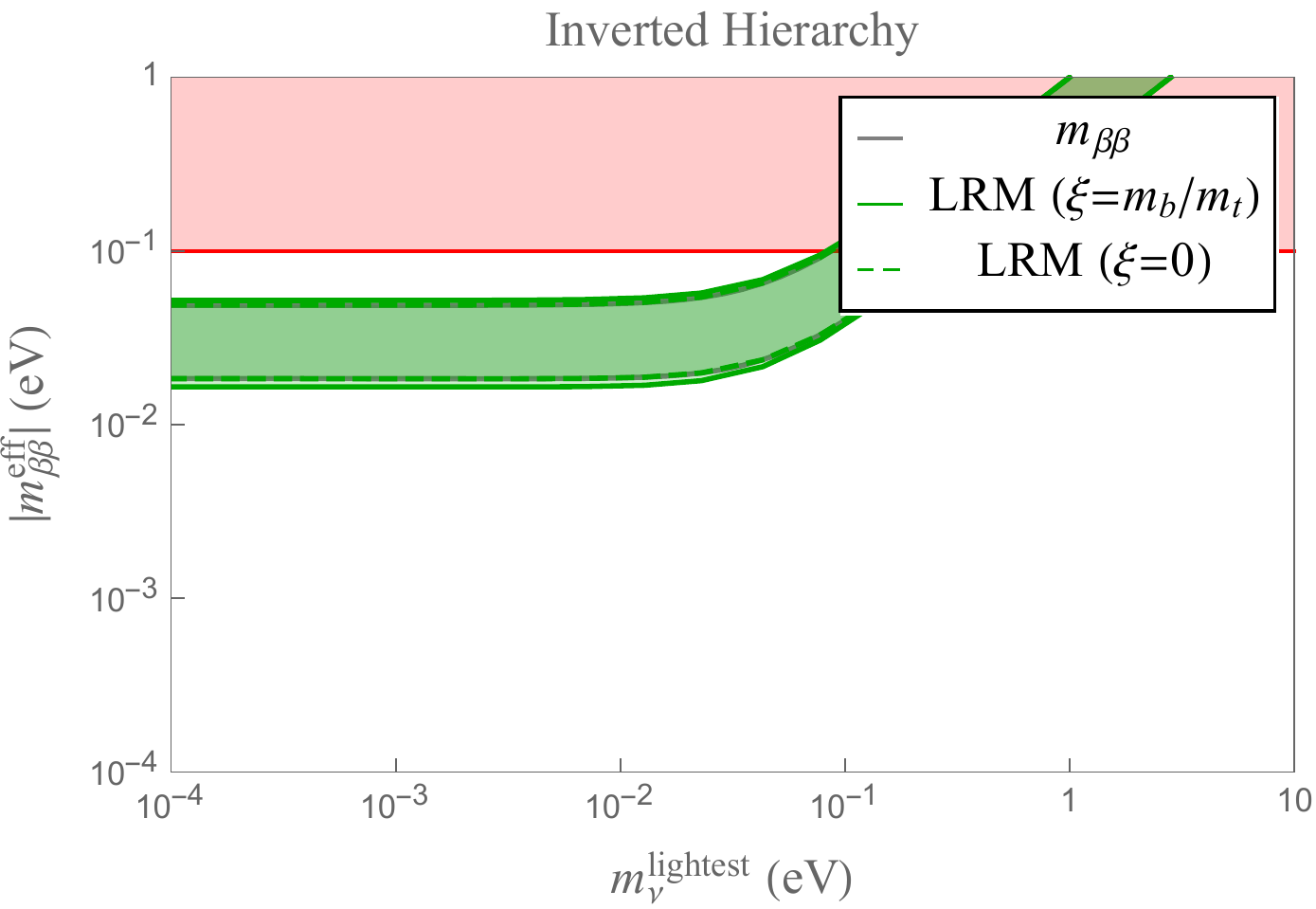}
\includegraphics[width=0.49\textwidth]{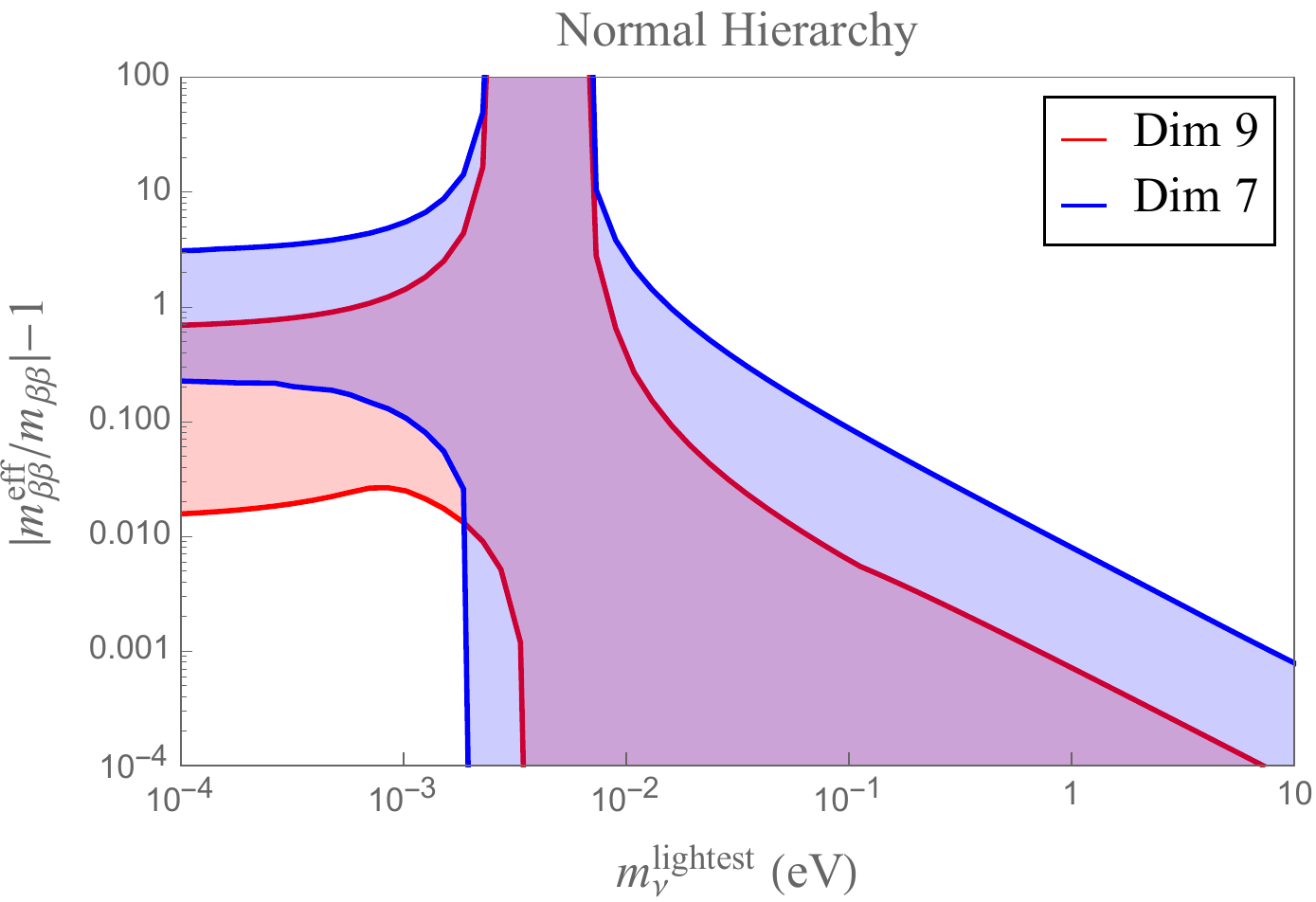}
\includegraphics[width=0.49\textwidth]{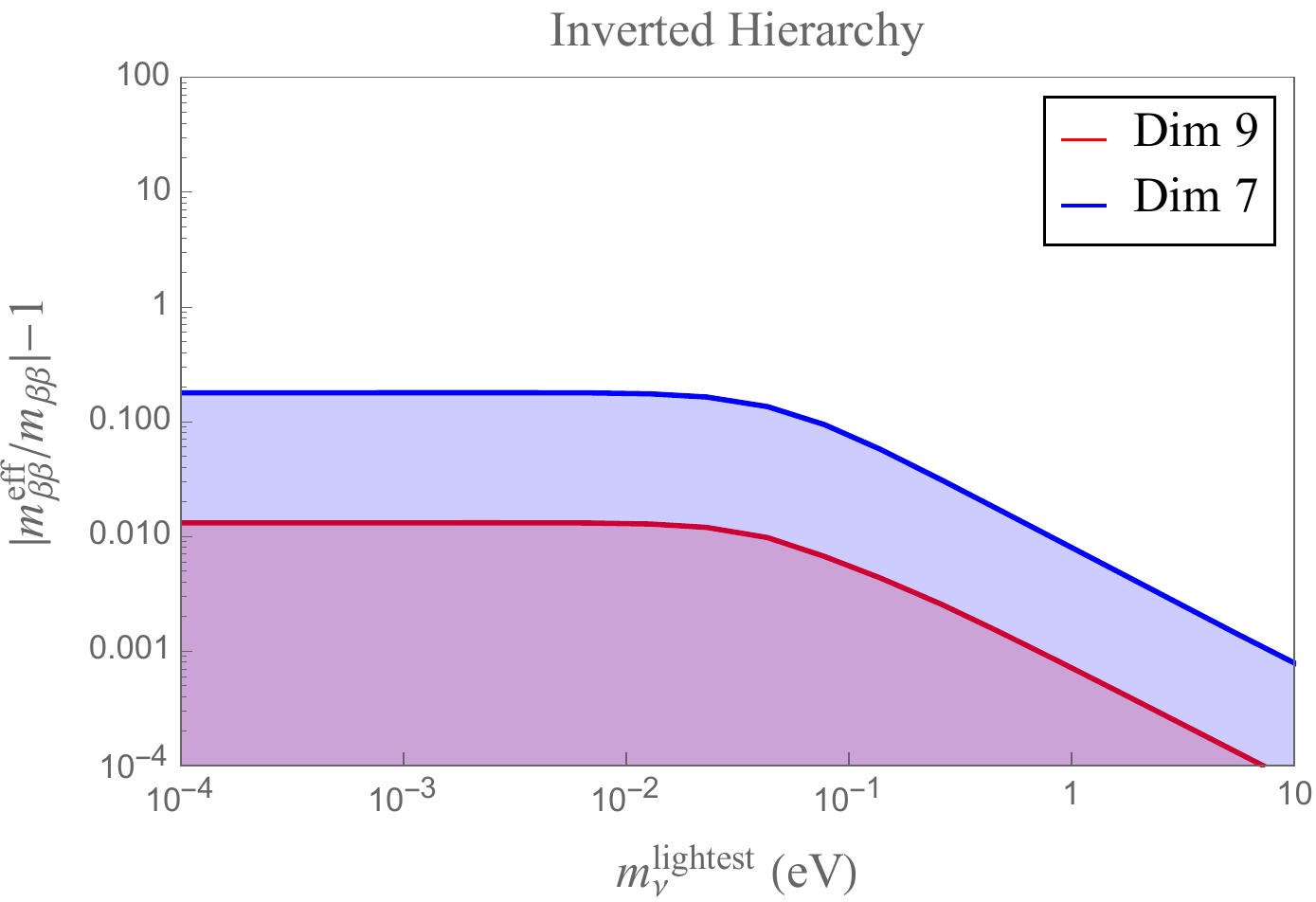}
\vspace{-0.75cm}
\end{center}
\caption{Same as Fig.\ \ref{LRplot1}, now with the choice of the model parameters in Eq.\ \eqref{eq:LRvalues2}.
}\label{LRplot2}
\end{figure}

To study a case where the \textoverline{dim-7} contributions dominate over the \textoverline{dim-9} terms, we perform a similar analysis with the following values for the model parameters
\bea\label{eq:LRvalues2}
 \quad v_L&=&100 \, {\rm eV}\,, \quad v_R=10\, {\rm TeV}\, ,\quad V_R^{ud}=V_L^{ud}\,,\quad m_{\Dt_R} = 10\, {\rm TeV}\,,\nn\\
\quad m_{\nu_{R_1}}&=&10\,{\rm TeV}\, ,\quad m_{\nu_{R_2}}=12\,{\rm TeV}\, ,\quad m_{\nu_{R_3}}=13\,{\rm TeV}\, .
\eea
The main difference with Eq.~\eqref{eq:LRvalues} is the larger value of $v_L$. Such large values require a significant cancellation between the type-I and type-II seesaw mechanisms in order to keep the neutrinos light enough, making these regions of parameter space  less attractive. Nevertheless, we show the resulting plots in Fig.\ \ref{LRplot2} and find that the \textoverline{dim-7} contributions have significant impact in the normal hierarchy.

\begin{figure}[t]
\begin{center}
\includegraphics[width=0.49\textwidth]{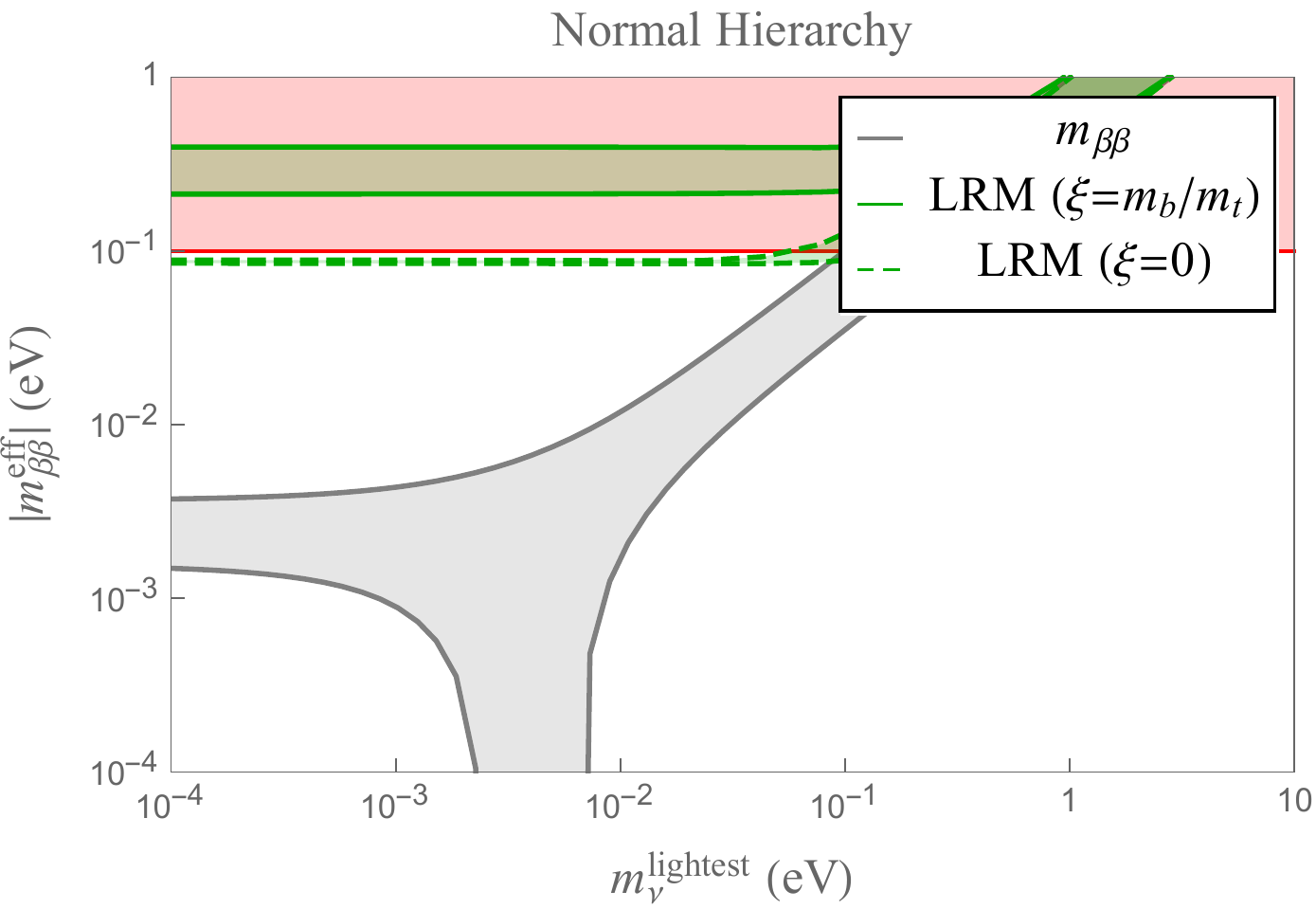}
\includegraphics[width=0.49\textwidth]{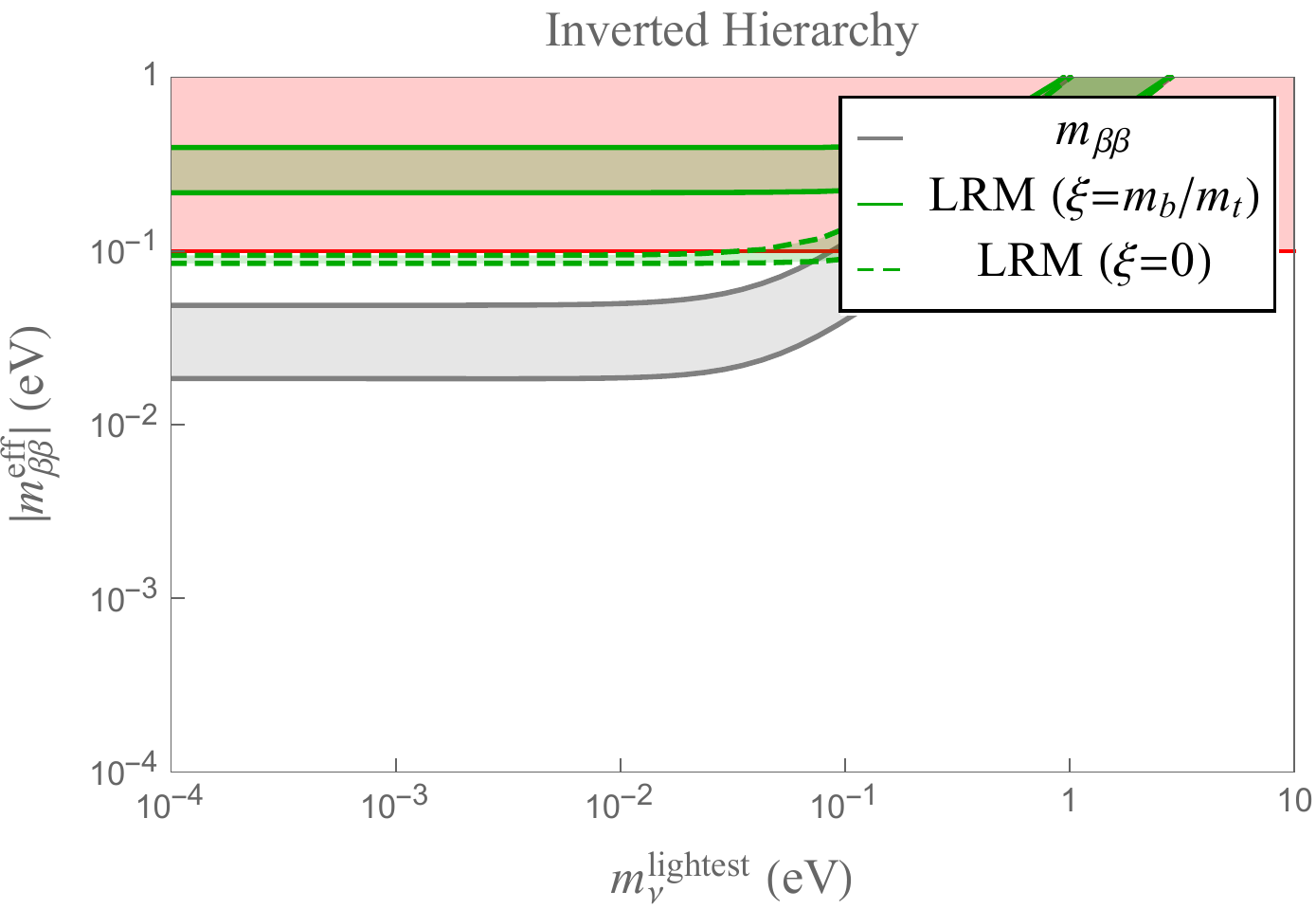}
\includegraphics[width=0.49\textwidth]{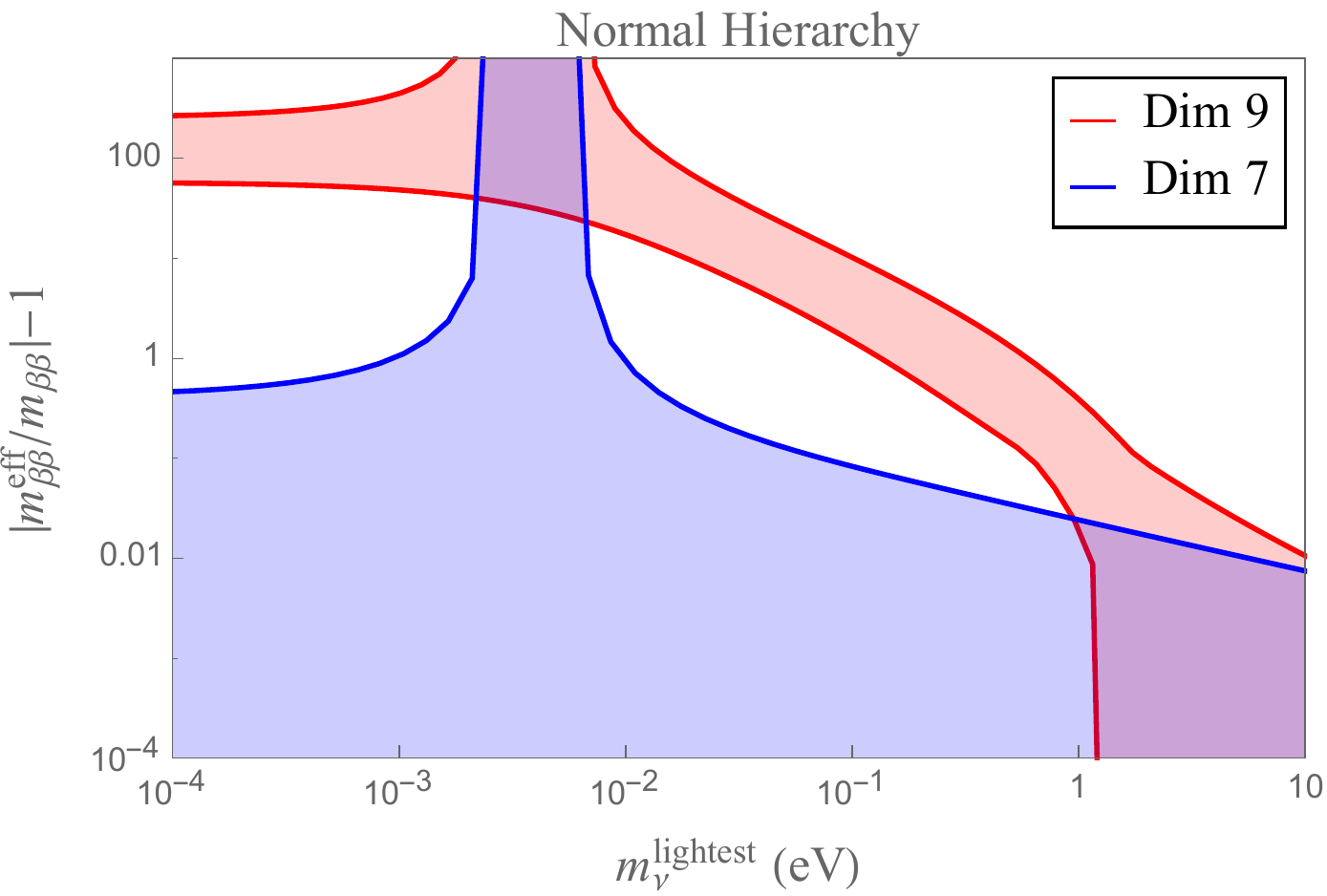}
\includegraphics[width=0.49\textwidth]{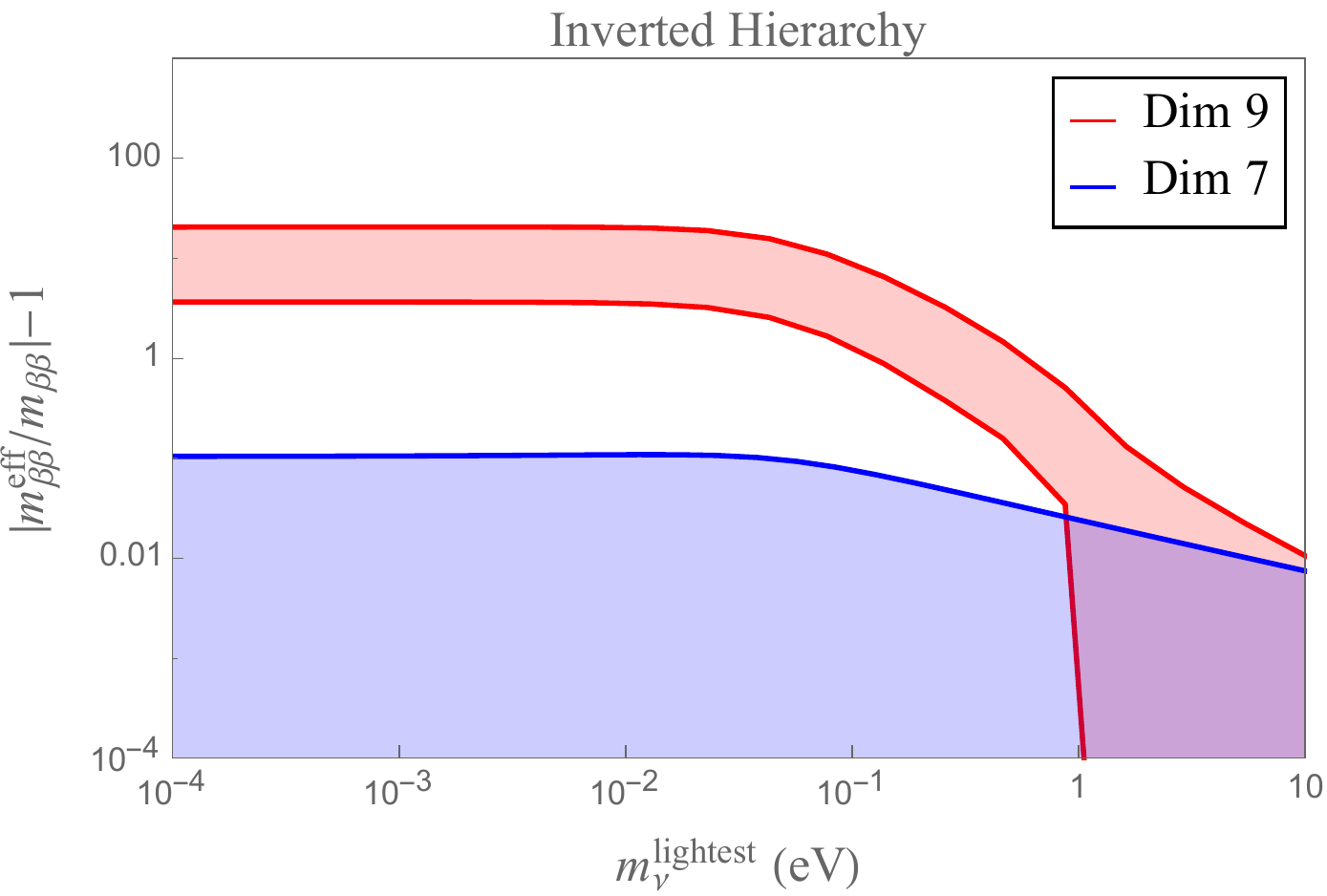}
\vspace{-0.75cm}
\end{center}
\caption{Same as Fig.\ \ref{LRplot1}, now with the choice of the model parameters in Eq.\ \eqref{eq:LRvalues3}. 
}\label{LRplot3}
\end{figure}

Finally, we investigate the impact of light right-handed neutrinos. As an example we take
\bea\label{eq:LRvalues3}
\quad v_L&=&0.1 \, {\rm eV}\,, \quad v_R=10\, {\rm TeV}\,,\quad V_R^{ud}=V_L^{ud}\,,\quad m_{\Dt_R} = 4\, {\rm TeV}\,,\nn\\
\quad m_{\nu_{R_1}}&=&10\,{\rm GeV}\,,\quad m_{\nu_{R_2}}=12\,{\rm GeV}\,,\quad m_{\nu_{R_3}}=13\,{\rm GeV}\,.
\eea
The difference with respect to Eq.~\eqref{eq:LRvalues} is the mass scale of the right-handed neutrinos which we now take to be $\Or(10)$ GeV. The results are shown in Fig.\ \ref{LRplot3}. Clearly, in the case of light right-handed neutrino masses the impact of higher-dimensional contributions on $m_{\bt\bt}^{\rm eff}$ can be significantly enhanced with respect to the case of heavy $m_{\nu_R}$. 
In fact, in this case the \textoverline{dim-9} contributions dominate the \textoverline{dim-5} terms for most of the range of $m_\nu^{\rm lightest}$ in both hierarchies. In addition, as shown in the figure,  such large contributions are (nearly) excluded for $\xi = m_b/m_t$ ($\xi=0$) by the non-observation of $0\nu\beta\beta$ decay.

A similarly large effect for light right-handed neutrinos was found in Ref.~\cite{Tello:2010am}. That work however assumed a type-II mechanism for generating neutrino masses, which together with the assumed 
charge-conjugation invariance, implies $M_L=M^\dagger_R$.  The right- and left-handed neutrinos are therefore diagonalized by the same matrix and, up to an overall 
scaling, have the same spectrum. In this situation, as the lightest active neutrino becomes lighter, the lightest right-handed neutrino mass decreases,  leading to a large 
effect in both the normal and inverted hierarchies. By contrast, in our scenario type-I seesaw dominates the neutrino mass mechanism, and the right-handed neutrino masses 
remain fixed as the lightest active neutrino decreases in mass. Our results for this parameter set generalize the conclusions of  Ref.~\cite{Tello:2010am}, showing that even for type-I neutrino mass mechanisms the effects of the left-right-symmetric model on $0 \nu \bt \bt$ can be significant. At the level of the amplitude, these new effects can easily dominate over the active Majorana mass contribution by a factor of 10 or 100, in the inverted or normal 
hierarchies, respectively.

To our knowledge such light right-handed neutrinos are not obviously excluded by other considerations. At the LHC, the right-handed neutrino can be produced with an electron, through an off-shell $W^*_R$. This leads to different signatures depending on whether the right-handed neutrino decays inside the detector or not.

Assuming first that it escapes the detector, the signature is $e+X+MET$ where $MET$ refers to the missing transverse momentum. It is well known that here the signal produces an edge in the 
transverse mass $m_T$ distribution at high values for $m_T$, and this can be used to obtain strong constraints on the masses of such resonances or the size of the effective higher dimension operator when below threshold. Here we update the 
analyses of Refs.~\cite{Cirigliano:2012ab,Gonzalez-Alonso:2013uqa,Gonzalez-Alonso:2018omy}, and use the $m_T$ search to place bounds on the Wilson coefficient $C^{(6)}_R$ of the effective Lagrangian given by 
Eq.~\eqref{eq:EFTnuR}. The CMS search at 13 TeV with 35.9 fb$^{-1}$ \cite{Sirunyan:2018mpc} considers various values for cutting on $m_T$. We find that in the electron channel, a transverse mass cut of  $m_T> 2$ TeV gives
the strongest bound, where $n=2$ events are observed and
 5 SM backgrounds expected. Using Bayesian statistics, this leads to a 90\% credibility level upper bound of 3 signal events. Assuming a total acceptance of around 90\%, this gives the bound 
 \beq
 |\tilde{\epsilon}_R |< 6.5 \times 10^{-4}
 \eeq
 in the notation of \cite{Cirigliano:2012ab}, where $C^{(6)}_R=-4 G_F V_{ud} \tilde{\epsilon}_R/\sqrt{2}$. A similar limit can be obtained 
 from an ATLAS search \cite{ATLAS:2018MT} using $79.8$ fb$^{-1}$, where no events are seen above $m_T > 3$ TeV. We note in passing that 
 this new LHC bound is roughly a factor of 8 stronger than that previously obtained using 7 TeV LHC data, and by a factor of 40 from neutron decay~\cite{Gonzalez-Alonso:2018omy}.
 In our case, the left-right model gives $C^{(6)}_R=- 1/v_R\sq V_R^{ud}$, or
 $\tilde{\epsilon}_R=v^2/2v_R^2=3 \times 10^{-4}$ for $v_R=10$ TeV. This is just below the current LHC limit, but recall we assumed all produced $\nu_R$'s escape the detector. 

The right-handed neutrinos are unstable as they can decay through off-shell right-handed gauge boson exchange, or by off-shell $W$ and $Z$ exchange through mixing with the active neutrinos.
To simplify the discussion, we consider only decays mediated by the $W_R$. Then by comparing to muon decay, we find $c \tau \sim \Or($cm$)(10$ GeV$/m_{\nu_R})^{5} (m_{W_R}/4.5$ TeV$)^{4} (9/N_{eff})$, where $N_{eff}$ counts the number of open decay channels including color factors. Thus the decay of $\nu_R$, if it occurs in the detector, produces displaced vertices comprising either two jets and a charged lepton or neutrinos, 
or three leptons \cite{Graesser:2007yj,Graesser:2007pc}. 

This is a complicated signal and we will not attempt to obtain any rigorous bounds, as this would require an involved simulation of the signal acceptance and recasting of existing displaced 
vertex searches, which is well beyond 
the scope of this paper \footnote{An impressive recasting of the
CMS search for displaced $e \mu$ pairs  \cite{Khachatryan:2014mea} to constrain long-lived superpartners of the tau can be found in Ref.\ \cite{Evans:2016zau}.}. 
Here we survey several displaced vertex searches and find none of them out-right exclude light right-handed neutrinos with masses of $O(10$ GeV$)$, though further investigation is justified. 

CMS has a 
roughly model-independent search for directly produced long-lived neutral particles decaying to displaced jets, using 2.6 fb$^{-1}$ of data \cite{Sirunyan:2017jdo}. The CMS analysis considers 
 $p p \rightarrow  N N $ where $N$ is long-lived and its decays produce displaced jets. We find this search to not be constraining as we now briefly discuss.
In our case, we have two dominant production modes. For $p p \rightarrow Z^*_R  \rightarrow \nu_R \nu_R \rightarrow$ displaced jets, this production 
channel directly maps onto the topology of the CMS search. However, the obtained limit for $m_{N} = 50$ GeV is 1000 fb and not constraining.
The other dominant production process is $p p \rightarrow W^*_R  \rightarrow e_R \nu_R \rightarrow e_R$ + displaced jets/leptons, but here we cannot directly translate the CMS limit, since the kinematics are not exactly the same, though we expect it not to deviate too strongly from the previous limit. CMS also has a search for displaced jets with a strong limit of $\sigma \cdot BR^2 \lsim \Or(0.1$ fb$)$ for $c\tau_0 \sim \Or($cm$)$ and 38.5 fb$^{-1}$ \cite{CMS:LongLivedParticles}. This limit is sensitive to the trigger efficiency, as they require $H_T>1000$ GeV. To crudely estimate $H_T$, we observe that it is 
correlated with $m_T$, and we find that for $m_T>1000$ GeV and $m_{W_R}=4.5$ TeV, $\sigma (pp \rightarrow \nu_R \nu_R) \sim \Or(0.5$ fb$)$ which is close to the observed limit. However, the CMS analysis 
additionally requires 4 jets each with $p_T > 40$ GeV. Right-handed neutrinos having masses $\Or(10$ GeV$)$ will be highly boosted, which further reduces
the signal efficiency, for the two displaced jets produced 
in each $\nu_R$ decay will tend to merge into a single jet. While without a simulation we cannot say what the acceptance is, we expect a further suppression of at least 10\%. This would make this CMS search possibly constraining. 

It would be interesting to further explore LHC limits on light
right-handed neutrinos that lead to displaced vertices.

\section{Discussion and conclusion}\label{sec:conclude}
In many well-motivated scenarios of physics beyond the Standard Model, 
lepton number is violated  at an  energy scale, $\Lambda$, well above the electroweak 
scale~\footnote{Exceptions to this include low-scale see-saw models \cite{deGouvea:2005er}.}.
Yet,  the  probes of LNV with the broadest (in terms of mechanisms) and strongest sensitivity
are searches for neutrinoless double beta decay,  which are
associated with typical nuclear scales. This large scale separation suggests that 
 the phenomenology of $0\nu\beta\beta$  
is best tackled by EFT methods, 
describing in a systematically improvable  way the 
LNV dynamics both at high energy and 
at hadronic and nuclear scales. 
We stress here that an EFT  approach, in conjunction with improved many-body methods, 
is  the only path towards reaching controlled uncertainties in \NLDBD\  calculations. 
This end-to-end EFT framework has been developed in several recent papers,  namely 
Refs.~\cite{Cirigliano:2017tvr,Cirigliano:2018hja} for the dimension-five Weinberg operator    
and  Ref.~\cite{Cirigliano:2017djv} for dimension-seven LNV operators.
The current  paper summarizes and finalizes this effort by including the effects of LNV operators of dimension nine and correcting several omission in the previous papers with respect to short-distance $0\nu\beta\beta$  mechanisms. 

Our work distinguishes itself from the previous literature on \NLDBD\ through the ``end-to-end"  EFT treatment, 
starting from the high-scale SM-EFT all the way to chiral EFT for the hadronic and nuclear aspects of the problem.   
For  similar approaches to  dimension-nine LNV operators  we point to Refs.~\cite{Prezeau:2003xn,Graesser:2016bpz}.  
Other approaches often use EFT methods only to classify high-scale LNV operators and at the hadronic scale  employ models and approximations 
that in certain cases lead to  the wrong scaling of the  \NLDBD\ amplitude.   
In Appendix~\ref{comparison-to-lit} we provide a comparison with previous approaches, highlighting the discrepancies with the EFT approach. 
\\
\\
We now summarize  the EFT framework and the main results of this paper:
\begin{itemize}
\item Assuming a high-scale origin, the effects of LNV dynamics
 at scales below $\Lambda$ can be expressed in an expansion in $1/\Lambda$ by considering gauge-invariant LNV operators of various dimension. After electroweak symmetry breaking via the Higgs vacuum expectation value, $v$, and integrating out heavy SM fields, such as W-bosons and Higgs fields,  the operators can be further expanded in  powers of $1/v$. Finally, at energies where QCD becomes non-perturbative the various effective LNV operators can be matched to chiral EFT, the low-energy EFT of QCD, resulting in an expansion in $m_\pi/\Lambda_\chi$ where $\Lambda_\chi \sim 1$ GeV is the chiral-symmetry-breaking scale. In total, the $0\nu\beta\beta$ rate is expanded as $(v/\Lambda)^\alpha (\Lambda_\chi/v)^\beta(m_\pi/\Lambda_\chi)^\gamma$, where the exponents $\alpha$, $\beta$, and $\gamma$ depend on the LNV source and the required accuracy of the final expression. In this work, we have identified the $0\nu\beta\beta$ rate for the gauge-invariant \textoverline{dim-5} ($\alpha=1$), \textoverline{dim-7} ($\alpha=3$), and a subset of \textoverline{dim-9} ($\alpha=5$) operators up to leading order in the $(\Lambda_\chi/v)$ and $(m_\pi/\Lambda_\chi)$ expansions.
 
 \item The main outcome of this work is the master formula in Eq.~\eqref{eq:T1/2} which is graphically depicted in Fig.~\ref{landscape}. This formula allows one to calculate the  $0\nu\beta\beta$ decay rate for $0^+ \rightarrow 0^+$ transitions in various  isotopes as a function of effective $\Delta L=2$ Wilson coefficients. Hadronic, nuclear, and atomic information is captured by LECs, NMEs, and phase space factors given in Tables \ref{Tab:LECs}, \ref{tab:comparison}, and \ref{Tab:phasespace}, respectively. Perturbative corrections due to QCD renormalization-group evolution are discussed in Section~\ref{sec:2} and Appendix \ref{App:RG}. Surprisingly we find that at leading order in the chiral expansion, all nuclear matrix elements that are necessary to describe $0\nu\beta\beta$ arising from higher-dimensional $\Delta L=2$ operators already appear for $0\nu\beta\beta$ induced by light Majorana neutrino exchange (corresponding to the $\Delta L=2$ \textoverline{dim-5} operator) and heavy Majorana neutrino exchange. That is, all matrix elements (apart from $M_T^{AA}$) can be lifted from the existing literature
and we strongly encourage future many-body calculations of light and heavy Majorana neutrino exchange to be organized in terms of the NMEs in Table \ref{tab:comparison}.   
While different calculations vary by factors of 2-3,   
 recent and future progress in nuclear many-body theory will  allow to reduce these uncertainties 
\cite{Barea:2015kwa,Hyvarinen:2015bda,Barea,Menendez:2017fdf,Horoi:2017gmj,Jiao:2017opc,Iwata:2016cxn,Vaquero:2014dna,Yao:2014uta,Holt:2013tda,Simkovic:2013qiy,Fang:2018tui,Wang:2018htk,Engel:2016xgb}.

 \item While all the nuclear input can be lifted from the literature, the same cannot unfortunately be said for the LECs connecting $\Delta L=2$ operators at the quark-gluon level to the hadronic level. Many of the LECs associated with the \textoverline{dim-9} operators are unknown and here we estimated them by naive dimensional analysis, or through arguments based on renormalization, leading to significant uncertainties. The number of unknown LECs grows once non-perturbative renormalization due to the nucleon-nucleon interaction of the LNV nucleon-nucleon operators is considered \cite{Cirigliano:2018hja}.
In particular, 4 LECs -- namely $g^{NN}_{2,3,4,5}$ -- that are induced by $O_{2,3,4,5}$ and $O^\prime_{2,3}$ and are
 ${\cal O}(1)$ in Weinberg's counting are promoted to ${\cal O}((4 \pi)^2)$ after non-perturbative renormalization of the nucleon-nucleon operators. This enhancement implies that for these  quark-level operators, the induced nucleon-nucleon operators contribute to the transition amplitude at the same order as the pion-range  contributions -- i.e., at leading order. However,  little is known about the exact values of the nucleon-nucleon LECs and in our numerical analyses we have set these effects to zero by hand, though we expect that they can affect the amplitude at the ${\cal O}(1)$ level. Lattice QCD calculations of the $nn\rightarrow pp ee$ transition are direly needed to improve this situation. We have summarized the state-of-the-art values of the LECs in Table~\ref{Tab:LECs}.
 
 \item An advantage of our EFT approach is that our work can be extended to higher orders in the various expansions. The largest corrections are due to higher-order chiral corrections that are suppressed by powers of $m_\pi/\Lambda_\chi$. Such corrections were calculated for the \textoverline{dim-5} operator in Ref.~\cite{Cirigliano:2017tvr}\,\footnote{These higher-order corrections did not include the effects of non-perturbative renormalization and the resulting enhancement of short-range operators. A study of higher-order corrections that does include this is in progress.},  and up to next-to-next-to-leading order ($\Or(\epsilon_\chi^2)$) include  (i)  factorizable corrections, encoded by  form factors in the nucleon currents; 
(ii) corrections due to genuinely  new two-body  operators,  involving
additional (unknown)  LECs;  and  (iii)  so-called   ``closure'" corrections arising from the exchange of ultrasoft neutrinos that depend on the nuclear excited states. Similar higher-order corrections can be calculated for the higher-dimensional LNV operators. 

\item Another advantage of the EFT approach is that it can be matched to \textit{any} specific UV-complete model of LNV. 
The power counting (see Table~\ref{TabPC}) allows one to isolate the important contributions and $0\nu\beta\beta$ decay rates can be immediately expressed in terms of model parameters. 
This allows for easy derivation of bounds on the specific high-energy model. 
For instance, our bounds on operators in the simplified circumstance that a single operator dominates the decay rate are given in Table \ref{tab:limits}.
As a more  realistic example, we have studied the minimal left-right-symmetric model in which various operators of different dimension 
contribute to $0\nu\beta\beta$, with the results of our analysis shown in Figs.\ \ref{LRplot1} and \ref{LRplot2}  for particular choices of model parameters.
We showed in Section \ref{LR-lightNuR} and illustrated in Fig.\ \ref{LRplot3} how to easily adapt our effective theory framework to 
include states with masses between the GeV scale and the scale of LNV physics.

\item Our work essentially provides a connection between high-scale sources of LNV and low-scale experiments. As such, the framework can be used in future work to study the \NLDBD\  phenomenology in  other models for neutrino masses,  
as well as to compare \NLDBD\  and  collider processes such as $pp \to ee jj$  at the LHC as probes of $\Delta L = 2$ dynamics. Other extensions are the connection between models of leptogenesis and  \NLDBD\  data \cite{Deppisch:2017ecm}. Finally, our framework is not applicable for non-SM states with masses below the GeV scale that occur, for example, in models with additional light sterile neutrinos. It would be interesting to extend the derivation of the $0\nu\beta\beta$ potential to include such states.

\end{itemize}

\section*{Acknowledgements}
We are grateful to  Jose Barea for providing us unpublished results for the nuclear matrix elements in the interacting
boson model. We thank Javier Men\'endez and Andr\'e Walker-Loud for several interesting conversations, and Bira van Kolck for discussion on the non-perturbative renormalization of the 
LNV $NN$ couplings. 
VC, WD,  and EM  acknowledge support by the US DOE Office of Nuclear Physics and by the LDRD program at Los Alamos National Laboratory.
MG acknowledges support by the US DOE Office of High Energy Physics and by the LDRD program at Los Alamos National Laboratory.
WD and JdV  acknowledge  support by the Dutch Organization for Scientific Research (NWO) 
through a RUBICON  and VENI grant, respectively.

\newpage

\appendix
\section{Phase space factors and nuclear matrix elements}\label{sec:app}

\subsection{Phase space factors}\label{Phase}
The definitions of the phase space factors appearing in Eq.\ \eqref{eq:T1/2} are given by,
\bea\label{eq:PhaseSpace}
G_{0k}=\frac{1}{\ln 2}\frac{G_F^4m_e\sq}{64\pi^5 R_A\sq}\int  dE_1 dE_2 |\vec k_1| |\vec k_2| d\cos\theta \,b_{0k} \,F(Z,E_1)F(Z,E_2) \dt(E_1+E_2+E_f-M_i)\,.
\eea
Here $\theta$ is the angle between the momenta of the outgoing electrons and we followed the standard normalization of Ref.~\cite{Doi:1985dx}. 
The Fermi functions $F(Z,E_{1,2})$ take into account the fact that the outgoing electrons interact with the Coulomb potential of the daughter nucleus, 
and their wavefunctions are not plane waves. Their expressions are given by
\bea \label{fermi}
F(Z,E) &=& \left[\frac{2}{\Gamma(2\g+1)}\right]\sq (2|\vec k|R_A)^{2(\g-1)}|\Gamma(\g+i y)|\sq e^{\pi y}\,,\nn\\
\g &=&\sqrt{1-(\al Z)\sq}\,,\qquad y = \al Z E/|\vec k|\,,\quad |\vec k | = \sqrt{E^2-m_e\sq}\,,
\eea
where $R_A = 1.2 \,A^{1/3}$ fm and $Z$ are, respectively, the radius and atomic number of the daughter nucleus. 
The Fermi functions describe the  Coulomb corrections in the assumption of a uniform charge distribution in the nucleus and only account for the lowest-order terms in an  expansion in the electron position. 
It is possible to go beyond these approximations by using exact Dirac wave functions \cite{Stefanik:2015twa} and including the effect of electron screening \cite{Kotila:2012zza}. The use of exact wave functions leads to  smaller phase space factors, with a reduction of  up to $30\%$ for the heaviest nuclei. The effects of electron screening are at the percent level \cite{Stefanik:2015twa}. 
The phase space factors in Table~\ref{Tab:phasespace} do not rely on Eq.\ \eqref{eq:PhaseSpace}, but reflect the more accurate results of Refs.~\cite{Horoi:2017gmj,Stefanik:2015twa}.  
Eq.\ \eqref{eq:PhaseSpace} can be used to get a quick estimate of the half-life, and of the differential decay rates we discuss below.

The $b_{0k}$ factors are obtained from the electron traces that result from taking the square of Eq.\ \eqref{eq:TotAmp}. Here we follow the notation of Ref.\ \cite{Cirigliano:2017djv}, in which these factors take the following form
\bea\label{eq:b0k}
b_{01}&=&E_1 E_2-\vec k_1\cdot \vec k_2\,,\quad b_{02} = \left(\frac{E_1-E_2}{m_e}\right)\sq\frac{E_1 E_2 +\vec k_1\cdot \vec k_2 -m_e\sq}{2}\,,\quad b_{03} = (E_1-E_2)\sq\,,\nn\\
b_{04} &=& \left(E_1 E_2 -\vec k_1\cdot \vec k_2 -m_e\sq\right)\,,\quad  b_{06} = 2 m_e \left( E_1+E_2 \right)\,,\quad b_{09} = 2\left( E_1E_2+\vec k_1\cdot \vec k_2 +m_e\sq\right)\,. \nn\\
\eea
These definitions agree with those commonly used in the literature \cite{Doi:1985dx}, up to the trivial rescalings discussed in Ref.\ \cite{Cirigliano:2017djv}. 
With the definitions of Eq.~\eqref{eq:b0k}, the different phase space factors for a given isotope are all of similar size, with no parametric enhancement or suppression, such that the relative importance of different contributions is determined by the matching coefficients and by the nuclear matrix elements. We list the phase space factors for $^{76}$Ge, $^{82}$Se, $^{130}$Te and $^{136}$Xe in Table 
\ref{Tab:phasespace}, for which we use the calculation of Ref.\ \cite{Horoi:2017gmj}.

As discussed in Ref.\ \cite{Cirigliano:2017djv}, the measurement of the half-life in one or several isotopes will not by itself allow to disentangle 
the effects of \textoverline{dim-5}, \textoverline{dim-7} or \textoverline{dim-9} operators.  
Some additional information can in principle be extracted from the differential decay rate with respect to the energy difference  of the two electrons, $y = (E_1 - E_2)/Q$, and
the angle between the electron momenta, $\cos\theta$.
The differential version of the master formula \eqref{eq:T1/2} is 
\bea \label{eq:dT1/2}
\frac{d\, \Gamma}{d y\, d \cos\theta} &=& g_A^4 \Big\{ g_{01} \, \left( |\mathcal A_{\nu}|\sq + |\mathcal A_{R}|\sq \right)
- 2 (g_{01} - g_{04}) \textrm{Re} \mathcal A_{\nu}^* \mathcal A_{R} 
+ 4g_{02} \,|\mathcal A_{E}|\sq \nn \\ & & + 2 g_{04} \left[|\mathcal A_{m_e}|\sq+{\rm Re} \left(\mathcal A_{m_e}^* (\mathcal A_{\nu} + \mathcal A_{R})\right)\right]
-2 g_{03}\,{\rm Re}\left( (\mathcal A_{\nu} + \mathcal A_{R} )\mathcal A_{E}^*+2\mathcal A_{m_e} \mathcal A_{E}^*\right)
\nn\\
&&+ g_{09}\, |\mathcal A_{M}|\sq + g_{06}\, {\rm Re}\left( (\mathcal A_{\nu} - \mathcal A_{R} )\mathcal A_{M}^*\right) \Big\}\,,
\eea
where all the dependence on $y$ and $\cos\theta$ is encoded in the unintegrated phase space factors $g_{0k}$
\bea\label{eq:PhaseSpace2}
g_{0k}
&=&\frac{1}{\ln 2}\frac{G_F^4m_e\sq}{64\pi^5 R_A\sq} \left( \frac{Q}{2} \right)^5  \sqrt{  1 - y^2}  \sqrt{ \left( 1 + y + \frac{4 m_e}{Q} \right) \left(  1 - y + \frac{4 m_e}{Q}  \right) } \nn \\
& & \, \tilde{b}_{0k}(y,\cos\theta) \,F\left(Z,E_1\right)F\left(Z,E_2\right)\,.
\eea
The variable $y = (E_1-E_2)/Q \in [-1,1]$, and the dimensionless factors $\tilde b_{0k}$ are related to Eq.\ \eqref{eq:b0k} by  
$
\tilde b_{0k}(y,\cos\theta) = 4 b_{0k}/Q^2$, and $E_{1,2} =\frac{1\pm y}{2}Q+m_e$. 

As discussed in Ref.\ \cite{Cirigliano:2017djv}, $g_{02}$ is the phase space factor whose $y$ dependence is distinct from the standard light neutrino exchange.
A measurement of the electrons energy difference  could therefore be used to single out operators, like $C^{(6)}_{\rm VR}$, which mainly contribute to $\mathcal A_E$. Furthermore, 
the phase space factors in Eq.\ \eqref{eq:b0k} exhibit a dependence on $\cos\theta$ which is at most linear.  The slope of the $\cos\theta$ dependence of the decay rate could distinguish 
between the standard light neutrino exchange or the contributions of dim-9 scalar  operators, for which one expects a negative slope, 
and $C^{(6)}_{\rm VR, VL}$ or dim-9 vector operators, which should produce a positive slope.

\subsection{Nuclear Matrix Elements}\label{NME}
To describe the nuclear parts of the amplitude, we follow standard conventions, e.g.\ those of Ref.~\cite{Hyvarinen:2015bda}, and introduce the following definitions
\begin{equation}\label{eq:hK}
h^{ij}_K(r) = \frac{2}{\pi} R_A \int_0^{+\infty} d |\vec q|\,   h^{ij}_{K}(\vec q\sq) j_{\lambda} (|\vec q| r)\,,\qquad h^{ij}_{K,sd}(r) = \frac{2}{\pi} \frac{R_A}{m_\pi^2} \int_0^{+\infty} d |\vec q|\, \vec q\sq \,   h^{ij}_K(\vec q\sq) j_{\lambda} (|\vec q| r)\,,
\end{equation}
where $K \in \{F, GT, T\}$, while $j_{\lambda} (|\vec q| r)$ are  spherical Bessel functions, with $\lambda = 0$ for F and GT, and $\lambda =2$ for the tensor. The $h^{ij}_K(r)$ functions describe long-range contributions, while the $h^{ij}_{K,sd}(r)$ indicate short-range contributions. 
The factors of $R_A$ and $m_{\pi}$ have been inserted so that the $h^{ij}_{K, (sd)}$ are dimensionless. The $h^{ij}_K(\vec q\sq)$ are defined as follows
\begin{eqnarray}\label{smff}
h^{AA}_{GT,T}(\vec q^2) &=&  \frac{g_A^2(\vec q^2)}{g_A\sq}\,, \quad \, h_{GT}^{AP}(\vec q^2) = \frac{g_P(\vec q^2)}{g_A\sq} \, g_A(\vec q^2) \frac{\vec q^2}{3 m_N}\,,\quad  \, h_{GT}^{PP}(\vec q^2) = \frac{ g^2_P(\vec q^2)}{g_A\sq} \frac{\vec q^4}{12 m_N^2}\,, \nonumber \\
h^{MM}_{GT}(\vec q^2) &=&  g_M^2(\vec q^2) \frac{\vec q^2}{6g_A\sq m_N^2}\,,\quad h_F(\vec q\sq) = g_V(\vec q\sq)\,,
\end{eqnarray}
where $h^{AP}_T(\vec q^2) = -h^{AP}_{GT}(\vec q^2)$, $h^{PP}_T(\vec q^2) = - h^{PP}_{GT}(\vec q^2)$,  and $h^{MM}_{T}(\vec q^2) = h^{MM}_{GT}(\vec q^2)/2$. In addition, at LO in $\chi$PT, 
$g_V(\vec q^2)=1$,
$g_A(\vec q^2)=g_A\simeq 1.27$, $g_M(\vec q\sq) = 1+\kappa_1\simeq 4.7$,  and 	$g_P(\vec q^2) =-g_A\frac{2m_N}{\vec q\sq+m_\pi\sq}$. 
The NMEs computed in the literature \cite{Hyvarinen:2015bda,Menendez:2017fdf,Barea:2015kwa,Barea} adopt the dipole parameterization of the vector and axial form factors 
\begin{equation}\label{dipole}
g_V(\vec q^2) =  \left(1 + \frac{\vec q^2}{\Lambda_V^2}\right)^{-2}\,, \qquad  g_A(\vec q^2) =  g_A \left(1 + \frac{\vec q^2}{\Lambda_A^2}\right)^{-2}\,, 
\end{equation}
with vector and axial masses $\Lambda_V = 850$ MeV and $\Lambda_A = 1040$ MeV. 
The magnetic and induced pseudoscalar form factors are then assumed to be given by
\begin{equation}\label{assumption}
g_M(\vec q^2) = (1 + \kappa_1) g_V(\vec q^2)\,, \qquad g_P(\vec q^2) = -\frac{2 m_N g_A(\vec q^2)}{\vec q^2 + m_\pi^2}\,.
\end{equation}

Using the above definitions, we express the NMEs as
\begin{eqnarray}\label{MEdef}
M_{F,(sd)}         &=&  \langle 0^+ | \sum_{m,n} h_{F,(sd)}(r) \tau^{+(m)} \tau^{+(n)}  | 0^+ \rangle\,, \nn\\
M^{ij}_{GT,(sd)} &=&  \langle 0^+ | \sum_{m,n} h^{ij}_{GT,(sd)}(r) \, \boldsigma^{(m)} \cdot \boldsigma^{(n)} \, \tau^{+(m)} \tau^{+(n)} | 0^+ \rangle\,, \nn\\
M^{ij}_{T,(sd)} &=&  \langle 0^+ | \sum_{m,n} h^{ij}_{ T,(sd)}(r) \, S^{(mn)}(\hat{\vec r}) \, \tau^{+(m)} \tau^{+(n)} | 0^+ \rangle\,, 
\end{eqnarray}
where the position-space tensor is defined by $S^{(mn)}(\hat{\vec r}) = \left( 3\,\boldsigma^{(m)} \cdot  \hat{ \vec r} \, \boldsigma^{(n)} \cdot  \hat{ \vec r} - \boldsigma^{(m)}\cdot \boldsigma^{(n)} \right)$. 
The matrix elements defined in Eq.~\eqref{MEdef} are all expected to be $\mathcal O(1)$ in the $\chi$PT power counting, apart from  $M^{MM}_{GT}$ and  $M^{MM}_{T}$ that are suppressed by 
$\mathcal O(\epsilon_\chi^2)$. However, these NMEs are proportional to the  large isovector magnetic moment of the nucleon which numerically scales as $(1+\kappa_1)\epsilon_\chi \simeq \mathcal O(1)$,
reducing the actual suppression.
 
The NME of the potential $V_9(\vec q^2)$ can be expressed in terms of the $h_{K}^{ij}$ defined in Eq.\ \eqref{smff}.
At LO in chiral EFT we can write 
\bea\label{scalarold}
V_{9}(\vec q^2) &=& -( \tau^{(1) +} \tau^{(2) + }) \, g_A\sq \,  \frac{4 G_F^2}{v} \,\bar u(k_1)P_R C\bar u^T(k_2) \nonumber \\
&&\times \Bigg\{\frac{  C^{(9)}_{\pi \pi\, \rm L}}{2m_\pi\sq}  \bigg[\left(h_{GT}^{PP}(\vec q\sq)+\frac{h_{GT}^{AP}(\vec q\sq)}{2}\right) \, \boldsigma^{(1)} \cdot \boldsigma^{(2)}+\left(h_{T}^{PP}(\vec q\sq)+\frac{h_{T}^{AP}(\vec q\sq)}{2}\right) \, S^{(12)}  \bigg]\nn\\
& & +     C_{\pi N\, \rm L}^{(9)} \left( h_{GT}^{AP}(\vec q\sq) \, \boldsigma^{(1)} \cdot \boldsigma^{(2)}+h_T^{AP}(\vec q\sq)\, S^{(12)}\right)+ C_{NN\, \rm L}^{(9)}\frac{2}{g_A^2} h_F(\vec q\sq) \Bigg\} + (L\leftrightarrow R)\, \nn \\
& & -( \tau^{(1) +} \tau^{(2) + }) \, g_A\sq \,  \frac{4 G_F^2}{v} \,\bar u(k_1)\gamma_0 \gamma_5 C\bar u^T(k_2) \nonumber \\
&&\times \bigg\{ - \frac{1}{2}\left( g_V^{\pi N}C_V+\tilde g_{V}^{\pi N}\tilde C_V^{}\right)\left( h_{GT}^{AP}(\vec q\sq) \, \boldsigma^{(1)} \cdot \boldsigma^{(2)}+h_T^{AP}(\vec q\sq)\, S^{(12)}\right)\nonumber \\
&& + \frac{2}{g_A^2}\left(g_6^{NN}C_{V}+g_7^{NN} \tilde{C}_{V}\right)  h_F(\vec q\sq) \bigg\}\,.
\eea
Eq.\ \eqref{scalarold} differs from Eq.~\eqref{scalar} only by the momentum dependence of the axial and vector form factors $g_{A,V}(\vec q^2)$, which is a subleading effect in $\chi$PT.

\section{Non-perturbative renormalization of LNV $N\!N$ couplings}\label{NonWeinberg}

\begin{figure}
\includegraphics[width=\textwidth]{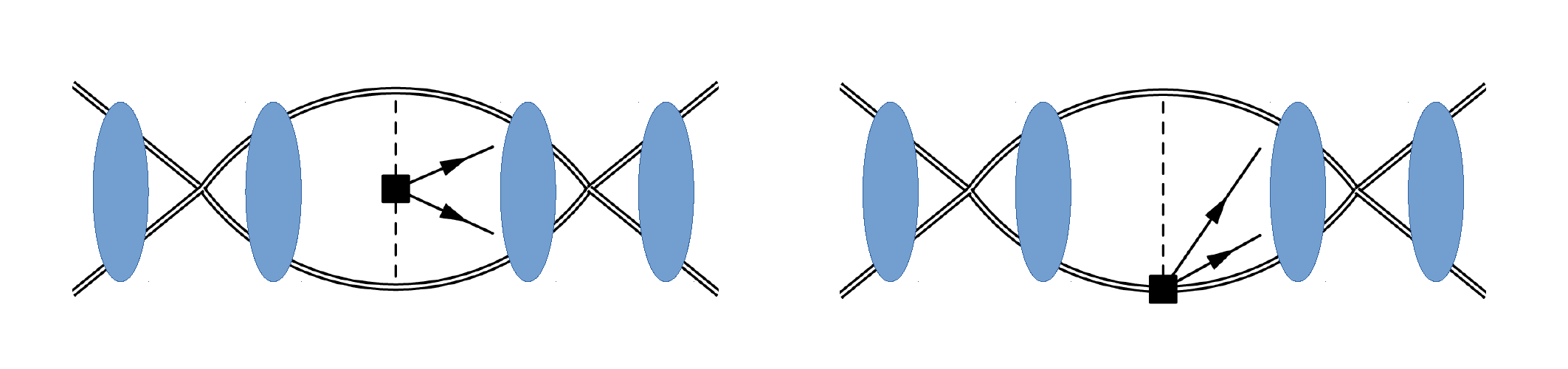}
\caption{Examples of ultraviolet divergent contributions  to the $n n \rightarrow pp e e$ scattering amplitude.  The blue ellipse denotes iterations of the Yukawa potential $V_\pi$, defined in Eq.\ \eqref{strong}.
}\label{AmpDim9}
\end{figure}

In Weinberg's power counting $\Delta L =2$ contact $N\!N$ interactions are only relevant for the scalar operator $O_1$
and the vector operators $O_{6}$--$O_9$, whose chiral properties forbid a LO non-derivative $\pi \pi e e$ coupling. 
Weinberg's counting predicts that for the scalar operators $O_{2}$--$O_5$
the $0\nu\beta\beta$ half-life is dominated by the contribution of the pionic operators in Eq.\ \eqref{eq:dim9pipi}, whose LEC  $g^{\pi\pi}_{2,3,4,5}$ are now well determined \cite{Cirigliano:2017ymo,Nicholson:2016byl,Nicholson:2018mwc}.
However, the $\pi\pi ee$ couplings  induce ultraviolet divergences in $nn \rightarrow p p e e$ scattering amplitudes
and, consequently, in the $0\nu\beta\beta$ nuclear matrix elements. These divergences are analogous to those induced by the exchange of a light-Majorana neutrino, discussed in Ref.\ \cite{Cirigliano:2018hja},
and can be absorbed by promoting the $N\!N$ counterterms to LO.

This can be seen by  repeating the analysis of Ref.\ \cite{Cirigliano:2018hja} in the presence of dim-9 LNV operators. 
At LO in  chiral EFT the strong interaction potential in the $^1S_0$ channel has a short-range and a Yukawa component
\begin{equation}\label{strong}
V_0(\vec q) = \tilde C + V_{\pi}(\vec q)\,, \qquad V_{\pi}(\vec q) = -\frac{g_A^2}{4 F_\pi^2} \frac{m_\pi^2}{\vec q^2 + m_\pi^2}\,.
\end{equation}
$\tilde C$ is the LO $N\!N$ coupling in the $^1S_0$ channel, $\tilde C \sim \{1/F_\pi^2,m_\pi^2/F_\pi^4\}$.
The LNV $n n \rightarrow p p e e$ scattering amplitude in the presence of dim-9 operators 
is obtained by sandwiching the $\Dt L=2$ potential $V_9(\vec q)$, defined in  Eq.\ \eqref{scalar}, between 
the incoming and outgoing scattering wavefunctions, $\psi_{\vec p}^{\pm}(\vec r)$,  which are obtained by solving the Schr\"odinger  equation with the potential $V_0$. The $\Dt L=2$ amplitude is thus given by
\begin{equation}\label{AL2}
\mathcal A_{\Delta L =2} =  - \int d^3 \vec r \, \psi^-_{\vec p^\prime}(\vec r)^* V_9(\vec r)  \psi^+_{\vec p} (\vec r)\,,
\end{equation}
where $V_9(\vec r)$ is the Fourier transform of the potential in Eq.\ \eqref{scalar}.

In the case of the operators $O_{2,3,4,5}$, in Weinberg's power counting $V_9$ reduces to the pion-exchange contribution
\begin{equation}\label{ALNV}
V_9(\vec r) \propto \, \boldsigma^{(1)} \cdot \boldsigma^{(2)} \, \frac{C^{(9)}_{\pi \pi\,\rm L} + C^{(9)}_{\pi \pi\,\rm R} }{6}   \frac{\left(2 - m_\pi r\right) e^{- m_\pi r}}{8 \pi r}  + \ldots\,,
\end{equation}
where the dots include the tensor component $S^{(12)}$, which does not contribute to scattering in the $^1S_0$ channel.
This potential leads to a matrix element in Eq.\ \eqref{AL2} that is regulator dependent. 
Because of the short-range component $\tilde C$, the scattering wavefunctions for the potential $V_0$  
go as $1/r$ for $r \rightarrow 0$, implying that the integral in Eq.\ \eqref{ALNV} is logarithmically divergent. 
In momentum space, the UV sensitivity arises from diagrams such as the first in Fig.\ \ref{AmpDim9},
which, when the blue ellipse is replaced by free nucleon propagators, are UV divergent.
By computing the two-loop diagrams in dimensional regularization, we  find an RGE that links the $\pi\pi$ and $NN$ couplings induced by $O_{2,3,4,5}$
\begin{eqnarray}\label{npRGE}
\frac{d}{d \log \mu} \tilde{g}^{NN}_{4,5} =  \frac{g_A^2}{4} g^{\pi\pi}_{4,5}\,,  \qquad 
\frac{d}{d \log \mu} \tilde{g}^{NN}_{2,3} = - \frac{g_A^2}{4} g^{\pi\pi}_{2,3}\,,  
\end{eqnarray}
where $g^{NN}_i = (m_N \tilde C/(4\pi))^2 \tilde g^{NN}_i$. Since the $\pi\pi$ couplings on the r.h.s.\ of these equations scale as $\Lambda^2_\chi$, 
the non-perturbative renormalization of the scattering amplitude implies $g_{i}^{NN} \sim \Lambda^2_\chi/F_\pi^2 \sim (4\pi)^2$.
Using regulators that are more suitable for few-body nuclear physics calculations does not change this conclusion   \cite{Cirigliano:2018hja}.

The $\pi\pi$ and $\pi N$ couplings induced by $O_1$ and the $\pi N$ couplings that arise from the dim-9 vector operators 
also cause divergences in the scattering amplitude that need to be absorbed by a contact operator. 
Defining
\begin{equation}
\tilde g_{1}^{NN} =  \left(\frac{4\pi}{m_N \tilde C}\right)^2  \left[ g_{1}^{NN}  + \frac{g_A^2}{2} \left(\frac{5}{6} g^{\pi\pi}_{1} - g^{\pi N}_{1}\right)\right],\,\,\, \tilde{g}_{6,7,8,9}^{NN}  = \left( \frac{4\pi}{m_N \tilde C}\right)^2  \left( g_{6,7,8,9}^{NN}  - \frac{g_A^2}{2} g_{6,7,8,9}^{\pi N} \right)\,,
\end{equation}
we find 
\begin{eqnarray}\label{RGE2}
\frac{d}{d\log \mu} \tilde g_{1}^{NN} = \frac{g_A^2}{2} m_\pi^2 \left( \frac{5}{3} g^{\pi\pi}_{1} - g^{\pi N}_{1}\right)\,, \qquad 
\frac{d}{d\log \mu} \tilde g_{6,7,8,9}^{NN} &=& - \frac{g_A^2}{2} m_\pi^2 g_{6,7,8,9}^{\pi N}\,.
\end{eqnarray}
In this case, however, the scaling of the nucleon-nucleon operators is not affected since both sides of Eq.\ \eqref{RGE2} scale as $F_\pi^2 \sim m_\pi^2$, so that the size of $g_{1,6,7,8,9}^{NN}$ agrees with Weinberg's power counting.

\section{$\Dt L=2$ potentials induced by dim-3, -6 and -7 operators} 
\label{app:neutrinopotentials567}

The chiral Lagrangians and the $0\nu\beta\beta$ potentials induced by the light neutrino Majorana mass
and by dim-6 and -7 operators are given in Refs.\ \cite{Cirigliano:2017tvr,Cirigliano:2018hja}
and \cite{Cirigliano:2017djv}, respectively.  In this appendix we recall the main ingredients, and 
include a few missing operators.

The neutrino Majorana mass in Eq.\ \eqref{LagDeltaE0} induces $0\nu\beta\beta$ by coupling to  nucleons and pions via the vector and axial weak currents.
The pion and single nucleon currents can be straightforwardly built in $\chi$PT \cite{Gasser:1983yg,Bernard:1995dp}, 
and they give rise to the long-range component of the neutrino potential, as reviewed in Refs.\ \cite{Cirigliano:2017tvr,Cirigliano:2017djv}.
The exchange of hard neutrinos induces short-range contributions to the $0\nu\beta\beta$ potential, parametrized  by the operator  $g_{\nu}^{NN}$
\begin{equation}\label{gnuNN}
\mathcal L =  2 G_F^2 V_{ud}^2 m_{\beta\beta} g_{\nu}^{NN} \, \bar N u^{\dagger} \tau^+ u N \, \bar N u^{\dagger} \tau^+ u N\, \bar e_L C \, \bar e_L^T\,.
\end{equation}
The renormalizability of $n n \rightarrow p p e e$ scattering amplitude requires the LEC $g^{NN}_{\nu}$
to scale as $g_{\nu}^{NN} \sim 1/F_\pi^2$ \cite{Cirigliano:2018hja}, implying that $g_{\nu}^{NN}$ contributes to the neutrino potential at LO.
 
The  neutrino potential in momentum space from light Majorana-neutrino exchange is 
\begin{eqnarray}\label{nupot}
V_{3}(\vec q\sq) &=& - (\tau^{(1) +} \tau^{(2) +})  (4 \,g_A\sq G_F^2V_{ud}\sq)  \, m_{\beta\beta}  \,  \bar u(k_1) P_RC \bar u^T(k_2)  \\ & & \Bigg\{ \frac{1}{\vec q^2} \left(  - \frac{1}{g_A\sq}h_F(\vec q^2)  + \boldsigma^{(1)}\cdot \boldsigma^{(2)}  \, h_{GT}(\vec q^2)   
+  S^{(12)}\,  h_T(\vec q^2) \right)  
 + \frac{2 g_\nu^{NN}}{g_A^2} h_F(\vec q^2)
\Bigg\}\,.\nn
\end{eqnarray}
The LO neutrino potential is obtained by setting the single nucleon form factors in $h_{F,GT,T}$
to their LO values. As discussed in Ref.\ \cite{Cirigliano:2017djv,Cirigliano:2017tvr}, including the $\vec q^2$ dependence of the vector and axial form factors 
accounts for a subset of the N$^{2}$LO corrections.
$g_{\nu}^{NN}$ is at the moment unknown, but can be obtained by matching chiral EFT and lattice QCD calculations of the
$n n \rightarrow p p e e$ amplitude   performed at the same kinematic point. The value of $g_{\nu}^{NN}$ thus 
depends on the details of the strong interaction potential, and its use in many-body calculations requires a consistent treatment of short-range effects 
in the strong and weak sector.

The dim-6 and dim-7 operators in Eqs.\ \eqref{lowenergy6} and \eqref{lowenergy7}
contain neutrinos, which need to be exchanged between nucleon lines to cause $0\nu\beta\beta$.
In Ref.\ \cite{Cirigliano:2017djv} we considered the long-range $\Dt L=2$ potentials induced by these operators,
which are mediated by the pion and nucleon vector, axial, scalar, pseudoscalar, and tensor currents.

The $0\nu\beta\beta$ transition operators arising from dim-6 and dim-7 can be divided in four components
\begin{equation}
V_6(\vec q^2) + V_7(\vec q^2) = V_a(\vec q^2) + V_b(\vec q^2) + V_c(\vec q^2) + V_d(\vec q^2)\,.
\end{equation}
$V_a$ is the largest piece and is dominated by the contribution of the pseudoscalar operators $C^{(6)}_{\rm SL, SR}$ 
\begin{eqnarray}\label{ps.1}
V_a(\vec q^2) =  & &  \tau^{(1) +} \tau^{(2) +} \,  \, 4g_A\sq G_F^2V_{ud} \, \left(  B  \left( C^{(6)}_{\textrm{SL}} - C^{(6)}_{\textrm{SR}} \right)  + \frac{m^2_\pi}{ v}  \left( C^{(7)}_{\textrm{VL}} - C^{(7)}_{\textrm{VR}} \right)  \right)  \frac{1}{\vec q^2} \,   \bar u(k_1) P_RC \bar u^T(k_2) \nonumber \\ 
& & \Bigg\{   \boldsigma^{(1)}\cdot \boldsigma^{(2)}  \, \left(\frac{1}{2} h^{AP}_{GT}(\vec q^2) + h^{PP}_{GT}(\vec q^2) \right)    + S^{(12)} \, \left(\frac{1}{2} h^{AP}_{T}(\vec q^2) + h^{PP}_{T}(\vec q^2) \right)  \Bigg\}\,\,.
\end{eqnarray}
Note that this potential 	goes like $1/(\vec q\sq)\sq$ for large $\vec q\sq $, ensuring that this potential does not induce divergences similar to those produced by $V_9$ as discussed in Sect.\ \ref{NonWeinberg}.

The long-range neutrino exchange  contributions of  the tensor operator, $C^{(6)}_{\rm T}$, and the  
right- and left-handed currents operators, $C^{(6)}_{\rm VR}$ and $C^{(6)}_{\rm VL}$, 
are chirally suppressed. For this reason,  short-range effects, arising from the exchange of neutrinos with momentum $\sim \Lambda_\chi$, could become important,
even in Weinberg's power counting. These effects can be accounted for by building hadronic operators with the same transformation properties as the tensor products 
of one non-standard current and the SM weak interaction. Carrying out the lepton contractions while neglecting the lepton momenta,  we find that the operators induced by $C^{(6)}_{\rm T, VL, VR}$ transform like the following non-local terms
\begin{eqnarray}
C^{(6)}_{\rm T} : & &  \quad  \Big(  \bar u_L \sigma^{\mu\nu} d_R \times   \partial_\mu \left( \bar u_L \gamma_\nu d_L \right) 
			             - \partial_\mu \left(    \bar u_L \sigma^{\mu\nu} d_R \right) \times   \bar u_L \gamma_\nu d_L  \Big)  \, \bar e_L C \bar e_L^T \,, \nn \\
C^{(6)}_{\rm VL} : & &  \quad  \Big( \bar u_L \gamma^\mu d_L \times  \partial^\rho ( \bar u_L \gamma^\nu d_L) -\partial^{\rho} (\bar u_L \gamma^\mu d_L)\times \bar u_L \gamma^\nu d_L   \Big)  \, L_{\mu\nu\rho\sigma}   \, \bar e_R \gamma^\sigma C \bar e_L^T \,, \nn \\
C^{(6)}_{\rm VR} : & &  \quad   \Big( \bar u_R \gamma^\mu d_R \times  \partial^\rho ( \bar u_L \gamma^\nu d_L) -\partial^{\rho} (\bar u_R \gamma^\mu d_R)\times \bar u_L \gamma^\nu d_L   \Big) \,  L_{\mu\nu\rho\sigma}   \bar e_R \gamma^\sigma C \bar e_L^T\,,
  \label{trans}
\end{eqnarray}
where $L_{\mu\nu\rho\sigma} = g_{\mu \rho} g_{\nu \sigma} + g_{\nu \rho} g_{\mu \sigma} - g_{\mu \nu} g_{\rho \sigma}  + i \varepsilon_{\mu\rho\nu\sigma} $.
From Eq.~\eqref{trans} we see that $C^{(6)}_{\rm T}$ induces operators that transform as a Lorentz scalar, but have the same chiral properties as the vector operators $O_{6,7,8,9}$ in Eq.~\eqref{LagVec}.
$C^{(6)}_{\rm VL}$  generates operators that are Lorentz vectors, but with the same chiral properties as $O_1$.  Similarly, short-range operators induced by  
$C^{(6)}_{\rm VR}$ are Lorentz vectors with the same chiral properties as $O_4$, but with negative parity.  

The form of short-range operators follows from the Lorentz, chiral, and parity transformation properties of the operators in Eq.\  \eqref{trans}.  We construct
\begin{eqnarray}\label{NNLag}
\mathcal L_{\pi\pi} &=&   \frac{F_0^2V_{ud}}{2 m_N}\,  \partial_\mu \pi^- \partial^\mu \pi^-{g^{\pi\pi}_{\rm T}C^{(6)}_{\rm T}}\frac{\bar e_L C \bar e_L^T}{v^4}\,,\nn \\ 
\mathcal L_{\pi N}  &=&  \sqrt{2}g_AV_{ud} F_0\, \bar p S\cdot (\partial \pi^-) n  \, \left (  \frac{g^{\pi N}_{\rm VL}}{m_N} C^{(6)}_{\rm VL}  v^\mu  \, \frac{\bar e \g_\mu \g_5 C\bar e^T}{v^4} 
+ \frac{g^{\pi N}_{\rm T}}{m_N} C^{(6)}_{\rm T} \frac{\bar e_L C \bar e_L^T}{v^4} 
\right)\,,  \nn \\
\vL_{NN} &=&   \bar p    n\, \bar p   n \left(   \frac{g^{N N}_{\rm VL}V_{ud}}{m_N} C^{(6)}_{\rm VL}  v^\mu  \, \frac{\bar e \g_\mu \g_5 C\bar e^T}{v^4} 
+ \frac{g^{N N}_{\rm T}V_{ud}}{m_N} C^{(6)}_{\rm T}\,  \frac{\bar e_L C \bar e_L^T}{v^4} \right)\,.
\end{eqnarray}
By NDA, the LECs  $g^{\pi\pi}_{\rm T}$ and $g^{\pi N, NN}_{\rm T, VL}$ are $\mathcal O(1)$.
The hadronic component of the operators in Eq.\ \eqref{NNLag} is parity-even. $C^{(6)}_{\rm VR}$, on the other hand, induces short-range operators that are parity-odd, if one neglects the lepton momenta, 
and thus does not contribute to Eq.\ \eqref{NNLag} or $0^+\to 0^+$ transitions.

From the discussion in Ref.\ \cite{Cirigliano:2017djv} and Eq.\ \eqref{NNLag}, it follows that the $0\nu\beta\beta$ potential induced by the tensor operator $C^{(6)}_{\rm T}$ is
\begin{eqnarray}\label{tensor}
V_b(\vec q^2) &=&  4 g_A\sq \tau^{(1) +} \tau^{(2) + } \, 2  G_F^2V_{ud} \,   m_N C^{(6)}_{\textrm{T}} \frac{1}{\vec q^2}\, \bar u(k_1) \,P_R C \bar u^T(k_2)\, 
\Bigg\{  \frac{ ( g^\prime_{T}(\vec q^2) - g_{\rm  T}^{NN} ) g_V(\vec q^2)}{g_A\sq} \frac{\vec q^2}{m_N^2} \nonumber \\
& & - 4\frac{ g_T(\vec q^2) }{g_M(\vec q\sq)}   \left( h_{GT}^{MM}(\vec q\sq)\boldsigma^{(1)} \cdot \boldsigma^{(2)} 
+ h_{T}^{MM}(\vec q\sq)S^{(12)} \right)  \nn 
\\ & & 
+  \frac{\vec q^2}{4m_N^2}  g^{\pi N}_{\rm T}\left(h^{AP}_{GT}(\vec q^2) {\boldsigma^{(1)} \cdot \boldsigma^{(2)} }+ h^{AP}_{T}(\vec q^2){S^{(12)} } \right) \nn\\
&&+\frac{\vec q^2}{4 m_N^2}  g_{ \rm T}^{\pi\pi} \left(  h^{PP}_{GT}(\vec q^2){\boldsigma^{(1)} \cdot \boldsigma^{(2)} }  + h^{PP}_{T}(\vec q^2){S^{(12)} } \right)
\Bigg\}\,. 
\end{eqnarray}
The contributions from $g^{NN,\, \pi N}_{\rm T}$ and $g^{\pi\pi}_{\rm T}$ were neglected in Ref.\ \cite{Cirigliano:2017djv}.

The leading potential induced by $C^{(6)}_{\rm VL}$ is 
\bea\label{eq:vl}
V_c(\vec q^2) &=&  
\, 4 \,  g_A\sq \tau^{(1) +} \tau^{(2) + }  G_F^2V_{ud}  \, m_N \,C^{(6)}_{\textrm{VL}}\,   \frac{1}{\vec q\sq} \bar u(k_1)  \gamma_0 \g_5 C \bar u^T(k_2) \nn \\
& & \Bigg\{ 
\,2  \frac{g_A(\vec q\sq)}{g_M(\vec q\sq)}\left(h_{GT}^{MM}(\vec q\sq)\,\boldsigma^{(1)} \cdot \boldsigma^{(2)}  +h_T^{MM}(\vec q\sq)S^{12} \right)
  \\ & & 
+ \frac{\vec q^2}{m_N^2} \left(- \frac{2}{g_A^2}g^{NN}_{\rm VL}  h_{F}(\vec q^2)
+ \frac{1}{2} g^{\pi N}_{\rm VL}\left( h^{AP}_{GT}(\vec q^2){\boldsigma^{(1)} \cdot \boldsigma^{(2)} } + h^{AP}_{T}(\vec q^2){S^{(12)}}\right)\right)
\Bigg\}\,,\nn
\eea
The short-range terms proportional to $g^{NN}_{\rm VL}$ and $g^{\pi N}_{\rm VL}$ are formally of the same order 
as first contribution in Eq.\ \eqref{eq:vl}, mediated by the nucleon magnetic moment. 

Finally, $C^{(6)}_{\rm VL, VR}$ induce contributions proportional to the lepton momenta. In the case of $C^{(6)}_{\rm VL}$, these contributions are suppressed by $\epsilon_\chi$ with respect to Eq.\ \eqref{eq:vl}, while they are the leading-order contribution in the case of $C^{(6)}_{\rm VR}$.
Also in this case additional contact interactions need to be introduced in order to make the amplitude regulator independent. These short-range interactions take the form,
\bea
\vL_{NN} &=&   \bar p    n\, \bar p   n \,v^\mu v^\nu i  \, \frac{\bar e \g_\mu (\partial_\nu -\overleftarrow\partial_\nu) C\bar e^T}{v^4} 
V_{ud}\left(   \frac{g^{E}_{\rm VL}}{m_\pi\sq} C^{(6)}_{\rm VL} 
+\frac{g^{E}_{\rm VR}}{m_\pi\sq} C^{(6)}_{\rm VR} \right)\,\nn\\
&&+  \bar p    n\, \bar p   n \,\, \frac{m_e \,\bar e C\bar e^T}{v^4} 
V_{ud} \left(   \frac{g^{m_e}_{\rm VL}}{m_\pi\sq} C^{(6)}_{\rm VL} 
+\frac{g^{m_e}_{\rm VR}}{m_\pi\sq} C^{(6)}_{\rm VR} \right)\,,
\eea
where $g_{VL,VR}^{E,m_e}$ are LECs of $\Or(1)$.
The combination of these contributions and the long-range terms give,
\bea\label{eq:vlvr}
V_d(\vec q^2) &=&  
\tau^{(1) +} \tau^{(2) + } \,  \,  g_A\sq G_F^2V_{ud} \,   \frac{1}{\vec q\sq} \Bigg\{  \bar u(k_1)  \gamma_0  C\bar u^T(k_2) \, (k_1^0 - k_2^0) \left[C^{(6)}_{\textrm{VL}}\, M^{(1)}_L+C^{(6)}_{\textrm{VR}}\, M^{(1)}_R\right] \nn \\
&& +2 m_e \, \bar u(k_1) C   \bar u^T(k_2) \, \left[C^{(6)}_{\textrm{VL}}\, M^{(2)}_L+C^{(6)}_{\textrm{VR}}\, M^{(2)}_R\right]
\Bigg\}\,\,,
\eea
where
\bea
M_{L,R}^{(1)} &=& - \frac{4}{3} \frac{g_V\sq }{g_A\sq}h_F(\vec q\sq)\mp \frac{8}{9} h_{GT}^{AA}(\vec q^2)\,  \boldsigma^{(1)} \cdot \boldsigma^{(2)}  \mp \frac{4}{9}\, h_{T}^{AA}(\vec q^2) \, S^{(12)}-\frac{8}{g_A\sq }\frac{\vec q\sq}{m_\pi\sq}h_F(\vec q\sq) g_{VL,VR}^{E}
\,\,,\nn\\
M_{L,R}^{(2)} &=&   \frac{1}{3} \frac{g_V\sq }{g_A\sq}h_F(\vec q\sq)\mp \left(\frac{1}{9}  h_{GT}^{AA}+h_{GT}^{AP}(\vec q^2)+h_{GT}^{PP}(\vec q^2)\right)\,  \boldsigma^{(1)} \cdot \boldsigma^{(2)}\nn\\  &\pm& \left(\frac{4}{9}\,  h_{T}^{AA}-h_{T}^{AP}(\vec q^2)-h_{T}^{PP}(\vec q^2)\right) \, S^{(12)}-\frac{4}{g_A\sq }\frac{\vec q\sq}{m_\pi\sq}h_F(\vec q\sq) g_{VL,VR}^{m_e}\,\,.
\eea

\section{Renormalization group equations}\label{App:RG}
In this appendix we briefly discuss the running between the electroweak scale and $\Lambda_\chi$, which follows from the RGEs  discussed in section \ref{sec:2}. The solutions of the RGEs for the  dimension-six operators are 
\bea
C_{\rm SL(SR)}^{(6)}(\mu) &=& \eta^{-3 C_F/\bt_0}C_{\rm SL(SR)}^{(6)}(m_W)\,,\qquad C_{\rm T}^{(6)}(\mu)=\eta^{C_F/\bt_0}C_{\rm T}^{(6)}(m_W)\,\,.\nn
\eea
where $\eta\equiv \frac{\al_s(m_W)}{\al_s(\mu)}$ and $\bt_0=\frac{1}{3}(11 N_c-2n_f)$, with $n_f$ the number of active flavors.

From the RGEs of the dimension-nine operators we obtain
\bea
C_1^{(9)}(\mu) &=& \eta^{3(1-1/N_c)/\bt_0} C_1^{(9)}(m_W)\,,\nn\\
\bma C_4^{(9)}(\mu)\\C_5^{(9)}(\mu)\ema  &=& 
\bma
\eta ^{3/(N_c \bt_0)} &0\\ \frac{1}{N_c}\left[\eta ^{-6C_F/ \bt_0}-\eta ^{3/(N_c \bt_0)}\right] & \eta ^{-6C_F/ \bt_0}
\ema
\cdot \bma C_4^{(9)}(m_W)\\C_5^{(9)}(m_W)\ema  \,,
\eea
with $C_F = (N_c\sq-1)/(2N_c)$, while the solutions to the RGEs for  $C_{2,3}^{(9)}$ and $C_{6,7}^{(9)}$ are more conveniently expressed as
\bea
\bma C_2^{(9)}(\mu)\\C_3^{(9)}(\mu)\ema  &=& R_{23}\cdot \bma
\eta^{\g_1^{(23)}/(2\bt_0)}&0\\ 0& \eta^{\g_2^{(23)}/(2\bt_0)}
\ema\cdot R^{-1}_{23}
\cdot \bma C_2^{(9)}(m_W)\\C_3^{(9)}(m_W)\ema  \,,\nn\\
\bma C_6^{(9)}(\mu)\\C_7^{(9)}(\mu)\ema  &=& R_{67}\cdot \bma
\eta^{\g_1^{(67)}/(2\bt_0)}&0\\ 0& \eta^{\g_2^{(67)}/(2\bt_0)}
\ema\cdot R_{67}^{-1}
\cdot \bma C_6^{(9)}(m_W)\\C_7^{(9)}(m_W)\ema  \,,
\eea
where $\g_{1,2}^{(i)}$ are the eigenvalues of the anomalous-dimension matrices and $R_i$ the matrices that diagonalize them. They can be written as
\bea
\g_1^{(23)} &=& 2\left(3-N_c+\frac{1}{N_c}- \bar\g\right),\qquad \g_2^{(23)} = 2\left(3-N_c+\frac{1}{N_c}+ \bar\g\right)\,, \nn\\
\g_1^{(67)} &=& 3-N_c+\frac{1}{N_c}- \bar\g\,,\qquad \g_2^{(67)} =3-N_c+\frac{1}{N_c}+ \bar\g\,,\nn\\
R_{23} &=& \bma \frac{-2N_c\sq+N_c-N_c \bar \g}{2(N_c-2)} &  \frac{-2N_c\sq+N_c+N_c \bar \g}{2(N_c-2)}\\ 1&1\ema\,,
\nn\\
R_{67} &=& \bma \frac{2N_c\sq-N_c-2+4/N_c+N_c \bar \g}{4(2-N_c)} &  \frac{2N_c\sq-N_c-2+4/N_c-N_c \bar \g}{4(2-N_c)}\\ 1&1\ema\,,
\eea
where $\bar \g =\sqrt{4N_c\sq-11+16/N_c\sq}$. As discussed in Section \ref{sec:2}, the remaining couplings either do not run, or have an equivalent evolution as the above couplings.

Using these solutions and taking into account the bottom-quark threshold, we obtain the following numerical relation for the dim-6 operators
\bea
C_{\rm SL(SR)}^{(6)}(\Lambda_\chi) &=& 1.5\, C_{\rm SL(SR)}^{(6)}(m_W)\,,\qquad C_{\rm T}^{(6)}(\Lambda_\chi)=0.87\,C_{\rm T}^{(6)}(m_W)\,\,.\nn
\eea
where we used $\Lambda_\chi\simeq 2$ GeV. Following the same procedure, we find for the dim-9 operators 
\bea
C_1^{(9)}(\Lambda_\chi) &=& 0.82\, C_1^{(9)}(m_W)\,,\nn\\
\bma C_2^{(9)}(\Lambda_\chi)\\C_3^{(9)}(\Lambda_\chi)\ema  &=& \bma
1.6 & -0.28\\
-0.07 & 0.59
\ema
\cdot \bma C_2^{(9)}(m_W)\\C_3^{(9)}(m_W)\ema\,, \nn\\
\bma C_4^{(9)}(\Lambda_\chi)\\C_5^{(9)}(\Lambda_\chi)\ema  &=& \bma
0.90 & 0\\
0.45& 2.3
\ema
\cdot \bma C_4^{(9)}(m_W)\\C_5^{(9)}(m_W)\ema\,, \nn\\
\bma C_6^{(9)}(\Lambda_\chi)\\C_7^{(9)}(\Lambda_\chi)\ema  &=& \bma
1.3 & -0.15\\
-0.07 &0.78
\ema
\cdot \bma C_6^{(9)}(m_W)\\C_7^{(9)}(m_W)\ema\,. \nn\\
\eea

\section{The left-right model}\label{App:LR}
In this appendix we discuss a few more details of the mLRSM that are needed to obtain the matching conditions in Eq.\ \eqref{eq:LREFT}. The main ingredients that are missing from Section \ref{sec:LR} are the masses of the heavy BSM fields. After the right-handed triplet field, $\Dt_R$, acquires a vev, several fields obtain masses proportional to $v_R$ and therefore become heavy. These fields are the right-handed gauge bosons, $W_R$ and $Z_R$, the right-handed neutrinos, $\nu_R$, and the left-handed triplet fields, $\Dt_L$. In addition, part of the right-handed triplet fields, namely $\dt_R^{++}$ and Re$\,\dt_R^0$, as well as part of $\phi$ become heavy. The $\dt_R^\pm$ and Im$\,\dt_R^0$ fields are `eaten' by the  $W_R$ and $Z_R$ gauge bosons, while the remaining part of $\phi$ stays light and can be interpreted as the SM Higgs doublet.

To integrate out these heavy fields we first have to rotate to the mass basis. To do so we   write $\phi$ in terms of two $SU(2)_L$ doublets, $\phi\equiv (\phi_1,\, \phi_2)$, and use the  relation to the mass eigenstates
\bea\label{HiggsRot}
\bma \tilde \phi_1\\\phi_2 \ema = \frac{1}{\sqrt{1+\xi\sq}}\bma -1 & \xi e^{-i\al}\\\xi e^{i\al} & 1\ema \bma \vp\\ \vp_H\ema \,,
\eea
which follows from the Higgs potential, see e.g.\ Refs.\ \cite{Duka:1999uc,Kiers:2005gh,Zhang:2007da,Dekens:2014ina}.
Here $\tilde \phi_i= i\tau_2 \phi_i^*$ and $\vp$ is the SM Higgs doublet, while $\vp_H$ obtains a mass of $\Or(v_R)$. 
The contributions to $\Dt L=2$  interactions arise from exchanges of $\nu_R$, $W_R$, $\dt^{++}_R$, and $\Dt_L$, so that the relevant masses are the following
\bea
m_{W_R}=\frac{g_R}{\sqrt{2}}v_R\,,\qquad m_{\nu_R} = \sqrt{2}v_R U M_R^\dagger U^T\,,
\qquad m_{\Dt_L}\sq = \frac{1}{2}(\rho_3-2\rho_1)v_R\sq\,,\qquad m_{\Dt_R}\sq = 2\rho_2 v_R\sq\,,
\eea
where $g_R$ is the $SU(2)_R$ gauge coupling, $U$ is the rotation matrix between the mass and flavor bases, $\nu_R^{\rm (mass)} = U\nu_R$, and 
 $\rho_{1,2,3}$ are parameters in the Higgs potential \cite{Zhang:2007da}. 
After integrating out these heavy fields, the induced $\Dt L=2$ interactions will depend on the masses of these fields. In our analysis we  choose values for $m_{\Dt_R}$, $v_R$ (which determines $m_{W_R}$), and $m_{\nu_R}={\rm diag}(m_{\nu_{R_1}},\,m_{\nu_{R_2}},\,m_{\nu_{R_3}})$.  
 
 Similarly,  integrating out $\Dt_L$ gives a contribution to $ \mathcal C^{(5)}$ which depends on  its mass. Explicitly, we have
 \bea\label{eq:dim5DeltaL}
 \mathcal C^{(5)}(\Dt_L) = -\frac{1}{m_{\Dt_L}\sq}\frac{v_R}{\sqrt{2}(1+\xi\sq)}\left[\xi \bt_1 e^{i(\dt_{\bt_1}+\al)}+\bt_2 e^{i\dt_{\bt_2}}+\xi\sq\bt_3 e^{i(\dt_{\bt_3}+2\al)}\right]M_L\,,
 \eea
 where $\bt_{i}$ and $\dt_{\bt_i}$ are parameters of the Higgs potential in the notation of Ref.\ \cite{Dekens:2014ina}.
However, we do not have to choose values for $\bt_i$, $\dt_{\bt_i}$, and $m_{\Dt_L}$ as we can trade the combination of these parameters for the vev of $\Dt_L$, $v_L$. The reason for this is that the  $\bt_i$ couplings represent the terms in the Higgs potential that are linear in $\Dt_L$. The same terms also appear in the minimum condition, $\frac{\partial V_H}{\partial \Delta_L}\big|_{S_i=\langle S_i\rangle} = 0$, where $S_i$ stand for  the scalars in the mLRSM. The only other (non-negligible) terms appearing in this condition are the terms quadratic in $\Dt_L$, which  determine the mass term, since $m_{\Dt_L}\sq\propto \frac{\partial\sq V_H}{\partial \Dt_L\sq}\big|_{S_i=\langle S_i\rangle}$. This leads to the following relation 
\bea
\frac{v_Le^{i\theta_L}}{v^2}=-\frac{v_R}{2(1+\xi\sq)}\frac{\xi \bt_1 e^{i(\dt_{\bt_1}+\al)}+\bt_2 e^{i\dt_{\bt_2}}+\xi\sq\bt_3 e^{i(\dt_{\bt_3}+2\al)}}{m_{\Dt_L}\sq}\,,
\eea
where we used $v^2 = \ka\sq(1+\xi\sq)$, and the right-hand side features the same combination that appear in Eq.\ \eqref{eq:dim5DeltaL}. After substituting the above relation in Eq.\ \eqref{eq:dim5DeltaL}, one  obtains the second term in Eq.\ \eqref{eq:LRdim5}.

Finally, as mentioned, we assume a charge-conjugation symmetry which acts on the fields as
\bea
C:\quad Q_{L,R}\to Q_{R,L}^c\,,\qquad  L_{L,R}\to L_{R,L}^c\,,\qquad \phi\to \phi^T\,,\qquad \Delta_{L,R}\to \Dt_{R,L}^*\,.
\eea
This symmetry  enforces a relation between the gauge couplings and Majorana mass matrices, $g_R = g_L$ and $M_L = M_R^\dagger$,  ensures the Dirac mass matrices are symmetric, and leads to  the relation in Eq.\ \eqref{Eq:MD}.

\section{Comparison with the literature}\label{comparison-to-lit}
The dim-9 operators of Eq.\ \eqref{eq:Lag} have been discussed in the context of \NLDBD\ in the literature before \cite{Gonzalez:2015ady,Faessler:1996ph,Pas:2000vn,Prezeau:2003xn,Graesser:2016bpz}. Since not all references apply the same framework based on $\chi$PT that is employed here, there are several instances were our results are significantly different with respect to older findings. Here we give a brief overview of the most significant discrepancies and focus on the comparison with Refs.\ \cite{Pas:2000vn,Gonzalez:2015ady}.

Starting with the basis of operators, the dim-9 $SU(3)_c \times U(1)_{\rm em}$-invariant four-quark two-lepton operators were first cataloged in Ref.\ \cite{Pas:2000vn},
while Refs.\ \cite{Graesser:2016bpz,Prezeau:2003xn} removed several redundancies. 
Compared to Ref.\ \cite{Pas:2000vn} our basis does not include tensor operators of the form $(\bar u_L \sigma^{\mu\nu} d_R)\times$ $(\bar u_R \sigma_{\mu\nu} d_L)$, which vanish due to the identity $\left[\sigma_{\mu\nu}(1\pm \g_5)\right]_{ij} \left[\sigma^{\mu\nu}(1\mp \g_5)\right]_{kl} = 0$. In contrast, we do include the operators $O_3$, $O_5$, and $O_7$, which are related to color-octet interactions of the form $\sim (\bar u\Gamma t^ad )\, (\bar u\Gamma t^ad )$. Such color-octet terms were neglected in Refs.\ \cite{Prezeau:2003xn,Graf:2018ozy}, however, the matrix elements of these operators are expected to be of the same size as their color-singlet cousins, $O_2$, $O_4$, and $O_6$. This is borne out by the lattice QCD results for the $\pi\pi$ LECs in Table \ref{Tab:LECs}. Furthermore, compared to Ref.\ \cite{Graesser:2016bpz} we find additional $SU(3)_c\times U(1)_{\rm em}$-invariant operators that can be induced by operators of \textoverline{dim-9}. These operators are not induced by the $SU(2)_L$-invariant two-lepton four-quark operators considered in Ref.\ \cite{Graesser:2016bpz}, but by additional $SU(3)_c\times SU(2)_L\times U(1)_Y$-invariant operators involving gauge- and Higgs-bosons. Finally, we note that although we expect the  $SU(3)_c\times U(1)_{\rm em}$-invariant operators included here to capture the dominant effects in most LNV scenarios, we have not included all \textoverline{dim-9} operators as a complete basis is currently unavailable.

To perform a quantitative comparison with Ref.\ \cite{Pas:2000vn}, we note that their Wilson coefficients are related to the ones in Eq.\ \eqref{eq:Lag} by
\begin{eqnarray}
C^{(9)}_{\rm 1L} &=& \frac{2 v}{m_N} \epsilon^{LLL}_3\,, \quad  C^{(9)}_{\rm 2L} = \frac{2 v}{m_N} \left( \epsilon^{LLR}_1 - 4\epsilon^{LLR}_2 \right)\,, \quad
C^{(9)}_{\rm 3L} =- \frac{16 v}{m_N}  \epsilon^{LLR}_2\,, \nn \\ 
C^{(9)\,\prime}_{\rm 1L} &=& \frac{2 v}{m_N} \epsilon^{RRR}_3\,, \quad  C^{(9)\, \prime}_{\rm 2L} = \frac{2 v}{m_N} \left( \epsilon^{RRR}_1 - 4 \epsilon^{RRR}_2 \right)\,, \quad
C^{(9)\,\prime}_{\rm 3L} =- \frac{16 v}{m_N}  \epsilon^{RRR}_2\,, \nn \\ 
C^{(9)}_{\rm 4L} &=& \frac{2 v}{m_N} \epsilon_3^{LRR}\,, \quad C^{(9)}_{\rm 5L} = -\frac{v}{m_N} \epsilon_1^{LRR}\,,\nn\\
C_6^{(9)} &=& \frac{v}{m_N} \left[\epsilon_5^{LRR}-i\frac{N_c+2}{N_c}\epsilon_4^{LRR}\right],\quad C_7^{(9)} = -4i\frac{v}{m_N} \epsilon_4^{LRR}\,,\nn\\
C_8^{(9)} &=& \frac{v}{m_N} \left[\epsilon_5^{LLR}-i\frac{N_c+2}{N_c}\epsilon_4^{LLR}\right],\quad C_9^{(9)} = -4i\frac{v}{m_N} \epsilon_4^{LLR}\,.
\label{supermatch}
\end{eqnarray}
The operators with subscript R are obtained by sending $L$ to $R$ in the last index of the $\epsilon$ coefficients in Eq.\ \eqref{supermatch}.

After using the translation of  Eq.\ \eqref{supermatch}, our results for the RGEs of the dimension-nine operators do not agree with parts of the literature. In particular, the RGEs for the vector operators, $C_{6,7,8,9}^{(9)}$, in Eq.\ \eqref{RGE9vector} differ from those derived in Ref.~\cite{Gonzalez:2015ady}. 
However, since the running gives rise to $\Or(1)$ effects, as illustrated by Table \ref{tab:limits}, these discrepancies have a relatively mild effect on resulting constraints.

More significant discrepancies arise from the different approaches to the hadronization of the quark-level operators. In this work we perform the matching of the quark-level operators to interactions in the chiral Lagrangian, in which the non-perturbative nature of QCD is captured by the LECs in Table \ref{Tab:LECs}. Instead, Ref.\ \cite{Pas:2000vn} relies on factorization to estimate the required matrix elements and neglects the effects of $\pi\pi$ and $\pi N$ operators, effectively matching onto $NN$ interactions only. This approach gives comparable results to the ones obtained here for several couplings,  namely $C^{(9)}_{1,6,7,8,9}$. The reason for this is that the corresponding operators induce  $NN$ interactions at leading order, which do not require enhancement with respect to Weinberg's counting (i.e.\ they follow NDA). 
However, there are several operators, namely $C^{(9)}_{2,3,4,5}$, for which the leading contributions come from $\pi\pi$ interactions and the $NN$ terms are enhanced compared to Weinberg's counting. Both of these effects are missed by the approach of Ref.\ \cite{Pas:2000vn} and lead to underestimates of the contributions to {{\em the amplitude}} of \NLDBD\ by  a factor of $\Or(16\pi\sq)$ compared  to the results found here.

These significant differences can be made explicit by comparing the different estimates for the LECs. As mentioned Ref.\ \cite{Pas:2000vn} does not include $\pi N$ and $\pi \pi$ interactions, effectively setting the corresponding LECs to zero. The factorization estimates for the $NN$ interactions give in our notation
\bea
\label{NNsuper}
g^{NN}_{1} &=& \frac{1}{4} ( 3 g_A^2 + 1)\,, \quad g^{NN}_{2} = \frac{g_S^2}{4}\,, \quad g^{NN}_{3} =  \frac{3 g_T^2}{4} - \frac{g_S^2}{8}\,, \nn \\
g_{4}^{NN} &=& \frac{1}{4} ( 3 g_A^2 - 1)\,, \quad g_{5}^{NN} = \frac{g_S^2}{2}\,,\nn\\
\left| g_{6,8}^{NN}\right| &=& \frac{1}{2\sqrt{2}} g_Ag_T \,,\quad \left|g_{7,9}^{NN}\right| = \frac{1}{8\sqrt{2}}\left|\frac{N_c+2}{N_c} g_S-3 g_T\right|\,,
\eea
where $g_S$ and $g_T$ are the isovector scalar and tensor densities. Ref.\ \cite{Pas:2000vn} employs the MIT-bag model estimates $g_S = 0.48$ and $g_T = 1.38$ of Ref.\ \cite{Adler:1975he}, for which  lattice QCD determinations are  available, see Table \ref{Tab:LECs}.

Note that the values used in Ref.\ \cite{Pas:2000vn}, together with Eq.\ \eqref{NNsuper}, imply $\Or(1)$ values, or slightly smaller, for the $g_{i}^{NN}$. As mentioned above, this leads to similar, though slightly weaker, limits for the $C^{(9)}_{1,6,7,8,9}$ operators. Instead, our estimates for $g_{2,3,4,5}^{NN}$ are a factor $\Or(16\pi^2)$ larger than those used in Ref.\ \cite{Pas:2000vn}. In addition, the $\pi\pi$ interactions, for which the LECs are known, give rise to contributions at the same order in these cases. As a result, we find $\Or(10-100)$ stronger limits on $C^{(9)}_{2,4,5}$ compared to  Ref.\ \cite{Pas:2000vn}.
These different values for the LECs  also led to the conclusion in Ref.\ \cite{Gonzalez:2015ady} that the limits on the scalar and tensor couplings $\epsilon_{1,2}^{LLR}$, equivalent to our $C_{\rm 2L,3L}^{(9)}$, are significantly affected by the RGEs. The reason for this is that Refs.\  \cite{Pas:2000vn,Gonzalez:2015ady} employ a matrix element for $\epsilon_{1}^{LLR}$ which is  smaller than that for $\epsilon_{2}^{LLR}$ by a factor of $\Or(100)$. This implies that the largest contribution from $\epsilon_{1}^{LLR}$ actually arises from the mixing  into $\epsilon_{2}^{LLR}$, making the running a large effect. 
However, since the two operators transform in the same way under the symmetries of QCD we estimate $O_{\rm 2L,3L}^{(9)}$ to have matrix elements of similar size, so that the running is not more than an $\Or(1)$ effect (as can be seen from Table \ref{tab:limits}). We stress that this expectation is borne out by the lattice QCD determinations of the $\pi\pi$ matrix elements in Table \ref{Tab:LECs}.

This significant discrepancy in the limits is in part due to the fact that  Ref.\ \cite{Pas:2000vn} neglects the $\pi\pi$ interactions. The importance of these terms was already pointed out in Refs.\  \cite{Faessler:1996ph} and \cite{Graesser:2016bpz,Prezeau:2003xn}. In fact, the latter two references  work within the framework of $\chi$PT, and therefore obtain very similar results as we do here. The main difference between this work and  Refs.\ \cite{Graesser:2016bpz,Prezeau:2003xn} is that the latter references assumed Weinberg's power counting and therefore did not take into account the enhancement of $g_{2,3,4,5}^{NN}$. However, the consequences of this assumption are much less severe since the $\pi\pi$ interactions contribute at the same order, and the LECs $g_{2,3,4,5}^{NN}$ are currently unknown.
An additional difference  is that  Ref.\ \cite{Prezeau:2003xn} neglected the color-mixed operators $O_3$, $O_5$, $O_7$ and $O_9$ 
and thus considered an incomplete low-energy basis \cite{Graesser:2016bpz}.

Finally, we stress that although the \NLDBD\ expressions in this work depend on several unknown LECs, the $\chi$PT framework allows for systematic improvement upon the current situation. In particular, the LECs which can  only be estimated through NDA and RG arguments at the moment, namely $g^{\pi N}_i$ and $g_{i}^{NN}$ in Table \ref{Tab:LECs}, are in principle amenable to lattice QCD determinations. Such lattice QCD calculations would provide control over the hadronic uncertainties of the contributions to \NLDBD\, and are a necessary ingredient to obtain reliable limits on the dim-9 operators.

\bibliographystyle{h-physrev3} 
\bibliography{bibliography}

\end{document}